\documentclass[acmlarge,authorversion,nonacm]{acmart}
\AtBeginDocument{%
  \providecommand\BibTeX{{%
    \normalfont B\kern-0.5em{\scshape i\kern-0.25em b}\kern-0.8em\TeX}}}

\setcopyright{acmcopyright}
\copyrightyear{2021}
\acmYear{2021}
\acmDOI{10.1145/1122445.1122456}

\acmJournal{IMWUT}
\acmVolume{0}
\acmNumber{0}
\acmArticle{0}
\acmMonth{0}

\usepackage{wrapfig}
\usepackage{multirow}
\usepackage{enumitem}
\usepackage{comment}
\usepackage{longtable}
\usepackage{subfig}

\begin{document}

\title{Deep Learning Techniques for Compressive Sensing-Based Reconstruction and Inference -- A Ubiquitous Systems Perspective}

\author{Alina L. Machidon}
 \email{alina.machidon@fri.uni-lj.si}
 \affiliation{%
   \institution{Faculty of Computer and Information Science, University of Ljubljana}
   \country{Slovenia}
 }
 \author{Veljko Pejović}
 \email{veljko.pejovic@fri.uni-lj.si}
 \affiliation{%
   \institution{Faculty of Computer and Information Science, University of Ljubljana}
   \country{Slovenia}
 }


\begin{abstract}
Compressive sensing (CS) is a mathematically elegant tool for reducing the sampling rate, potentially bringing context-awareness to a wider range of devices. Nevertheless, practical issues with the sampling and reconstruction algorithms prevent further proliferation of CS in real world domains, especially among heterogeneous ubiquitous devices. Deep learning (DL) naturally complements CS for adapting the sampling matrix, reconstructing the signal, and learning form the compressed samples. While the CS-DL integration has received substantial research interest recently, it has not yet been thoroughly surveyed, nor has the light been shed on practical issues towards bringing the CS-DL to real world implementations in the ubicomp domain. In this paper we identify main possible ways in which CS and DL can interplay, extract key ideas for making CS-DL efficient, identify major trends in CS-DL research space, and derive guidelines for future evolution of CS-DL within the ubicomp domain.

\end{abstract}



\maketitle

\section{Introduction}
\label{sec:introduction}

Apollo 11's landing on the Moon represents an unrivaled feat of engineering and a milestone achievement in the history of mankind. The trip to the Moon, however, was anything but smooth sailing. During Apollo's lunar descent the guidance computer sounded a 1202 alarm indicating a computing overflow. The alarm was caused by a sensing system -- a rendezvous radar -- rapidly requesting computing cycles. At this segment of the approach, the processor was already at the 85\% utilization and the additional requests for processing sensor data lead to failed scheduling requirements. Luckily, the Apollo ground team and the astronaut Buzz Aldrin successfully diagnosed the design limitation and safely made it to the history books. The anecdote, however, remains a very important lesson on both the possibilities and the challenges of sensing-computing integration.

The capacity to sense the world around them represents the key affordance of computing devices nowadays found under popular terms, such as the Internet of Things (IoT), cyber-physical systems, and ubiquitous computing. Devices falling under the above categories are characterized by the inclusion of sensors that convert physical measures to electric signals and processing hardware that enables these signals to be interpreted and actioned upon. This integration of computing and sensing was essential for achieving such early milestones as the already mentioned Moon landing and the first humanoid robot in late 1960s~\cite{kato1973development}. The inclusion of wireless transmission capabilities has further exacerbated the growth of sensing systems and wireless sensors networks soon emerged as an essential tool in areas as diverse as environmental monitoring for agriculture, structure monitoring, and military applications. Yet, the moment when the first iPhone hit the shelves in 2008 marked the start of a new era of sensor-computing integration, the one in which compact mobile computing devices equipped with an array of sensors will soon outnumber people on this planet. The ever-increasing range of sensors available on mobile devices, nowadays including multiple cameras, microphones, accelerometers, gyroscopes, location, light and temperature sensors, and wireless gesture recognition sensors, to name a few, enabled revolutionary new services to be provided by the mobiles. Consequently, devices such as smartphones and smartwatches have became indispensable for tasks as diverse as real-time navigation, voice recognition and interaction, online translation, and health-related physiological signal monitoring. Furthermore, in parallel to the rise of smartphone and wearable sensing, the advances in embedded computing have propelled the use of sensors in systems ranging from unmanned areal vehicles (UAVs, i.e. ``drones'') over factory robots and IoT devices, to self-driving cars. Consequently, the spectrum of applications relying on integrated sensing and computing has already expanded to cover anything from wildfire monitoring to vacuuming a home, and with the increase in number of deployed devices showing no signs of waning, we can safely assume that the true potential of sensing-computing integration is yet to be observed. 

Widening the range and the sophistication of applications led to increased expectations with respect to the complexity of computation that should be supported by sensing-computing systems. High-level inferences from sensor data are possible, but only if complex data processing, in the form of data filtering, extracting features from raw sensor data, and machine learning modeling are supported. Recent advancements in the area of deep learning pushed the complexity of the requested computation even further. Models containing millions of parameters can process high-resolution camera images and detect objects present in these images; samples taken from on-body accelerometers can, with the help of long short-term memory (LSTM) models infer a wearer's physical activity. The Apollo 11 anecdote, however, reminds us that robust and reliable sensing applications can be supported only if the information about the environment, i.e. sensor samples, do not overwhelm the processing part of the pipeline. Unfortunately, resource-constrained sensing devices, such as various microcontrollers, wearables, IoT devices and similar devices predominant in today's ubiquitous computing deployments often cannot cope with the processing burden of modern sensing and machine learning. Equipped with multicore CPUs and GPUs and relatively large storage space, modern smartphones can run certain deep learning models. However, even these high-end devices support only sporadically used and carefully optimized models processing sensor data streams of relatively modest sampling rates~\cite{wu2019machine}. 

Supporting advanced inference applications, while reducing the sampling and processing burden appears unsolvable at the first glance. According to the Nyquist theorem, signals can be reliably captured only if sensor sampling rates are as twice as high as the highest frequency expressed in such signals. Real-world phenomena, however, tend to be fast changing. It appears that  objects can be recognized only in images of sufficient resolution, wireless radars can detect small movements only if the signal is sampled millions of times per second, and a user's physical activity can be detected only if an on-body accelerometer is sampled at sub-Hz frequency. 

In 2000s a series of papers by Donoho, Candes, Romberg, and Tao~\cite{candes2006robust,donoho2006compressed,candes2006quantitative,candes2006stable}, investigated the properties of signals that can be successfully reconstructed even if sampled at rates lower than prescribed by the Nyquist theorem. Signal sparsity, i.e. a property that in a certain projection most of the signal's components are zero and incoherence, a condition of low correlation between the acquisition domain and the sparsity domain, are needed in order for the signal to be fully preserved with only about \(K log (N/K)\) samples taken from the original $K$-sparse $N$-dimensional signal. \textit{Compressive sensing (CS)}\footnote{also known as \textit{compressed sensing} and \textit{sparse sampling}} involves drastically reduced sampling rate, administered when the above conditions are fulfilled, and subsequent signal reconstruction via signal processing, often reduced to finding solutions to underdetermined linear systems. In the last two decades, CS has been successfully demonstrated in image processing, wireless ranging, and numerous other domains.

The benefits of reduced sampling rate does not, however, come for free, as CS remains challenging to implement. First, not all reduced-rate sampling is equal. Compression rates interplay with the properties of the input to impact the ability to successfully reconstruct the signal from limited samples. Furthermore, while the early implementations focused on random sampling, recent research advances demonstrate the utility of carefully crafted reduced sampling strategies~\cite{wang2019deterministic}. Second, reconstructing the signal from sparse measurements, in theory, requires solving an NP hard problem of finding non-zero signal components. In practice, the problem is solved through iterative solutions that are nevertheless computationally demanding. Finally, high-level inference, not signal reconstruction, is often the key aim of sensing. Thus, it is essential to construct a full machine learning pipeline that natively supports CS. 


\subsection{Towards Deep Learning-Supported Compressive Sensing}

Interestingly, the above challenges have a joint characteristic -- for a specific use case, a suitable sampling strategy, a reliable reconstruction algorithm, and a highly accurate inference pipeline can be \textit{learned} from the collected sensor data and data labels. Machine learning methods, therefore, naturally augment CS. The last decade witnessed an explosive growth of one particular such technique -- deep learning (DL). DL harnesses a large number of adaptable parameters to construct a non-linear model explaining high-level abstractions through low-level data. The availability of suitable computing hardware, such as GPUs and TPUs, even in embedded and mobile computers, made DL truly pervasive.

\begin{wrapfigure}[16]{r}{0.55\textwidth}
 \begin{center}
    \includegraphics[width=0.5\textwidth]{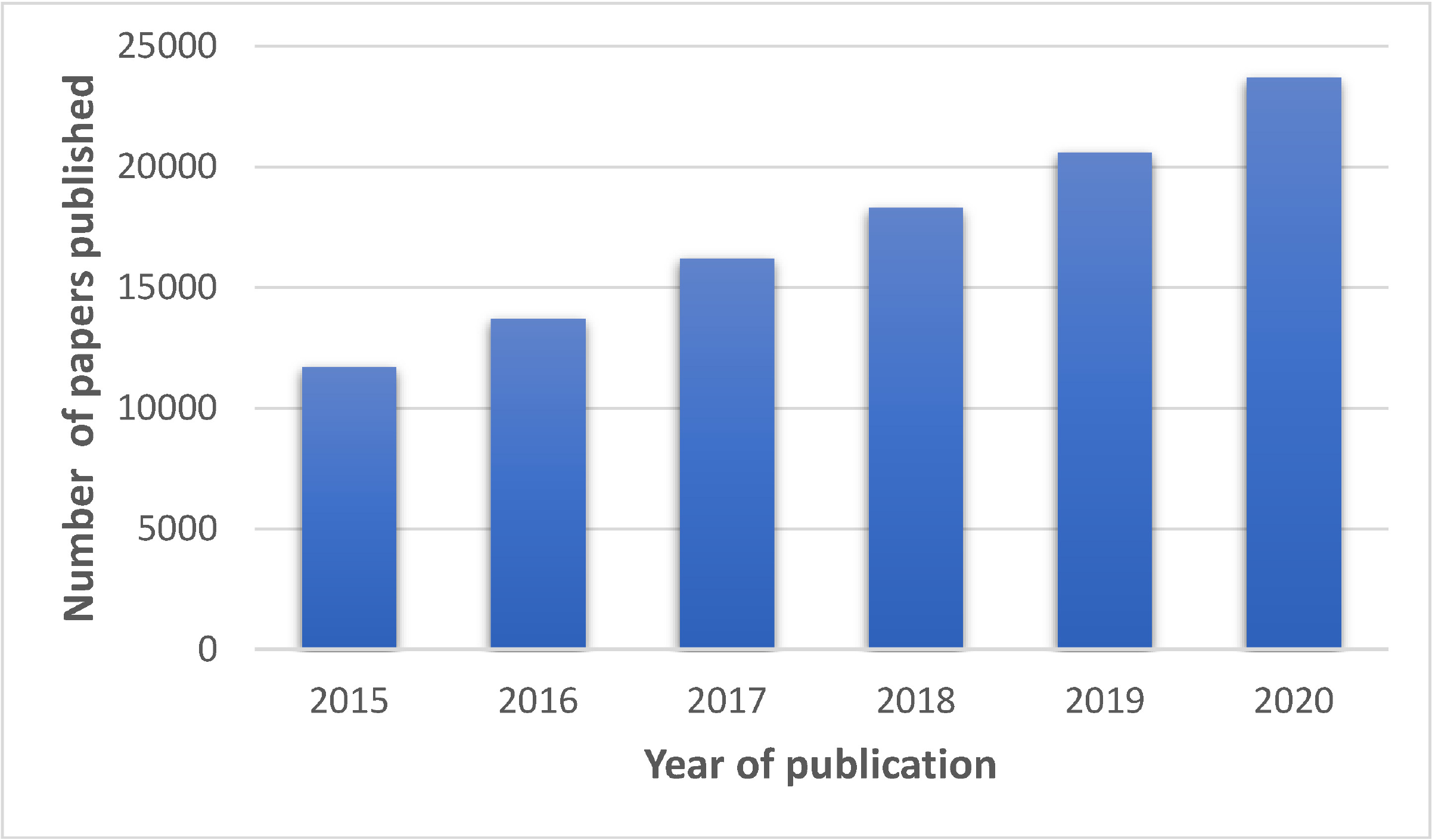}
  \end{center}
  \vspace{-14pt}
  \caption{Number of scientific papers on the topic CS-DL published between 2015 and 2020 (Data from Google Scholar using the search terms ``deep learning'' and ``compressed sensing'')}
  \label{fig:CSDLpapers}
\end{wrapfigure}

\begin{figure}
\centering
\includegraphics[width=0.85\textwidth]{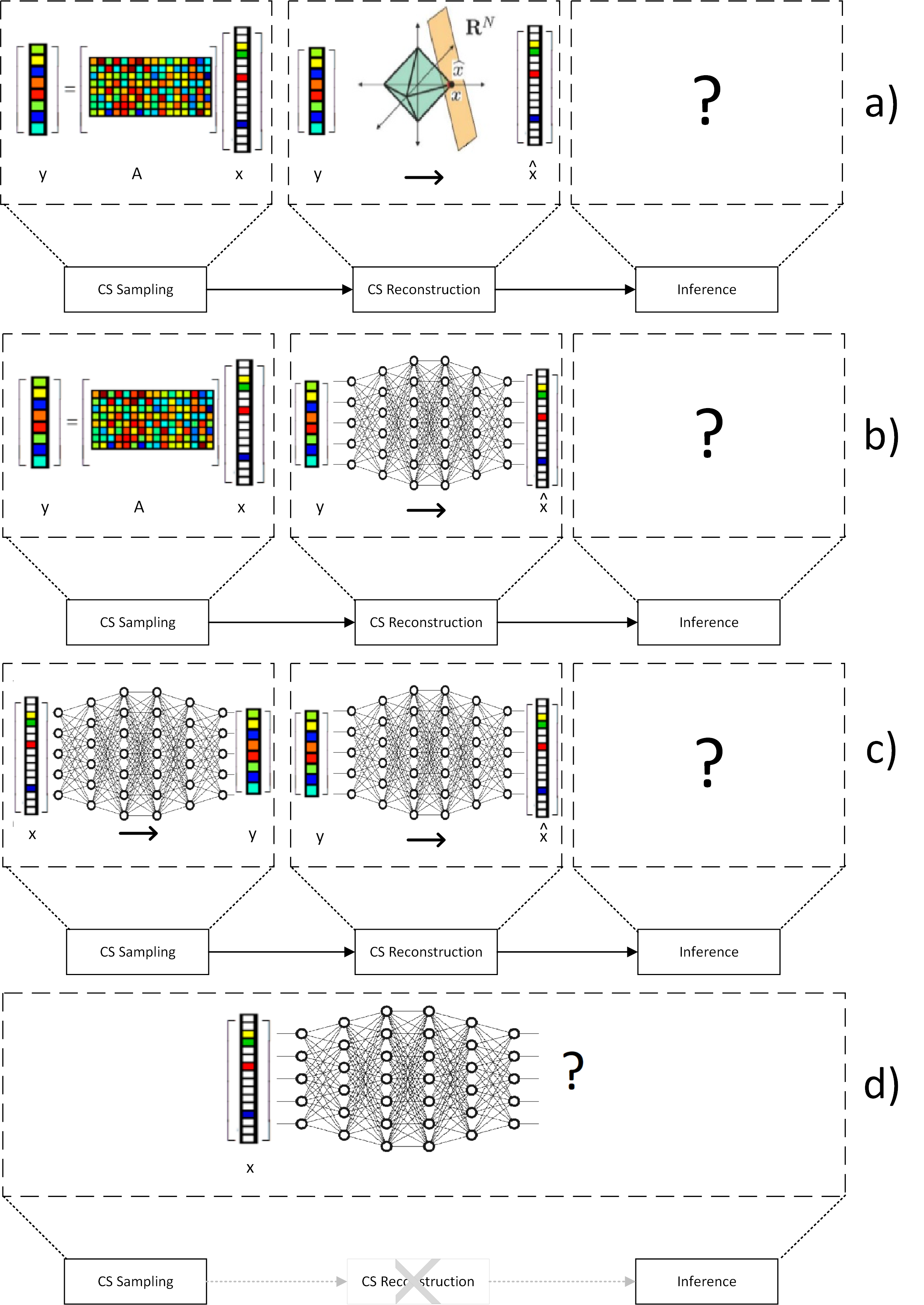}
\vspace{-5pt}
\caption{CS approaches: a) conventional CS; b) DL for CS reconstruction; c) DL for CS sampling and reconstruction; d) DL for CS direct inference}
\label{fig:CSDL}
\end{figure}

Figure~\ref{fig:CSDL} depicts possible ways deep learning and compressive sensing can interplay. A common CS pipeline~(a) consists of the reduced-frequency sampling, followed by signal reconstruction, from which high-level inferences are made, if needed. Iterative signal reconstruction algorithms, in particular, tend to represent a weak point in the pipeline due to their temporal requirements. Yet, with sufficient CS-sampled and original signal data available, a rather fast-to-query DL reconstruction model can be built. Using DL for signal reconstruction (b) has, thus been successfully demonstrated in numerous domains~\cite{kulkarni2016reconnet, iliadis2016deepbinarymask, wang2016accelerating, schlemper2017deep, han2018sparse, kim2020compressive}. The performance of such solutions not only matched, but also significantly exceeded the performance of the standard reconstruction approaches as additional signal structure can be captured by generative DL models~\cite{polania2017exploiting, ma2017extraction, nguyenlearning, grover2019uncertainty}. The sampling matrix, too can be adapted to a problem at hand thanks to DL (c). Often an encoder-like structure is trained to guide the sampling in the most efficient manner. Finally, as the reconstructed signal is usually used as a basis for high-level inference, DL allows us to short-circuit the expensive reconstruction step and train a network that provides high-level inferences directly from the CS-sampled data (d).

The above-identified natural links between efficient sampling embodied in CS and powerful learning enabled by DL have recently been recognized by the research community. Tremendous research interest that has spurted is evident in a steady increase in the number of scientific papers published on the topic yearly from 2015 to 2020 (see Figure~\ref{fig:CSDLpapers}). The exploration is far from theoretical with a range of application fields, including magnetic resonance imaging (MRI), ultra wideband (UWB) radar ranging, human activity recognition, and numerous other domains benefiting from the CS-DL integration. 

Especially promising is the revolutionising potential of CS-DL integration in the area of ubiquitous computing. Here, devices are characterized by wide heterogeneity, and limited computational and battery resources. Besides the general benefits of accelerating signal reconstruction, fine-tuning the sampling matrix, and improving the high-level inference, in the ubiquitous computing domain CS-DL integration can reduce the energy, storage, and processing requirements. As a result, there is a potential for previously prohibitively demanding continuous sensing and inference to finally be realized in certain domains. Furthermore, graceful degradation in end-result quality can be supported with the CS-DL pipeline. Through reduced CS sampling and reduced accuracy DL inference we can, in a controlled manner, trade result quality for resource usage. This allows seamless adaptation of the sensing-inference pipeline, so that complex applications can run on low-resource devices, albeit with limited accuracy. Finally, mobile devices operate in dynamic environments. With the environment so can vary the signal properties (i.e. its sparsity) as well as a user's requirements with respect to the calculated result quality. A seamless adaptation offered through CS sampling rate adjustment, for example, would enable CS-DL solutions to perfectly match the operating conditions and a user's requirements with the minimal use of a device's resources.

\subsection{Survey Rationale and Organization}

In this survey we explore deep learning-supported compressive sensing, an area that despite the rapidly gaining popularity (see Figure~\ref{fig:CSDLpapers}) has not been systematically studied before. 

Compressive sensing is the main subject of a few monographs (e.g.~\cite{eldar2012compressed}) that introduce the topic from the historical perspective, present key theoretical postulates, and discuss open challenges in the area of signal sampling and processing. These almost exclusively focus on the mathematical issues and remain implementation platform-oblivious. The volume by Khosravy et al. investigates the use of CS in healthcare~\cite{khosravy_dey_duque_2020} and considers applications, such as electrocardiogram (ECG) and electroencephalogram (EEG) sensing, that are, with the expansion of wearable computing capabilities, highly relevant for the ubicomp domain. Yet, the book focuses on the sensing part and does not discuss potential integration with deep learning. Focused survey papers cover compressive sensing applications in different domains, for instance wireless sensor networks (WSNs)~\cite{wimalajeewa2017application,yin2016compressive}, IoT~\cite{djelouat2018compressive}, and EEG processing~\cite{gurve2020trends}, to name a few. Our survey is orthogonal to these, as we do not focus on a particular domain, but rather merge contributions made in different fields on the common ground of using both the DL techniques and the CS concepts. We opt for this approach with the intention to inform future research in any domain, by providing researchers and practitioners with a toolbox of CS-DL implementations that may be transplanted to any domain. Finally, to the best of our knowledge, ours is the first paper that provides in-detailed consideration of practical issues of CS-DL integration on \textit{ubiquitous} computing devices. It is our hope that the guidelines on efficient implementations presented in this survey will serve as a foundation for practical realizations of deep learning-supported ubiquitous compressive sensing systems of the future. 

An extremely popular research area, deep learning is not short of textbooks, surveys, and tutorial on the topic. For the completeness sake, here we list books by the most notable authorities on the subject~\cite{goodfellow2016deep,aggarwal2018neural,bishop1995neural}. From a range of DL survey papers, we find~\cite{cheng2017survey} and~\cite{choudhary2020comprehensive}  particularly relevant. These surveys focus on techniques that prune, compress, quantize, and in other manners reshape powerful DL models so that they can be ran on resource-constrained devices. The expansion of context awareness and artificial intelligence over a wide range of devices and applications is our guiding vision. Reduced sampling rates afforded by CS, together with powerful but computationally light inference enabled by (compressed) DL pave the path towards our vision.

Guided with the above vision, the objective of this paper is threefold: 
\begin{itemize}
    \item Present CS fundamentals, DL opportunities, and ubiquitous computing constraints to previously disparate research communities, with a goal of opening discussion and collaboration across the discipline borders;
    \item Examine a range of research efforts and consolidate DL-based CS advances. In particular, the paper identified signal sampling, reconstruction, and high-level inference as a major categorization of the reviewed work;
    \item Identify major trends in CS-DL research space and derive guidelines for future evolution of CS-DL within the ubicomp domain.
\end{itemize}




The remaining sections of this paper are organized as follows. Section~\ref{sec:primer} provides a crash-course overview of compressive sensing, highlighting the necessary conditions for successful sampling, as well as main signal recovery approaches, with an emphasis on algorithm efficiency. The section also briefly revisits key deep learning concepts relevant for CS-DL integration. Section~\ref{sec:reconstruction} discusses CS-DL efforts in the area of CS signal reconstruction. The advantages and limitations of different DL flavors with regard to the CS reconstruction challenges are exposed and analyzed. A unique affordance of DL is the ability to infer high-level concepts directly from CS-sampled signal without intermediate signal reconstruction. These, so-called, reconstruction-free approaches are presented in Section~\ref{sec:inference}. Section~\ref{sec:systems} critically examines the constraints that CS-DL implementations have to face once deployed in real-world ubiquitous computing environments, presents key lessons learned from different domains, and discusses potential future research in the CS-DL for ubiqutious computing. Finally, we conclude the paper with the key guidelines in Section~\ref{sec:conclusions}.

\section{Deep Learning and Compressive Sensing Primers}
\label{sec:primer}

In the first part of this section we aim to bring the area of compressive sensing closer to ubiquitous computing researchers and practitioners. Yet, we focus on the bare essentials and points relevant for real-world implementation of CS and direct an interested reader to more in-depth presentations of the subject, such as~\cite{eldar2012compressed}. Throughout the section we identify possibilities for deep learning (DL) within the CS domain and in the later part of the section we succinctly overview key DL concepts relevant for the remainder of the paper.

\subsection{Compressive Sensing Primer}

Despite the growth of both the range and the quality of sensors embedded in ubiquitous computing devices, sampling rates remain a limiting factor in a device's ability to reliably capture and reconstruct information about its environment. Analog-digital conversion is a costly operation, both in terms of the hardware needed for high-resolution sampling as well as in terms of the energy needed for frequent sampling. Furthermore, the amount of data produced by high-rate sampling can quickly overwhelm limited computational and storage resources of ubiquitous computing devices. 

\subsubsection{Theoretical basis}

Classical signal processing is based on a notion that signals can be modeled as vectors in a vector space. Nyquist sampling rate requirement was derived based on an assumption that signals may exist anywhere within the given vector space~\cite{nyquist2002certain} and requires that the sampling frequency is at least as twice as high as the highest frequency component present in the signal. In reality, however, signals exhibit structure that constrains them to only a subset of possible vectors in a certain geometry, i.e many real-world signals are naturally sparse in a certain vector space. Furthermore, if not truly sparse, or even if subject to noise, many signals are compressible -- i.e. a limited number of the strongest signal components tends to uniquely describe the signal. 

The above observations represent the intuition behind compressive sensing (CS). The idea of joint sensing and compression was theoretically developed in~\cite{candes2006robust,donoho2006compressed} by Emmanuel Cand\`{e}s, Justin Romberg, Terence Tao and David Donoho who also formalized the conditions need for efficient reconstruction of a signal from a significantly reduced number of samples, compared to the number of samples assumed under the Nyquist sampling criterion.


The main idea behind CS is that having a $k$-sparse signal vector $x \in \mathcal{R}^N$, an accurate reconstruction of $x$ can be obtained from the undersampled measurements taken by a sampling:

\[y = Ax \in \mathcal{R}^M \]

Where the $M\times N $ matrix $A$ is called \textit{the sensing matrix} (also\textit{the projection matrix}) and is used for sampling the signal. Since $M<N$, this linear system is typically under-determined, permitting an infinite number of solutions. However, according to the CS theory, due to the sparsity of $x$, the exact reconstruction is possible, by finding the sparsest signal among all those that produce the measurement $y$, through a norm minimization approach:


\begin{equation*}
\begin{aligned}
& {\text{minimize}}
& &  \|x\|_0\ \\
& \text{subject to}
& & Ax=y
\end{aligned}
\end{equation*}

where $\|\cdot\|_0$ is the \textit{$l_0$}-norm and denotes the number of non-zero components in $x$, i.e. the sparsity of $x$.

However, this is generally an NP-hard problem. An alternative solution is to minimize the \textit{$l_1$} norm, i.e. the sum of the absolute value of vector components:


\begin{equation*}
\begin{aligned}
& {\text{minimize}}
& &  \|x\|_1\ \\
& \text{subject to}
& & Ax=y
\end{aligned}
\end{equation*}

Since the \textit{$l_1$}-norm minimization-guided solution can be found through iterative tractable algorithms, if the solution to the \textit{$l_0$}-norm and \textit{$l_1$}-norm conditioned systems were the same, the CS-sensed signal could be perfectly reconstructed from $m$ measurements in a reasonable amount of time. 

Cand\`{e}s and Tao show that indeed in certain situations solutions to both problems are equivalent. The condition for the above to hold is that the signal's $x$ sparsity $k$ is sufficiently low with respect to the number of measurements $m$ and that the matrix $A$ to satisfies certain properties. One of these properties is the so-called Null Space Property (NSP), a necessary and sufficient condition for guarantying the recovery, requiring that every null space vector of the sensing matrix is not concentrating its energy on any set of entries. A stronger condition on the sensing matrix is the Restricted Isometry Property (RIP), which states that $A$ must behave like an almost orthonormal system, but only for sparse input vectors. More formally, matrix $A$ satisfies $k$-RIP with restricted isometry constant $\delta_{k}$ if for every $k$-sparse vector $x$: 

\begin{equation*}
(1-\delta_{k})\|x\|_2^2\leq\|Ax\|_2^2\leq(1+\delta_{k})\|x\|_2^2
\end{equation*}

A uniquely optimal solution for the the \textit{$l_0$}-norm and \textit{$l_1$}-norm conditioned signal reconstruction systems exists, if $\delta_{2k}+\delta_{3k}<1$. Intuitively, sampling matrices satisfying this condition preserve signal size and therefore do not distort the measured signal, so the reconstruction is accurate. 

In practice, however, assessing RIP is computationally difficult. Another related condition, easier to check, is the incoherence, or the low coherence, meaning than the columns of the sensing matrix should be almost mutually orthogonal. Additional mathematical properties that the sensing matrix should satisfy for ensuring the stability of the reconstruction have been introduced in \cite{donoho2006compressed} and \cite{donoho2009counting}. From the real world applications perspective, the sensing matrix should ideally fulfill constraints such as: optimal reconstruction performance (high accuracy), optimal sensing (minimum number of measurements needed), low complexity, fast computation and easy and efficient implementation on hardware.  Random sensing matrices such as Gaussian, Bernoulli or uniform were shown to satisfy the RIP and are widely used in compressed sensing. However, due to the unstructured nature of these matrices, difficulties arise for hardware implementation. Random matrices are more costly in terms of storage, processing time and energy consumption, when compared to matrices that following a given structure, which reduces the randomness, memory storage, and processing time. In an application running in a resource constrained environment, such as those for wearable wireless body sensors, this is of great importance. 
Another class of measurement matrices is represented by the deterministic matrices which are constructed in a deterministic manner to have a small mutual coherence or satisfy the RIP. Examples of matrices of this type are the Circulant and Toeplitz matrices. Although this category of matrices was shown to speed up the recovery process, they lack the flexibility in term of size and are not universal because they are designed for specific applications.

Finally, real-world data often hide structures, beyond sparsity, that can be exploited. By learning these regularities from the data through the sensing matrix design, the salient information in the data can be preserved, leading to better reconstruction quality.
In addition, most of the existing recovery algorithms, rely on the prior knowledge of the degree of sparsity of the signal to optimal tune their parameters. Difficulties might arise especially when the signal is very large or it exhibits great variations in terms of sparsity. In these cases, the conventional CS approaches cannot perform the optimal reconstruction, but a data-driven approach could learn important signal features and design the signal sampling to work optimally, even for varying sparsity levels.

\subsubsection{Signal reconstruction approaches}

The effective and efficient recovery of the original signal from the compressed one is crucial for CS to become a practical tool. Approaches to signal reconstruction from the undersampled measurements can be roughly grouped into convex optimization, greedy, and non-convex minimization algorithms. 

The class of \textit{convex optimization} algorithms solve a convex optimization problem, i.e. the \textit{$l_1$}-norm conditioned system, through linear programming to obtain the reconstruction. Basis Pursuit (BP)~\cite{chen2001atomic}, Basis Pursuit De-Noising (BPDN)~\cite{chen2001atomic}, Least Absolute Shrinkage and Selection Operator (LASSO)~\cite{tibshirani1996regression}, the Iterative Shrinkage/Thresholding Algorithm (ISTA)~\cite{beck2009fast}, the Alternating Direction Method of Multipliers (ADMM) ~\cite{boyd2011distributed}, the Dantzig Selector ~\cite{candes2007dantzig}, the Gradient Projection for Sparse Representation ~\cite{figueiredo2007gradient}, the Gradient Descent ~\cite{garg2009gradient} or Total Variation Denoising ~\cite{rudin1992nonlinear} are some examples of such algorithms. The most common convex optimization algorithm is Basis Pursuit, which attempts to generate a sparse decomposition of the signal into an ``optimal'' superposition of dictionary elements, where optimal means having the smallest \textit{$l_1$} norm of coefficients among all such decompositions. One of the advantages of these algorithms is the small number of measurements required for achieving an exact reconstruction. However, their computational complexity is high -- BP for example runs in $\mathcal{O}(n^3)$, where $n$ is the original vector space dimensionality. 

\textit{The greedy approach} for CS involves a step-by-step method for finding the support set of the sparse signal by iteratively adding nonzero components, and reconstructing the signal by using the constrained least-squares estimation method. Examples of these algorithms include Matching Pursuit (MP)~\cite{mallat1993matching}, Orthogonal Matching Pursuit (OMP)~\cite{pati1993orthogonal}, Compressive Sampling Matching Pursuit (CoSAMP)~\cite{needell2009cosamp}, Stagewise Orthogonal Matching Pursuit (StOMP)~\cite{donoho2012sparse}, Generalized Orthogonal Matching Pursuit (GOMP)~\cite{wang2012generalized}, to name a few. Matching Pursuit (MP) is one of simplest algorithms in this class. Its principle is to approximately decompose a signal into a weighted sum of atoms. Starting with an initial approximation and a residual, at each iteration the atom that best correlates with the residual is added to the current approximation and the residual is updated, and this process is repeated until the signal is satisfactorily decomposed.
The thresholding algorithms such as  Iterative Hard Thresholding (IHT)~\cite{blumensath2009iterative} are another subclass of greedy algorithms that work by thresholding at each iteration, setting all the largest component (in magnitude) of a vector to zero and leaving the remaining components untouched. These algorithms are characterized by low implementation cost and improved running time, but their performance is highly constrained by the level of sparsity of the signal and in general, their theoretical performance guarantees remain weak.

\textit{The Non Convex Minimization Algorithms} implies the use of non convex minimization algorithms to solve the convex optimization problem of signal recovery.
Algorithms such as the Bayesian compressive sensing (BCS) algorithm~\cite{ji2007bayesian}, the Focal Underdetermined System Solution (FOCUSS) ~\cite{gorodnitsky1997sparse} and the Iterative Reweighted Least Squares (IRLS) algorithm~\cite{chartrand2008iteratively} fall into this category. Among those algorithms the Bayesian subclass of algorithms harnesses the probabilistic information of the sparse signal and noise for the signal support recovery. These methods can show better recovery probability than the convex optimization algorithms, but are restricted to a certain distribution of the sparse signal. Furthermore, these algorithms may or may not produce the desired solutions.

A joint property of all the above reconstruction algorithms is their high computational cost dictated by the iterative calculations these algorithms rely on. 
In order to achieve the goal of incorporating CS in real-world ubiquitous computing applications, fast and efficient reconstruction algorithms need to be developed. Deep Learning emerged as an unexpected candidate for such an algorithm. While DL usually requires substantial computing resources and significant memory space for hundreds of thousands of network parameters, the most burdensome computation is still performed during the algorithm training phase and the inference time remains lower than the time needed for running the conventional iterative reconstruction algorithms. 

\subsubsection{From samples to inferences}

In the last 15 years, compressive sensing transitioned from a theoretical concept to a practical tool. One of the first demonstrations of CS was the so called \textit{one-pixel camera}. Here, a digital micromirror array is used to optically calculate linear projections of the scene onto pseudorandom binary patterns. A single detection element, i.e. a single ``pixel'' is then evaluated a sufficient number of times, and from these measurements the original image can be reconstructed. This early success set the trajectory of practical CS, which is nowadays used for a range of image analysis tasks. Thus, CS is used for fMRI image sampling and reconstruction \cite{chiew2018recovering,li2020deep},
ultrasound images \cite{kruizinga2017compressive,lorintiu2015compressed,kim2020signal}, remote sensing images \cite{zhao2020adaptive,wang2017compressive}, and other image-related domains.
WSN data sub-sampling and recovery represents another significant area for CS \cite{xiao2019compressed,liu2017efficient,qie2020wireless} as does speech compression and reconstruction \cite{shawky2017efficient,al2017combined}.

Characteristic for most of the signal processing applications listed above, is the departure from signal reconstruction, as the key aim of CS, towards higher level inference, i.e. detection, classification, or prediction.
In such cases, signal reconstruction may represent an unnecessary step and may even be counterproductive. Studies~\cite{davenport2007smashed,calderbank2009compressed,lohit2015reconstruction} have shown theoretical guarantees that the compressed measurements can be directly used for inference problems without performing the recovery step. Therefore, an increasing number of research works aims to solve the problem of learning directly from sparse samples. This is yet another area where neural networks shine. Driven by the fact is it possible to learn directly in the compressed domain, and that neural networks have a inherent ability to extract hidden features, deep learning can be successfully used to infer from the compressed measurements. 


\subsection{Augmenting Compressive Sensing with Deep Learning -- Approaches in Nutshell}


Deep learning usually refers to neural networks with more than two layers, and defines a set of algorithms that attempt to model high-level abstractions in data using multiple layers and non-linear transformations. The capability of deep neural network to solve complex problems and learn hierarchical features from various types of data, made them widely used today for tasks such as classification, regression or recognition, in a range of domains. In recent years, innovative deep learning-based paradigms have been introduced at an increasing rate, overcoming the limitations of the traditional neural networks. From the vast deep learning toolbox, here we present a few key paradigms that have been proven useful in combination with compressive sensing.



\subsubsection{Dense Networks}

Feedforward neural networks (FFNNs) represent the most straightforward NN architecture where neurons remain connected in a cycle-free topology. The most commonly used subclass of FFNNs is a Multilayer Perceptron (MLP), where multiple layers of neurons are interconnected in a feedforward manner, usually with non-linear activation functions (sigmoid, rectifier linear unit -- ReLU, or tanh) following. As such, MLP represents a universal function approximator and is, with sufficient training data, an attractive option for solving general regression and classification problems. Stacked MLPs are also regraded as fully connected (FC) networks or dense networks, because of their very dense web of weights, as a result of connecting every neuron in one layer to the neurons in the next layer.

\subsubsection{Convolutional Neural Networks}

Harnessing the structure that is inherent for many inference problems, especially in the area of image analysis, Convolutional Neural Networks (CNNs) use smaller groups of neurons (filters) to detect patterns in the data. Multiple layers of neurons enable CNNs to detect progressively more complex patterns, i.e. from edges in an image, to kinds of animals in the image. CNNs borrow their name from  convolution that is performed between the input data and the filters. 
A common CNN architecture of consists of an input and an output layer, and several hidden layers, among which a series of convolutional layers, pooling layers, fully connected layers and normalization layers are placed.

\subsubsection{Recurrent Neural Networks}

Recurrent neural networks (RNNs), are a type of neural networks for processing sequential data, having the connections between the nodes organized as a directed graph along a temporal sequence. Unlike CNNs which are specialized for processing images, RNNs are specialized for processing a sequence of values, such as timeseries, and use the temporal information in the reconstruction process and construct the data based on the information contained in both the current sequence and also taking into consideration the temporal dependencies among sequences. A particular type of RNN is the Long Short-Term Memory (LSTM) designed to overcome the vanishing gradient problem by summarizing or forgetting the old states before moving on to the next subsequence.

\subsubsection{Residual Networks}

Residual Neural Networks (ResNets) are network architectures with skip connections (or residual connections). Skip connections allow gradient information to pass through the layers, adding the output of a previous layer to the output of a deeper layer, maintaining data propagation in deeper networks and thus avoiding the vanishing gradient problem.

\subsubsection{Autoencoders}

An autoencoder (AE) is a feedforward neural network (FFNN), trained to encode the input into a lower space representation and then reconstruct it using this learned representation.
The internal structure of the AE consists of an input layer, an output layer and one or more hidden layers between them. The AE consists of two parts: an encoder and a decoder. Autoencoders have been successfully applied to dimensionality reduction and feature extraction, improving performance of different tasks, due to their ability to capture important properties of the data. 

A wide range of autoencoders have been proposed, including the denoising autoencoder (DAE), the variational autoencoder (VAE) or the stacked autoencoder.
The denoising autoencoder (DAE) is a particular type of AE trained to reconstruct and denoise corrupted versions of the inputs.
Variational Autoencoders (VAE) are a another version of AE whose training is regularized to ensure learning a latent representation, and not just specific encodings in the latent space, thus allowing the decoding of a range of variations of the encodings. 
Another extension of the standard AE topology is the Stacked Autoencoder, consisting of several layers of AE where output of each hidden layer is connected to the input of the successive hidden layer.

Similar to the stacked autoencoder model, the Deep Belief Networks (DBN) are also composed of stacked modules, which can be trained in an unsupervised manner. DBNs are best known for their ability to model and approximate any discrete statistical distribution.

\subsubsection{Generative Adversarial Networks}

Generative Adversarial Networks (GANs) represent a deep learning approach that aims to learn to generate new data from a a training set, having the same particularities as the training set. Two models are trained simultaneously in an adversarial setting: a generative (generator - G) model that emulates the data distribution, and a discriminative (discriminator - D) model that predicts whether a certain input came from real data or was artificially created. The generative model learns a mapping from a low-dimensional vector to the high dimensional space.
GANs unique ability to mimic data distribution made them successfully applied to model data distributions within the low-dimensional latent spaces, and thus to improve the reconstruction quality in the CS domain.
In CS, the G network usually takes the undersampled CS data and generates corresponding reconstruction. The discriminator network D, takes the outputs of the G and classifies it as real (fully sampled data) or generated (reconstructed).

\section{Reconstructing Compressive Sensing Signal with Deep Learning}
\label{sec:reconstruction}

The reconstruction of compressively sensed signal can be reduced to solving, via convex optimization, an \textit{$l_1$}-norm conditioned under-determined system of equations (Section~\ref{sec:primer}). Neural networks have been used for solving various optimization problems for the last four decades \cite{tank1986simple} and different neural network models have been developed to solve convex optimization problems ~\cite{bian2012neural,hosseini2013recurrent}. Within the CS literature, DL solutions for signal reconstruction can be classified into those that follow the general philosophy set by the traditional iterative reconstruction algorithms (discussed in Section~\ref{sec:reconstruction:iterative}) and those that harness the modeling power of DL directly (Section~\ref{sec:reconstruction:direct}).

\subsection{Deep Learning for Iterative Reconstruction}
\label{sec:reconstruction:iterative}
\begin{figure}
\centering
\includegraphics[width=0.6\textwidth]{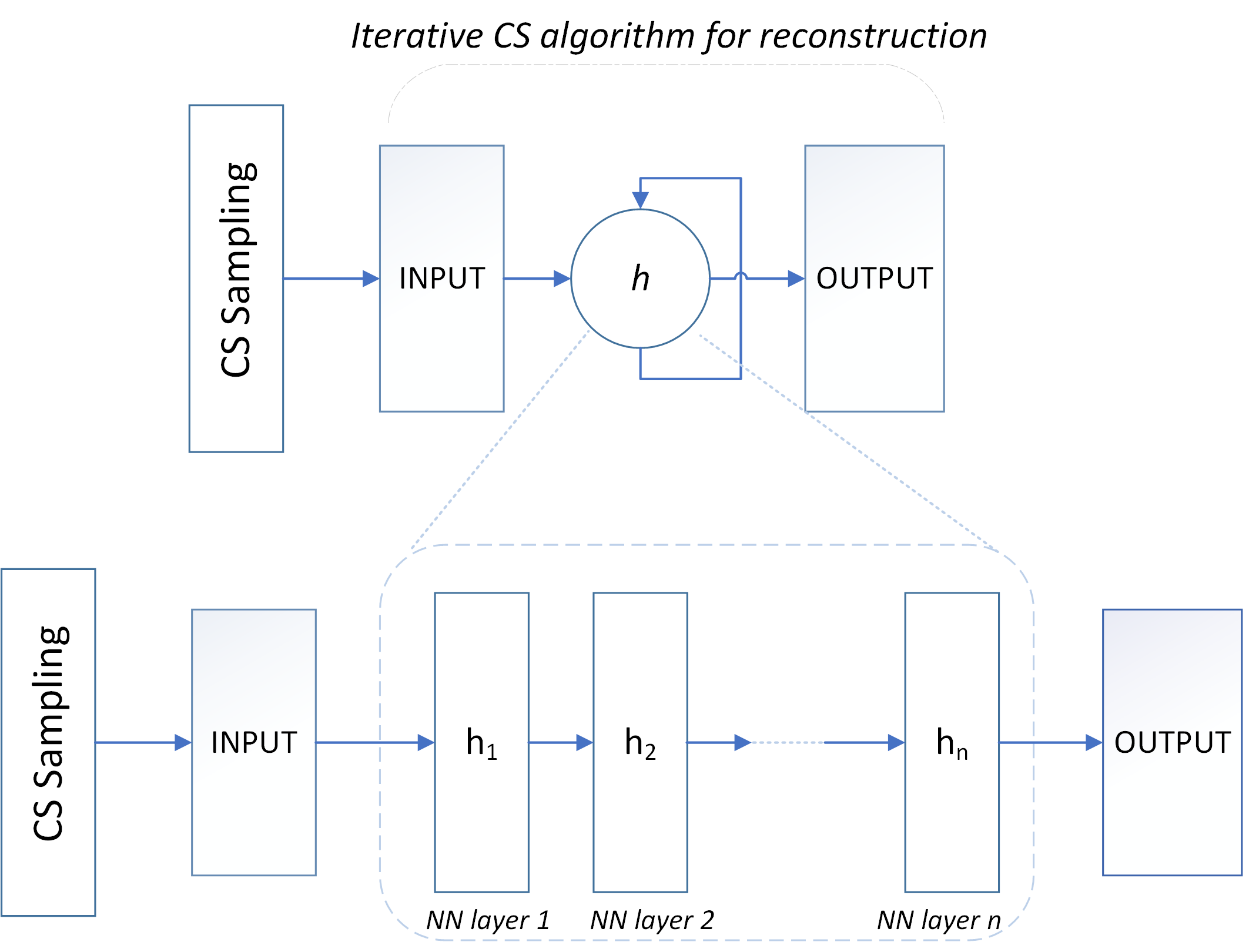}
\caption{CS signal reconstruction using an iterative algorithm unrolled over a deep neural network.}
\label{fig:unrolling}
\end{figure}

The first group of methods for CS signal reconstruction consists of methods designed to mimic the iterative CS algorithms using dedicated neural network architecture. Most of these methods are based on the technique called algorithm unrolling (or unfolding) that maps each iteration into a network layer, and stack a determined number of layers together (Figure \ref{fig:unrolling}). The parameters of the algorithm are weights to be learned and after the unrolling, the training data is fed through the network, and stochastic gradient descent is used to update and optimize its parameters. A summary of the most relevant methods in this category is provided in Table ~\ref{tab:unrolling}.

ISTA (The Iterative Shrinkage/Thresholding Algorithm) \cite{beck2009fast} is a prominent traditional class of iterative algorithms used for solving the CS reconstruction problem. ISTA algorithms rely on a gradient based approach, where at each iteration the gradient of the differentiable term is projected and then it is thresholded (or shrinked) by a certain value. The first unfolded approach of the ISTA algorithm called Learned ISTA (LISTA)~\cite{gregor2010learning} was proposed in the area of sparse coding, i.e. finding a sparse representation of a given signal in a given dictionary. LISTA uses a deep encoder architecture, trained using stochastic gradient descent to minimize a loss function defined as the squared error between the predicted code and the optimal code averaged over the training set. ISTA-Net, too, proposes a DL network mimicking ISTA, but moves from the sparse coding to the problem of CS reconstruction~\cite{zhang2018ista}. In the proposed deep network all parameters (including the transformation matrix) are discriminately learned instead of being hand-crafted or fixed. The authors show that ISTA-Net reduces the reconstruction complexity more than 100 times compared to the traditional ISTA. TISTA is another sparse signal recovery algorithm inspired by ISTA and based on deep learning~\cite{8695874}. The notable difference between ISTA-Net and TISTA is that the latter uses an error variance estimator, which improves the speed of convergence. The work in \cite{song2020learning} also exploits the idea of unfolding the classic iterative algorithm ISTA as a deep neural network, but to deal with nonlinear cases and to solve the sparse nonlinear regression problem, using the Nonlinear Learned Iterative Shrinkage Thresholding Algorithm (NLISTA). For further enhancements to LISTA, a reader is referred to Step-LISTA \cite{ablin2019learning}, LISTA-AT \cite{kim2020element}, and GLISTA \cite{wu2019sparse}.

A modified version of ISTA, with a better sparsity–undersampling tradeoff, is the Approximate Message Passing (AMP)~\cite{donoho2009message} algorithm, inspired by the message passing (or belief propagation) algorithms on graphs.
A neural network model for unfolding the iterations of the AMP algorithm was proposed in \cite{borgerding2017amp}. This AMP variant, called Learned AMP (LAMP) unfolds the AMP algorithm to form a feedforward neural network whose parameters are learned using a variant of back-propagation. The performance of LAMP (as well as AMP) is restrained to i.i.d. Gaussian matrices, hence the authors also propose an additional version, dubbed Learned Vector Approximate Message Passing (LVAMP). LVAMP is build around the Vector Approximate Message Passing (VAMP) algorithm ~\cite{schniter2016vector}, which extends AMP's guarantees from i.i.d. Gaussian and works well with a much larger class of matrices. Another extension of AMP is the Denoising-based Approximate Message Passing algorithm (D-AMP),  ~\cite{metzler2014denoising}, which is based on the denoising perspective of the AMP algorithm, that considers the non-linear operations in each iteration as a series of denoising processes. A deep unfolded D-AMP is implemented in ~ \cite{metzler2017learned} as the Learned D-AMP (LDAMP), and also in ~\cite{zhang2020amp} as the AMP-Net. While both implementation are designed as CNNs, AMP-Net has an additional deblocking module (inspired by ResNet~\cite{he2016deep}) to eliminate the block-like artifacts in image reconstruction. In addition, AMP-Net also uses a sampling matrix training strategy to further improve the reconstruction performance.

Besides building upon ISTA and AMP, ADMM~ \cite{boyd2011distributed} is another algorithm that can be used for CS reconstruction. This convex optimization approach breaks problems into smaller pieces, each of which are then easier to handle and was used for sparse signal recovery in CS \cite{pham2013efficient,zhang2020signal,feng2019robust}. The authors in~\cite{yang2017admm} propose ADMM-NET, a deep architecture based on CNN and inspired by the ADMM algorithm for reconstructing high-quality magnetic resonance images (MRI) from undersampled data. A more general and powerful unrolled version of the ADMM algorithm, for CS imaging of both MRI and natural images is the ADMM-CSNet~\cite{8550778}. The  ADMM-CSNet discriminatively learns the imaging model and the transform sparsity using a data driven approach, which enhances the image reconstruction accuracy and speed.

Remaining in the MRI area, another unrolling model, inspired by the Total Variation algorithm ~\cite{krahmer2017total}, and dubbed as TV-Inspired Network (TVINet) is presented in ~\cite{zhang2020deep}. TVINet applies the CS reconstruction framework and solves the TV regularization by combining the iterative method with a CNN deep neural network. 
The iterations of the Landweber algorithm ~\cite{landweber1951iteration} are unfolded into the steps of a variational autoencoder for multi-coil MRI, in  ~\cite{hammernik2018learning}. The Landweber algorithm belongs to the class of gradient descent-based methods for solving inverse problems and consists of a combination of the steepest descent method with a stopping rule.

\begin{figure}
\centering
\includegraphics[width=0.6\textwidth]{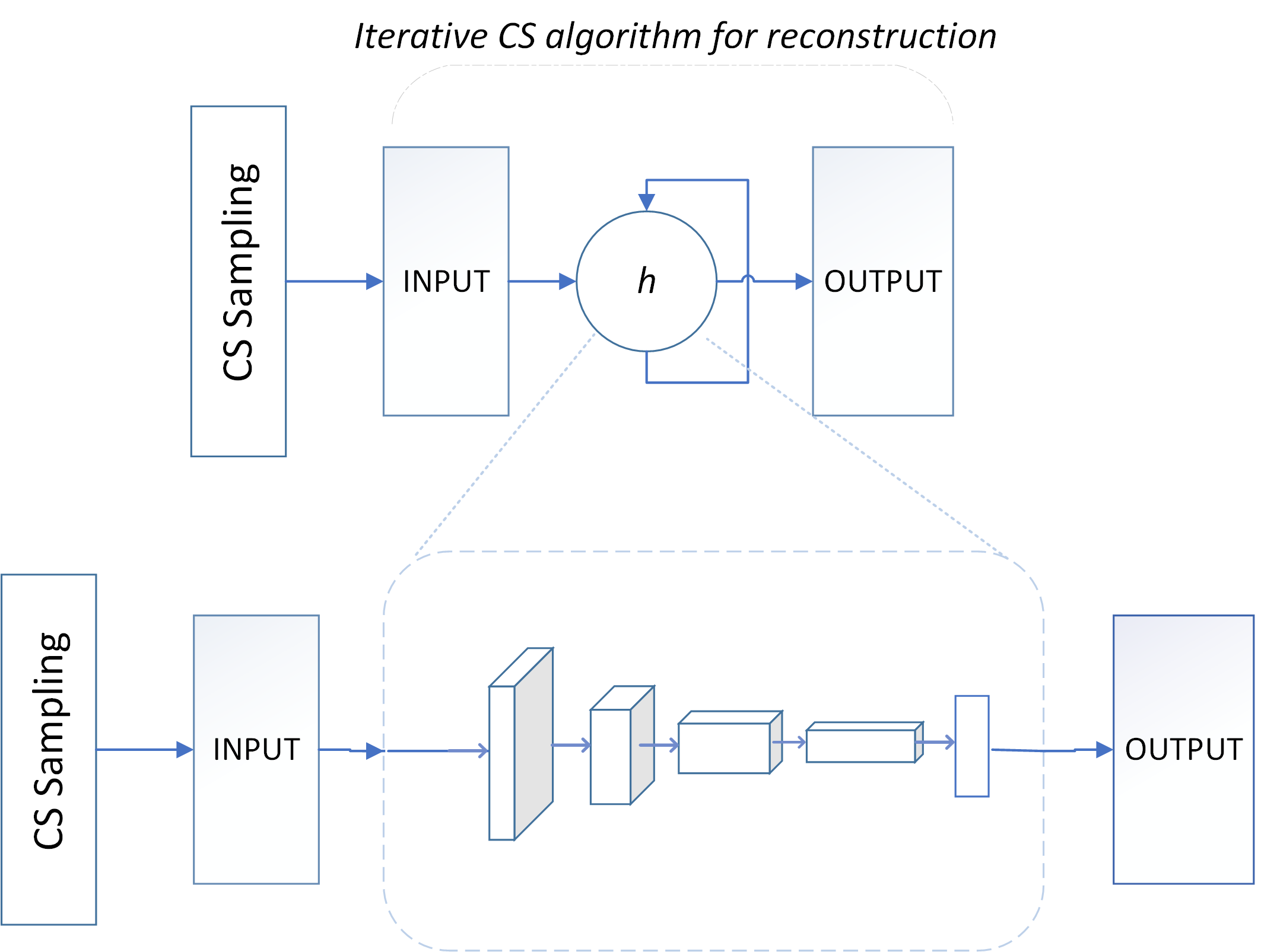}
\caption{CS signal reconstruction where a traditional iterative algorithm is fully (or partly) replaced using a deep neural network.}
\label{fig:customNN}
\end{figure}

Unrolling is not the only method, instead, iterative reconstruction algorithms cam be enhanced by replacing various steps in the algorithm with a NN (Figure \ref{fig:customNN} shows the borderline case where the whole reconstruction algorithm is replaced by a NN). In \cite{6009227}, the correlation step of the Orthogonal Matching Pursuit (OMP)~\cite{pati1993orthogonal} algorithm is replaced with a three-layer fully connected feed forward network trained to give an estimation of the unidentified nonzero entries of the original signal vector. The complexity overhead for training and then integrating the network in the sparse signal recovery is only justified in the case when the signal has an added structure, e.g. the zero coefficients of the sparse signal follow a certain spatial probability density function.

\begin{table}[H]
  \caption{A summary of CS methods employing algorithm unrolling}
  \label{tab:unrolling}
  \begin{tabular}{cccccc}
    \toprule
    Reference & Method & Year & Application & Network type & Iterative Algorithm \\
    \midrule
    \cite{gregor2010learning} & LISTA & 2010 & images & AE & ISTA \\
    \cite{zhang2018ista} & ISTA-Net & 2018 & images & CNN & ISTA \\
    \cite{8695874} & TISTA & 2018 & images & FFNN & ISTA \\
    \cite{song2020learning} & NLISTA & 2018 & generic & RNN & ISTA \\
    \cite{borgerding2017amp} & LAMP & 2017 & generic & FFNN & AMP \\
    \cite{zhang2020amp} & AMP-Net & 2020 & images & CNN & AMP \\
    \cite{metzler2017learned} & LDAMP & 2017 & images & CNN & D-AMP \\
    \cite{yang2017admm} & ADMM-Net & 2017 & MRI & CNN & ADMM \\
    \cite{8550778} & ADMM-CSNet & 2020 & images & CNN & ADMM \\
    \cite{zhang2020deep} & TVINet & 2020 & MRI & CNN & TV \\
     \cite{hammernik2018learning} & VN & 2018 & MRI & AE & Landweber\\
    \cite{6009227} & NNOMP & 2011 & images & FFNN & OMP \\

 \\
  \bottomrule
\end{tabular}
\end{table}

Compressive sensing evolved from theoretical studies and its initial practicality was predominantly limited by the time complexity of the reconstruction algorithms. Deep learning brought tremendous improvements on that front, enabling real-time reconstruction in certain applications. Compared with the conventional iterative reconstruction methods, the DL-supported algorithm unrolling brings a consistent computational speed up (the computational savings were in fact the motivation for unrolling). For example, ADMM-CSNet \cite{8550778} can be about four times faster than the BM3D-AMP algorithm \cite{7351377}. LISTA  \cite{gregor2010learning} may be 20 times faster than ISTA \cite{beck2009fast} after the training phase, while LDAMP \cite{metzler2017learned} can achieve a 10 times speedup, when compared to BM3D-AMP \cite{7351377}. This is due to the fact that it is faster to process data through neural network layers, especially since special operations such as convolutions can been highly optimized. In addition, the number of layers in a deep network is smaller than the number of iterations required in an iterative algorithm used for CS reconstruction. Interestingly, DL approaches mimicking the unrolled algorithms can also be faster even that the classic neural networks implementations aiming to replace the whole algorithm. For example, ADMM-CSNet \cite{8550778} can be about twice as fast as ReconNet \cite{kulkarni2016reconnet}, a pioneering NN reconstruction approach that does not mimic a known iterative algorithm. Nevertheless, the comparison might depend on the efficiency of the implementation of individual network layers.

The true potential of DL for signal reconstruction, however, is observed if we compare the reconstruction accuracy of DL approaches with the accuracy accuracy achieved by conventional iterative algorithms. ADMM-CSNet produces the highest recovery accuracy in terms of PSNR (Peak Signal-to-Noise Ratio) and SSIM (Structural Similarity Index Measure). AMP-Net~\cite{zhang2020amp} and ISTA-Net \cite{zhang2018ista} improve the reconstruction accuracy over both D-AMP \cite{metzler2014denoising} and a NN-approach ReconNet \cite{kulkarni2016reconnet}. By learning parameters in each iteration, instead of keeping them fixed for the whole network, unrolled methods are able to extend the representation capacity over iterative algorithms, thus are more specifically tailored towards target applications.

Finally, if from the efficiency and performance point of view, the unrolled approach often remains superior to both iterative and neural network approaches not based on algorithm unrolling, from other perspectives, such as parameter dimensionality and generalization, unrolled approaches remain in an intermediate spaces between iterative algorithms and more general DL-based solutions~\cite{monga2019algorithm}.

\subsection{Deep Learning for Direct Reconstruction} 
\label{sec:reconstruction:direct}

Harnessing neural networks in unrolled iterative approaches provides a certain level of intuition, yet, such intuition is not necessary for the optimization to work well. In the rest of the section we describe CS reconstruction (and sampling) approaches that were not inspired by the traditional optimization algorithms. These approaches, together with specific affordances they bring, are summarized in Table ~\ref{tab:csdl_rec}.  Free from any constraints, most of the research we discuss here optimizes both signal reconstruction as well as signal acquisition, i.e. the sampling matrix, essentially reflecting \textit{approach b)} in Figure~\ref{fig:CSDL}. This brings additional benefits, as many real-world signals, while indeed sparse when projected to a certain space, need not be sparse in the fixed domain we are observing them in -- learning the sampling matrix from the data often solves this issue.

\subsubsection{Autoencoder-based approaches}

A bird's eye perspective on CS reveals that, with its sampling (i.e. dimensionality reduction) and reconstruction pipelines, the method closely resembles a DL autoencoder (AE). Thus, it is not surprising that some of the early forays of DL in CS utilize AEs, so that the encoding process of the AE replaces the conventional compressed sampling process, while the decoding process of the AE replaces an iterative signal reconstruction process in CS. In such an arrangement the AE brings two immediate benefits. First, based on the training data it adapts the CS sampling matrix, which need not be a random matrix any more. Second, it greatly speeds the reconstruction process. Once trained, the AE performs signal reconstruction through a relatively modest number of DL layers, making it an attractive alternative to iterative reconstruction, even on ubiquitous computing devices, especially those embedded with GPUs. 

A pioneering approach in the area of AE-based CS is presented in~\cite{mousavi2015deep}, where Mousavi et al. propose the use of a stacked denoising autoencoder (SDAE). One of the challenges of using SDAE is that its network consists of fully-connected layers. Thus, as the signal size grows, so does the network, imposing a large computational complexity on the training algorithm and risking potential overfitting. The solution proposed in~\cite{mousavi2015deep} and adopted by similar approaches~\cite{kulkarni2016reconnet,liu2019compressed} is to divide the signal into smaller blocks and then sense/reconstruct each block separately. From the reconstruction time point of view, the simulation results show that this approach beats the other methods, whereas the of quality of the reconstruction does not necessarily overshadow that of other state-of-the-art recovery algorithms. Still, the SDAE was also used in practical applications. Majumdar et al. harness the fact that SDAE learns a non-linear mapping from  unstructured aliased images to the corresponding clean images and propose an MRI processing pipeline that independently reconstructs every sampled frame, faster than the data acquisition rate, thereby achieves real-time reconstruction~\cite{majumdar2015real}.

The amount of memory and processing power required by DL may prohibit CS-DL on ubiquitous computing devices. Therefore, reducing the number of DL parameters necessary for CS is highly desired. A sparse autoencoder compressed sensing (SAECS) approach is proposed in~\cite{han2018sparse}. The sparse autoencoder's loss function is constructed in a way that activations are penalized within a layer, resulting in fewer non-zero parameters, thus a ``lighter'' encoder, that is also less likely to overfit. Furthermore, combining SDAE and SAECS, a stacked sparse denoising autoencoder for CS (SSDAE CS) is proposed in~\cite{zhang2019optimally}. The proposed model, consists of an encoder sub-network which performs nonlinear measurements (unlike the conventional CS approach that involves linear measurements) on the original signal and a decoder sub-network that reconstructs the original signals by minimizing the reconstruction error between the input and the output. The denoising autoencoder's main advantage is that it is robust to noise, being able to reconstruct the original signals from noise-corrupted input.

Signals in the AE-compressed space, by default, need not exhibit any regularities. The AE merely ensures that the encoding-decoding process is efficient. Variational autoencoder (VAE) is trained to ensure that the latent space exhibits suitable properties enabling generative decoder behavior. Thus, \cite{nguyenlearning} proposes a VAE for the CS recovery problem, presenting it as ``an intuitive choice''. 
A novel Uncertainty autoencoder (UAE) structure is proposed by Grover and Ermon~\cite{grover2019uncertainty}. Compared to VAE, UAE does not explicitly train for the minimization of the Kullback–Leibler divergence, but optimizes a variational lower bound to the mutual information between the datapoints and the latent representations. While not discussed in~\cite{grover2019uncertainty} it would be interesting to examine whether UAE enables faster training with less data -- a property that down the road could enable privacy-preserving on-device training in ubicomp environments.

The versatility of the autoencoder approach to CS led to its adaptation to a range of domains. Such adaptation is evident in~\cite{adler2016deep}, one of the early approaches geared towards image reconstruction. High dimensionality, and consequently high memory and computation load, incurred by the sheer size of images called for block-based reconstruction. This was later enhanced with a modified loss function -- in \cite{zur2019deep} the authors moved from a generally-applicable mean squared error to image-specific structural similarity index measure (SSIM) as a training loss function for the autoencoder. Adapting to the domain is also evident in \cite{iliadis2016deepbinarymask} where Iliadis et al. developed a DL architecture for compressive sensing and reconstruction of temporal video.
Another adaptation~\cite{9044777} was designed for the particularities of the biological signals. In the case of ECG or EEG for example, short acquisition windows are beneficial for reducing the computational complexity, the storage requirements and the latency at the encoder side. However, short windows are also out of the reach of classical CS mechanisms, because the sparsity constraint is no longer fulfilled in small sized measurements. To make the CS acquisition and recovery feasible even for short windows, the authors step away from the classical approach that directly reconstructs the input signal and propose a two stage approach: first guessing which components are non-zero in the sparse signal to recover and then computing their magnitudes.

The representational power of the DBNs was exploited in the context of CS in \cite{polania2017exploiting}, where a DBN is employed to model the prior distribution of the sparsity pattern of the signal to be recovered. This approach showed to improve the reconstruction performance and to require fewer measurements, compared to algorithms that do not exploit structural information beyond sparsity. 
The ability to model higher–order dependencies of DBNs can also improve the efficiency of learning recognition parameters. In \cite{ma2017extraction}, DBNs are used for extracting motion-onset visual evoked potential waveforms from the background EEG signal in a brain-computer interface application.

\subsubsection{Dense and convolutional network-based approaches}

Dense networks based on a standard multilayer perceptron (MLP) have also been considered a reasonable choice for CS reconstruction problem, because their ability to learn a nonlinear function that maps the compressed measurements to the the original signal, in a supervised manner, unlike the autoencoders. With the introduction of convolutional filters and pooling layers, CNNs can achieve improved performance for reduced memory storage and increased learning speed, compared to vanilla dense networks, thus, represent an attractive architectural choice especially for image reconstruction tasks. Furthermore, compared to autoencoders, these networks can often handle larger inputs, due to the reduced dimensionality of convolutional and sparse connected layers.

ReconNet~\cite{kulkarni2016reconnet} is considered to be the first work that employs CNN for compressive sensing. Inspired by the success of the CNN-based approach for image super-resolution, the authors proposed a fully connected layer along with a CNN, which takes in CS measurements of an image as input and outputs the reconstructed image. The quality of the reconstruction is superior to those of the traditional iterative CS reconstruction algorithms. From the time complexity perspective, this approach is about three orders of magnitude faster than traditional reconstruction algorithms. One of its drawbacks is the fact that this approach uses a blocky measurement matrix, to reduce the network complexity and hence, the training time, and therefore, ReconNet does not exploit potentially strong dependencies that may exist between the reconstructions of different blocks. CombNet provides improved quality of reconstruction by using a deeper network structure and a smaller convolution core~\cite{liu2019compressed}.

Going a step further from these block based approaches (\cite{kulkarni2016reconnet,liu2019compressed}), the work in~\cite{mousavi2017learning} introduces the first DL framework that works for images acquired with measurement matrices of arbitrary sizes, and not only with blocky matrices. This network, named DeepInverse, is a CNN architecture, with a fully connected linear layer, whose weights are designed to implement the adjoint operator (transpose) of the measurement matrix. Including this auxiliary information into the reconstruction process simplifies the ill-posed reconstruction problem, and allows the network to reconstruct signals, without subdivision, using just four layers.

DeepCodec~\cite{mousavi2017deepcodec} is a variant of DeepInverse that instead of using random linear undersampled measurements learns a suitable measurement matrix from the data. The experiments showed that this network is computationally cheaper than DeepInverse, enhances the overall recovery performance, speeds up training of recovery framework, and could lead to fewer model parameters as well. A similar solution is proposed in ConvCSNet~\cite{lu2018convcsnet} for both the sensing and the reconstruction phase. The first convolutional layer performs the sensing of the whole image using a set of convolutional filters and the following layers perform  nonlinear reconstruction. The differences between DeepCodec and ConvCSNet lie in the implementation details of both the sensing layer as well as the reconstruction part. For the sensing phase, DeepCodec uses a learned downsampling layer (instead of the hand designed ones, such as max pooling), while ConvCSNet relies on a linear layer to obtain the measurements. In the reconstruction part, DeepCodec has eight convolutional layers (with batch normalization), while ConvCSNet implies a reconstruction network containing two branches (and 15 layers). Another approach meant to alleviate the complexity of the high dimension measurements is presented in \cite{canh2018deep}, where the sampling and the initial reconstruction are performed with convolutional Kronecker layers \cite{zhou2015exploiting}, that decompose the large weight matrices of the convolutional layer into combinations of multiple Kronecker products of smaller matrices, thus reducing the number of parameters and the computation time.

Domain knowledge can greatly enhance signal reconstruction. In~\cite{lu2019wdlreconnet} a wireless deep learning reconstruction network (WDLReconNet) aiming to recover signals from compressive measurements transmitted over a WSN is proposed. To counter the effect of wireless distortions, the authors prefix a CNN with a dictionary learning-based feature enhancement layer. MRI is one of the key application areas of CS in general, thus a number of CNN-based solutions have been adapted to this domain. Schlemper et al. \cite{schlemper2017deep} propose a framework for reconstructing dynamic sequences of MR images from undersampled data using a deep cascade of CNNs to accelerate the data acquisition process. A simple CNN could show signs of overfitting, when not enough training data is available -- a situation often encountered in the medical imagery field. Therefore, the authors proposed to concatenate a new CNN on the output of the previous CNN to create a DNN that iterates between intermediate de-aliasing and the data consistency reconstruction. CNNs are also used in \cite{wang2016accelerating}, where a three-layer CNN is designed and trained to define a relationship between zero filled solution and high‐quality MRI. 

Finally, while CNNs remain the predominant approach for CS-DL integration, standard MLP-based dense networks also have their place in the literature. Iliadis et al. use such networks for the problem of video CS reconstruction~\cite{iliadis2018deep}. The authors motivated their choice of using fully-connected layers for the entire network to the fact that the first hidden layer is constrained to be a fully-connected layer, for providing a 3D signal from the compressed 2D measurements, and the subsequent layers do not allow for convolutions to be effective, due to the small size of the blocks. Different variants of MLP networks are tested in \cite{shrivastwa2018fpga} and shown to perform well for compressed electrocorticography (ECoG) signal reconstruction. High compression rates demonstrated in this work open way for interesting ubiquitous computing applications, such as using remotely sampled and compressed brain signals for remote prosthetic control.

\subsubsection{Recurrent Neural Networks and Long Short-Term Memory Networks}


Recurrent neural networks (RNNs), to the best of our knowledge first used for CS signal reconstruction in \cite{li2016signal}, found suitable ground in domains where the temporal dependencies in the data are essential for achieving a fast and accurate reconstruction. Consequently, domain-adapted versions of RNNs and LSTMs can be found mostly in video processing and speech analysis, but extend to other domains as well. 
In speech processing, RNNs were exploited for shaping the structural differences between voiced and unvoiced speech \cite{ji2019recurrent}. To mitigate the noisy distortions caused by unvoiced speech, the authors build an RNN dictionary learning module that learns structured dictionaries for both voiced and unvoiced speech. These learned codebooks further contribute to improving the overall reconstruction performance of compressed speech signals.  
RNNs are also appropriate for processing sequences of images, particularly if the images are temporally correlated. An RNN model adapted for the image domain can be found in \cite{qin2018convolutional}, where a convolutional recurrent neural network (CRNN) is used to improve the reconstruction accuracy and speed by exploiting the temporal dependencies in CS cardiac MRI data. By enabling the propagation of the contextual information across time frames, the CRNN architecture makes the reconstruction process more dynamic, generating less redundant representations. 

Long Short-Term Memory (LSTM) networks are similarly adapted to particular CS domains. In \cite{zhao2019deep} an LSTM harnesses a strong similarity among distributed signals collected by UWB soil sensors for efficient signal reconstruction. The field of distributed compressed sensing or Multiple Measurement Vectors (MMV) was also targeted in \cite{palangi2016distributed} and in \cite{palangi2016reconstruction}, where an LSTM and a Bidirectional Long Short-Term Memory (BLSTM), respectively were proposed for reconstruction. The LSTM and BLSTM models are good candidates for reconstructing multiple jointly sparse vectors, because of their ability to model difficult sequences and to capture dependencies (using both past and future information, in the case of BLSTM) among the sparse vectors of different channels. 
In the field of biological signals processing, a multilayer LSTM \cite{han2017new} was developed for CS reconstruction. The authors took advantage of the natural sparsity of the measurements and modeled the acquired compressed pressure data of a region of the human body as time sequences to enable an LSTM-based CS reconstruction.
In the video compression domain, a noticeable LSTM adaptation is the CSVideoNet framework \cite{xu2016csvideonet}. CSVideoNet employs an LSTM that combines the temporal coherence in adjacent (compressed sampled) video frames with the spatial features extracted by the CNN, for an enhanced reconstruction quality. The comparison with and without the LSTM network shows that exploiting the temporal correlations between adjacent frames can significantly enhance the CS performance of video applications in terms of the trade-off between compression rates and reconstruction quality.

Building upon these results, the field of ubiquitous computing can be further enriched with novel applications, such as those targeting smart city sensing or health telemonitoring with compressed electrocardiogram (ECG) and electroencephalogram (EEG) signals, where learning and incorporating temporal dependencies in the CS reconstruction algorithm might improve both the reconstruction accuracy and the response time. In addition, using network models that exploit the inherent temporal structure of the signals  may also reduce the number of measurements needed, which is again very important for applications using sensors with low power consumption and limited battery life. 

\subsubsection{Residual networks}

Increasing the depth of network by adding more layers may improve the performance of the CS reconstruction networks discussed above. However, training very deep networks poses difficulties due to the vanishing gradient problem, overfitting, and accuracy saturation. The solution to these problems came with the introduction of the residual learning. Residual networks prove to be able to alleviate this challenges, being easier to train, better optimized and having higher performances, as opposed to very deep CNN architectures. Moreover, residual learning architecture proved to be very useful not only to speed up the training process, but also for denoising and superresolution, thus improving the quality of the reconstruction.

Learning the sampling as well further improves the quality of the reconstruction. A jointly optimized sampling and reconstruction approach is presented in \cite{yao2019dr2}. This network (called the Deep Residual Reconstruction Network, or DR2-Net) has a fully connected layer to perform the sampling, followed by other linear mapping blocks (to obtain a preliminary reconstruction) and several residual learning blocks (to infer the residual between the ground truth image and the preliminary reconstruction). The experimental results showed that the DR2-Net outperforms other deep learning and iterative methods, being more robust for the CS measurement at higher compression rates.
Du et al.~\cite{du2019fully} propose a similar architecture, the main difference being the use of a convolutional layer for getting the adaptive measurements. A fully convolutional neural can deal with images of any size, breaking the limitation of fully-connected layers that are only capable of sampling fixed size images.
Another approach that learns both the sampling and the reconstruction, is the one in \cite{8765626}. The sampling network is designed to learn binary and bipolar sampling matrices tailored for easy storage and hardware implementation. Such matrices are then suitable for a range of ubicomp applications, e.g. for WSNs used for critical infrastructure health monitoring where signals are collected, compressed, and wireless transmitted by low-power and memory-limited WSN nodes.

Incorporating domain knowledge can improve the performance of the reconstructing model that is based on an RNN.  
In the field of computed tomography (CT), the use of residual networks was motivated mainly by the inherent presence of striking artifacts caused by the sparse projection that are  difficult to remove with vanilla CNNs architectures. Multi-scale residual learning networks based on the U-Net~\cite{ronneberger2015u} architecture aiming to remove these streaking artifacts was proposed in  \cite{han2016deep} and \cite{jin2017deep}. 
Aliasing artifacts are also common in CS MRI, so several approaches \cite{lee2017deep, ouchi2020reconstruction, han2018deep} introduced residual learning to enhance the reconstruction accuracy by learning and removing the aliasing artifacts patterns. This also accelerates the training process, since learning the aliasing artifacts is easier and faster than learning to predict the original aliasing-free MR images which posses a more complex topological structure. After learning the aliasing artifacts patterns, the aliasing-free image can be obtained by subtracting the estimated aliasing artifact from the original image with artifacts. 

In addition, Han et al. \cite{han2018deep} also addresses the issue of limited available data for training (frequent in the MRI field), proposing the technique of domain adaptation (pre-training the network with CT data for MRI), that also expands the applicability of the model.
Domain knowledge integration is also evident in \cite{kim2020compressive} where a residual network adapted for CS spectroscopy reconstruction (ResCNN) is proposed. The main challenge in this field is to identify a sparsifying basis that would work for the great variety of spectra available.
Having a residual connection between the input and the output of a CNN, ResCNN learns the spectral features and recovers the fine details for various kinds of spectra, without requiring a priori knowledge about the sparsifying basis or about the spectral features of the signal. 
Another domain adaptation targets the video compressed sensing reconstruction \cite{zhao2020hybrid}, and uses residual blocks for improving the recovery procedure in terms of both the quality and the speed of the reconstruction. The architecture proposed has a convolution layer for the sensing phase, a deconvolution layer for the preliminary recovery phase and several residual blocks for enhancing the preliminary recovery results.

These promising results achieved by residual networks, in ensuring a good balance between the number of network parameters and the model performance, can open new perspectives in the ubiquitous computing domain for new applications, such as using compressed sampled images for license plate recognition directly on edge devices with limited computing resources. 
In light of the proliferation of IoT devices in today’s smart cities, increased efforts are underway to transcend them into edge computing nodes where the processing takes place on-device without transferring the data \cite{chen2020binarized, chen2021edge}; as such, a CS-DL algorithm has the potential to act as a complementary approach to the above.
In addition, the promising results achieved by Kim et al. in~\cite{kim2020compressive} could be capitalized in the emerging field of small satellites on-board applications. Given their moderate depth and reduced number of parameters, residual CS networks can be successfully deployed in such resource-constrained environments. In addition, due to their proven ability to discriminate among various spectral signatures, residual CS networks could consequently provide real-time information in applications such as optical discrimination of vegetation species \cite{maimaitijiang2020soybean} or marine algal cultures \cite{deglint2019investigating}.

\subsubsection{Generative Adversarial Networks}


Generative Adversarial Networks (GANs) can facilitate the reconstruction and synthesis of realistic images and opened the door to innovative ways to approach challenging image analysis tasks such as image denoising, reconstruction, segmentation or data simulation. Inspired by the success of GANs in computer vision tasks, a modified version of the ReconNet, with adversarial loss (in addition to the Euclidean loss) and a jointly learning approach of both the measurement matrix and the reconstruction network is proposed in \cite{8379450}. In this network architecture, ReconNet acts as the generator, while another network, the discriminator, is trained to classify the input received as being a real image or a ReconNet reconstructed one. This GAN approach has sharper reconstructions, especially at higher measurement rates, compared to the original ReconNet. 

GANs ability to provide realistic, high texture quality images was exploited in \cite{mardani2017deep} for CS MR image reconstruction. A deep residual network with skip connections is trained as the generator that learns to remove the aliasing artifacts by projecting it onto a low-dimensional manifold containing the desired, high-quality data. The discriminator, in the form of a CNN-based architecture, is trained to assess the projection quality, scoring one if the image is of diagnostic quality, and, zero if it contains artifacts. This network model, dubbed GANCS, scores superior results in terms of both diagnostic quality of the reconstructed images, and running time, relative to the alternative, conventional CS algorithms.

These promising results are not easy to achieve, since training two competing neural networks is a challenging task, that requires extra care for designing the model and tuning the parameters, in order to ensure the stability and the proper convergence of the model. A solution for this issues was proposed in \cite{yu2017deep}, where the authors use refinement learning to stabilize the training of a GAN model for MRI CS. The generator was trained to generate only the missing details in the image, which proved to reduce the complexity of the network and lead to a faster convergence. 
In addition, the loss function was enriched with a perceptual loss and a content loss incorporating both pixel and frequency domain information to improve the reconstructed image quality in terms of anatomical or pathological details. 

The authors of \cite{kabkab2018task} introduce the task-aware GANs for CS, a model that allows optimizing the generator specifically for a desired task, be it reconstruction, classification, or super-resolution. Their solution improves the objective function by introducing the optimization of the latent code, or in other words of the compressed representation of the data, in addition to the optimizations of the generator and of the discriminator. In this manner, the compressed data obtained is more tailored towards a certain task and more beneficial for training the generator and the discriminator. This approach also addresses the cases where no or very little non-compressed data is available, and for that, the model is trained using a combination of compressed and non-compressed training data. This requires another discriminator to be added, hence the proposed model has one generator and two discriminators -- one for distinguishing between actual training data and generated data, and another for distinguishing between actual compressed training data and generated data.

All these approaches highlight GANs remarkable potential to enable realistic image reconstructions in the CS domain, but most importantly, reveal the inherent ability of this type of networks to emphasize what is indeed relevant for the end-user. Unlike other deep learning-based approaches, the GANs provide images that humans find visually realistic, and therefore their reconstructed images preserve more useful perceptual details. In the case of MRI, anatomical or pathological details for diagnosis, e.g., more detailed texture, sharper organ edges, better defined tumor textures and boundaries, and other factors beyond the commonly used image quality metrics (such as the Signal-to-Noise Ratio, Mean Squared Error, etc.) are elicited. 

Although most of the existing GAN based CS approaches addressed the MRI domain, there are still many yet unexplored fields where CS reconstruction (and sampling) may benefit from the advantages brought by the GANs architectures. In particular, the ability of GANs to reconstruct data with relevant features to the end-user application can be extremely valuable in  ubiquitous computing applications with budgeted acquisition and reconstruction speed, where not the accuracy of the reconstruction, but the usefulness for subsequent processing prevails. One such example is the use of drones equipped with hyperspectral cameras for capturing compressed images and using the salient information of the scene for wildfire monitoring.

\begin{longtable}{p{0.5cm}p{6cm}p{6.5cm}}
    \caption[A summary of the DL approaches for CS and their key affordances]{A summary of the DL approaches for CS and their key affordances}
   \label{tab:csdl_rec}\\
   \hline \multicolumn{1}{l}{\textbf{DL Approach}} & \multicolumn{1}{l}{\textbf{Study (domain, sensing matrix)}} & \multicolumn{1}{l}{\textbf{Affordances}} \\ \hline 
   \endfirsthead
    
    \multicolumn{3}{c}%
    {{\bfseries \tablename\ \thetable{} -- continued from previous page}} \\
    \hline \multicolumn{1}{l}{\textbf{DL Approach}} & \multicolumn{1}{l}{\textbf{Study (domain, sensing matrix)}} & \multicolumn{1}{l}{\textbf{Affordances}} \\ \hline 
    \endhead

    \multirow{7}{*}{AE} & \cite{mousavi2015deep}, images, learned & \multirow{7}{*}{
    \noindent
    \begin{minipage}[t]{0.4\textwidth}
    \raggedright
    \begin{itemize}[leftmargin=0cm,noitemsep,topsep=0pt]
        \item adapt the sampling matrix;
        \item speed up the reconstruction process;
        \item relatively modest number of layers;
    \end{itemize}
    \end{minipage}
    }\\
    & \cite{majumdar2015real}, MRI, random binary & \\
    & Uncertainty AE \cite{grover2019uncertainty}, images, learned &\\
    & \cite{iliadis2016deepbinarymask}, video, learned &\\
    & \cite{9073945}, ECG, learned &\\
    & TCSSO \cite{9044777}, ECG, learned &\\
    & SAECS \cite{han2018sparse}, bio signals, Gaussian & \\
    & SSDAE-CS \cite{zhang2019optimally}, images, learned & \\
   \midrule
   
    \multirow{2}{*}{DBF} & \cite{polania2017exploiting}, images, Gaussian & \multirow{2}{*}{
    \noindent
    \begin{minipage}[t]{0.4\textwidth}
    \raggedright
    \begin{itemize}[leftmargin=0cm,noitemsep,topsep=0pt]
        \item model higher-order dependencies;
        \item exploit structural information in data;
    \end{itemize}
    \end{minipage}
    }\\
    & \cite{ma2017extraction}, moVEP (EEG), echelon & \\  
    \midrule
    
    \multirow{9}{*}{CNN} & ReconNet(1)\cite{kulkarni2016reconnet}, images, Gaussian &  \multirow{9}{*}{
    \noindent
    \begin{minipage}[t]{0.4\textwidth}
    \raggedright
    \begin{itemize}[leftmargin=0cm,noitemsep,topsep=0pt]
        \item minimize number of parameters;
        \item reduce memory storage;
        \item increase learning speed;
        \item can better handle larger inputs;
        \item capture structure in images;
    \end{itemize}
    \end{minipage}}\\
    & DeepInverse \cite{mousavi2017learning}, images, Gaussian & \\
    & \cite{liu2019compressed}, images, Gaussian & \\
    & KCSNet \cite{canh2018deep}, images, learned & \\
    & WDLReconNet\cite{lu2019wdlreconnet},images, random & \\
    & \cite{wang2016accelerating}, MRI, 2D Poisson &  \\
    & \cite{mousavi2017deepcodec}, generic, learned &\\ 
    &\cite{schlemper2017deep}, MRI, Cartesian &\\
    & \cite{lu2018convcsnet}, image, learned &\\
    \midrule
    
    \multirow{5}{*}{MLP/FC} &\cite{adler2016deep}, images, learned & \multirow{2}{*}{
    \noindent
    \begin{minipage}[t]{0.4\textwidth}
    \raggedright
    \begin{itemize}[leftmargin=0cm,noitemsep,topsep=0pt]
        \item support supervised learning;
        \item lead to fewer operations, compared to CNNs;
    \end{itemize}
    \end{minipage}
    }\\
    & \cite{zur2019deep}, images, learned\\
    & \cite{shrivastwa2018fpga}, ECoG , learned &\\
    & \cite{7560597}, neural recordings, learned &\\
    & \cite{iliadis2018deep}, video, random binary & \\
    \midrule
    
    \multirow{3}{*}{RNN}& \cite{li2016signal}, generic, Gaussian & \multirow{3}{*}{
    \noindent
    \begin{minipage}[t]{0.4\textwidth}
    \raggedright
    \begin{itemize}[leftmargin=0cm,noitemsep,topsep=0pt]
         \item support CS sampling and reconstruction of temporal data; 
         \item improve reconstruction; 
    \end{itemize}
    \end{minipage}
    }\\
    & \cite{ji2019recurrent}, speech, learned &\\
    & CRNN-MRI\cite{qin2018convolutional}, MRI, learned & \\
    \midrule
    
    \multirow{3}{*}{LSTM} & CSVideoNet \cite{xu2016csvideonet}, video, learned & \multirow{3}{*}{
    \noindent
    \begin{minipage}[t]{0.4\textwidth}
    \raggedright
    \begin{itemize}[leftmargin=0cm,noitemsep,topsep=0pt]
        \item improve training convergence (compared to RNNs);
        \item take temporal coherence into account (especially if BLSTM is used);
    \end{itemize}
    \end{minipage}
    }\\
    & BLSTM-CS \cite{palangi2016reconstruction}, images, random & \\
    & LSTM-CS \cite{palangi2016distributed}, images, random &\\
    & \cite{han2017new}, biological signals, Gaussian &\\
    \midrule
    
    \multirow{9}{*}{Res} & \cite{du2019fully}, images, learned & \multirow{9}{*}{
    \noindent
    \begin{minipage}[t]{0.4\textwidth}
    \raggedright
    \begin{itemize}[leftmargin=0cm,noitemsep,topsep=0pt]
        \item alleviate the vanishing gradient problem, overfitting and accuracy saturation;
        \item are easier to train as opposed to very deep CNNs;
        \item provide balance between number of parameters and performance;
        \item accelerate training process;
        \item improve quality of the reconstruction;
        \item can be used for denoising or de-aliasing;
    \end{itemize}
    \end{minipage}
    }\\
    & \cite{han2016deep}, CT, - &  \\
    & FBPConvNet \cite{jin2017deep}, CT, Cartesian & \\
    & \cite{lee2017deep}, MRI, uniform random & \\
    &\cite{han2018deep}, CT, MRI, radial & \\
    & DR2-Net \cite{yao2019dr2}, images, learned & \\
    & DRL-CNN \cite{ouchi2020reconstruction} MRI, 1D,2D random\&non-random & \\
    & \cite{zhao2020hybrid}, video, learned & \\
    & ResCNN \cite{kim2020compressive}, spectroscopy, 2D filter-array &  \\
    \midrule
    
    \multirow{4}{*}{DBF} & ReconNet(2) \cite{8379450} , images, learned & \multirow{4}{*}{
    \noindent
    \begin{minipage}[t]{0.4\textwidth}
    \raggedright
    \begin{itemize}[leftmargin=0cm,noitemsep,topsep=0pt]
        \item can be tailored towards a certain task;
        \item more useful perceptual details preserved; 
        \item provide sharper, more realistic reconstructions;
    \end{itemize}
    \end{minipage}
    }\\
    & DAGAN \cite{8233175}, MRI, Gaussian & \\
    & GANCS \cite{mardani2017deep}, MRI, radial &\\
    & CSGAN \cite{kabkab2018task}, images, Gaussian & \\
  \bottomrule
\end{longtable}

\section{Inferring Higher-Level Concepts from Compressive Sensing with Deep Learning}
\label{sec:inference}

The recovery of data acquired through compressive sensing need not always be necessary, as we might be interested, not in the original signal, but certain inferences stemming from it. For instance, we might be concerned about whether certain objects are present in an image, what kind of a modulation technique is used in a wireless electromagnetic wave, or whether a certain pathology is present in a compressed ECG signal.  

Such a CS approach, without the actual signal reconstruction, is termed \textit{compressive learning}, and was first introduced in \cite{calderbank2009compressed}, where the authors demonstrated that learning using compressed data need not produce a substantial accuracy loss, compared to learning from the full uncompressed data. Many compressed learning approaches have been proposed since, expanding over domains such as compressed object tracking~\cite{kwan2020detection}, \cite{vargas2018object}, compressed hyperspectral image classification ~\cite{hahn2014adaptive}, or reconstruction-free single pixel image classification ~\cite{latorre2019online}.
The main advantage of compressive learning is that by skipping the reconstruction phase and extracting features from compressed measurements directly, the computation complexity and the processing time get significantly reduced. 
From the systems point of view, additional benefits are achieved by keeping the whole inference pipeline on a single device: the data transmission costs are reduced, the inference latency decreases, and data privacy is maintained. Furthermore, compressive learning may perform well even at very high compression rates where reconstruction based approaches would fail. This allows on-device learning implementations on resource-constrained systems that otherwise would not be able to support the inference task. Finally, in some cases the reconstruction phase may introduce artifacts and errors that can distort the reconstructed signal and therefore also the inference result. In many cases, compressed samples contain most of the relevant information, and thus can be considered as a comprehensive feature representation. This points to an interesting parallel between the traditional use of the encoder part of an autoencoder as a feature extraction tool, and its use for the sampling matrix adaptation in Section~\ref{sec:reconstruction:direct}.

The direct high-level inference from CS data using DL is depicted in Figure~\ref{fig:CSDL}d). Free from the need to reconstruct the signal, we can work directly in the compressed domain and harness the neural network's inherent ability to extract discriminative non-linear features, for which an intuitive explanation is not needed. Thus, Lohit et al.~\cite{lohit2016direct} proposed a DL approach for image classification compressive learning approach. This approach employed CNNs and random Gaussian sensing matrices~\cite{lohit2016direct}. Building upon this work, Adler et al.~\cite{adler2016compressed} demonstrate that by jointly learning both the sensing matrix as well as the inference pipeline, the image classification error can be reduced, which is especially evident when high compression rates are used.

In certain situations, however, signal reconstruction may be desired even when high-level inference remains the main goal of the CS system. For instance, a security camera may need to detect an intruder, yet, it would be desirable to reconstruct the original signal as a potential evidence of intrusion. A joint inference-reconstruction pipeline is proposed in~\cite{xuan2018deep}, where the authors optimize the DL pipeline, so that after a jointly learned sensing matrix, the two branches -- image reconstruction and image labeling -- continue their separate ways. Such a configuration is, in sum, more efficient than a solution that relies on separate pipelines for reconstruction and labeling. The work by Singahl et al. ~\cite{singhal2017semi} also combines the two stages in a single one and classifies compressed EEG and ECG signals at sensor nodes. Two different experiments are conducted, the first ones involves seizure detection from EEG compressed samples and the second one arrhythmia classification from ECG compressed samples. The authors show that by eliminating a separate reconstruction stage, upon which the inference would be done, the errors and artifacts are minimized and hence the results are improved. A joint construction of the two pipelines opens interesting opportunities for adaptive CS-DL deployment in heterogeneous systems. Different pipelines may have different processing and memory requirements, and the application packages could be made so that either the inference or the signal reconstruction, or both, are supported at different (edge) platforms the application is deployed to.

High compression rates are crucial in certain domains. Distributed video surveillance and UAV-based imaging are just two examples of applications that generate enormous volumes of data whose storage and wireless transmission is impractical. Instead, high compression rates are used, which leads to poor signal reconstruction quality. Optimizing the inference process over the compressed data, however, can provide better results compared to the case when the inference is performed after decompressing severely compressed data. A data-driven reconstruction-free CS framework for action recognition from video data is proposed in~\cite{gupta2019data}. This architecture consists of two separable networks: the convolutional encoder, which performs the sensing and generates undersampled measurements, and the classifier trained to classify the undersampled measurements. To ensure compatibility between the encoder and an existing DL classifier an upsampling layer is added after the encoding part. In this manner, the customized encoding sub-network can be jointly trained with a validated classifier for better results. 
This approach of resizing the encoded data before inference was also explored in ~\cite{bacca2020coupled}, where a CS reconstruction-free deep learning implementation for single pixel camera is proposed. Two network architectures are evaluated: one that re-projects the measurements to the original image size to preserve the image size for classification, and another that extracts features directly from the compressed measurements without learning an image size re-projection operator. Although the first approach achieves slightly better accuracy results on average, the second one has advantages in terms of smaller number of parameters and faster computation times with results comparable with those achieved by the first approach.

A common feature of most deep learning-based compressive sensing approaches is that they work only for the measurement rate that they have been trained on and cannot be used on other measurement rates without retraining. However, real-world scenarios often impose time-varying constraints on the measurement rates (e.g. the sparsity of the data fluctuate, the memory/energy/bandwidth limitations vary, the content is dynamically changing, etc.). This is especially symptomatic for mobile solutions, where the context of usage changes with the location of a smartphone, smartwatch, or any other device a user is carrying. Thus, is of practical importance to enable dynamically adaptive measurement rates. In practice, this could be realized via several different network models each trained for a separate measurement rate. Yet, this would incur potentially prohibitive additional storage and computation costs associated with training and storing multiple network configurations. Especially for the devices with limited resources, it is crucial to enable a single neural network to perform inference over a range of measurement rates. 

The first challenge for rate-changing scenarios, is finding the optimal measurement rate under fluctuating conditions. A preliminary approach addressing this issue was developed in ~\cite{sekine2019lacsle} where deep learning algorithms are used to estimate the optimal compression rate according to the data sparsity. This solution manages to maximize the data transmission efficiency, being thus very suitable for edge devices. The system was tested on a Raspberry Pi 3 Model B+, using vertical acceleration data of a domestic bridge. However, this solution relies on conventional CS algorithms for sampling and reconstruction, that can easily incorporate sparsity priors as their input parameters. The second challenge towards rate-adaptive CS-DL algorithms is developing neural networks that can work with different measurement rates. 

When it comes to the actual adaptable network implementations, such a solution was first proposed by Lohit et al.~\cite{lohit2018rate}. The authors first train the entire network for the highest desired measurement rate, and then in the second stage, all the parameters are frozen and the network is trained for the lowest measurement rate. Finally, in the third stage, the network is again optimized over a subset of parameters corresponding to adding an additional row at a time to the measurement matrix with the rest of parameters frozen. In the end, any subset of consecutive rows of the measurement matrix represents a valid measurement matrix corresponding to a different measurement rate in the range between the highest and the lowest specified measurement rates. To map the size-varying compressed inputs to the same inference network an additional conversion layer is needed that maps the inputs back in the original space by applying the pseudo-inverse of the measurement matrix. This approach was tested in an object tracking scenario in video sequences and the measurement rate was adapted based on the content evaluation among successive frames.
The adaptivity in the context of compressed learning has also been addressed by Xu et al. ~\cite{xu2020compressed}. In this work, CS measurement vectors of different lengths, corresponding to different measurement rates, and randomly shuffled, are provided as inputs to the neural network with a fixed input layer size and the network is trained on them. For handling the size mismatch between the size-varying CS measurement vectors and input layer two approaches are explored: one that zero-pads the measurement vectors so they are all of the maximum length, which is also the dimension of the input layer of network; and the other that projects back into the original space dimension the measurements, to get a pseudo-inverse of the measurement matrix,like in the previous work ~\cite{lohit2018rate}. The differences among these two approaches come from the different training process, which in one case consists of freezing part of the parameters while in the other case optimizes all parameters simultaneously, in the similarity between the weights of the multiple-rate network and those of a single-rate network, in the compatibility with different sensing matrices, the first approach is designed for learned matrices only, while the last one works with all kinds of matrices including random, structured, or learned matrices, etc.

Compressive learning can also be performed over a distributed system, such as a cloud computing network, raising concerns regarding the privacy protection of sensitive data. While compressive sensing can be seen as a form of data encryption, the compressed data is vulnerable to privacy attack, since the compressed sensing matrix could easily be decoded, by using a brute force attack of trial and error method, and that would expose the original data. 
Thapaliya et al.~\cite{thapaliya2020asymptotically} build a privacy-preserving predictive compressed learning model based on using a strong transformation matrix, instead of the attack vulnerable classic compression matrix.
Unlike other privacy preserving approaches, that are based on hiding the patterns existing in the data, the proposed approach perturbs the data using patterns that are not present in the data. In this manner, the data is perturbed enough to be robust to privacy attacks but the predictive accuracy of the model, trained to recognize the patterns of the data is still preserved.

\section{Towards Deep Compressed Sensing Systems}
\label{sec:systems}

The CS-DL approach has created hopes for the implementation of many practical ubiquitous computing applications, nevertheless, to the best of our knowledge, a large majority of the existing CS-DL approaches use pre-collected data (mostly images) and run on desktop computers or servers. In this section we first discuss real-world limitations that CS-DL is facing in the ubicomp domain and potential solutions addressing some of these limitations. We then, in selected application domains, present a few existing research efforts tackling particular domain challenges. Finally, we present a few interesting opportunities for future research in deep compressed sensing systems. 

\subsection{Challenges of Moving to the Edge}

\textit{Energy} is the most critical resource on ubiquitous computing devices. Portability implies that devices often run on limited capacity batteries with few opportunities for charging. Both sensor sampling and deep learning incur a significant energy cost. Limited \textit{communication} capacity is another constraint. Wirelessly connected devices have to cope with intermittent and varying quality links. For CS-DL applications this calls for a careful adaptation of the sampling rate, as higher sampling rates produce more data, which might need to be processed remotely, while lower sampling rates may reduce the quality of the reconstructed signal. \textit{Computational and storage} limitations, especially in terms of the GPU support for deep learning models on embedded architectures and the ability to store large deep networks in memory, are characteristic for ubiquitous computing devices. Moreover, from the software side, DL often harnesses dedicated libraries, which can be difficult to migrate to such devices. 

The above challenges, however, did not prevent researchers from proposing practical CS-DL solutions. For instance, Shen et al.\cite{shen2018cs} proposed incorporating the theory of compressive sensing at the input layer of a CNN model to reduce the resources consumption for IoT applications. However, the authors offer no actual evaluation of their model on resource constrained devices. Lee et al. \cite{lee2019deep} developed a joint transmission-recognition CS-DL framework with low complexity for IoT devices to effectively transmit data to a server for recognition. The proposed model involves a DNN architecture at the transmitter (the IoT device). Nevertheless, no actual implementation using IoT devices is provided. Sun et al. \cite{7560597} present a deep compressed sensing framework for wireless neural recording, with potential applications for real-time wireless neural recording and for low-power wireless telemonitoring of physiological signals. However, the framework is only assessed from the time and accuracy point of view, without any energy/power usage evaluation. Related to CS-DL efforts are solutions proposed for resource-efficient DL. To enable models running on edge devices, efforts are currently being made towards devising optimization techniques which aim to trim down the network complexity and reduce the redundancy without significant degeneration in performance. Such neural networks optimization techniques include: quantization ~\cite{gupta2015deep}, low-rank compression~\cite{novikov2015tensorizing}, pruning ~\cite{han2015learning}, slimmable neural networks ~\cite{yu2018slimmable}, and early exiting ~\cite{teerapittayanon2016branchynet}, to name a few. Crucial from the mobile systems perspective is the fact that these techniques can also be used in a dynamic context and by taking advantage of the variations in the input’s complexity for example, important resources can be saved with minimal impact on the accuracy \cite{laskaridis2020spinn}.

\subsection{Lessons from Domain-Specific CS-DL}

\textit{Magnetic resonance imaging (MRI)} has revolutionized medical diagnosis in the late $20^{th}$ century. Nevertheless, conducting MRI scans requires significant time when a subject needs to be immobilized. Reducing the number of samples, while preserving the visual quality of the image, would greatly improve the utility of MRI. CS-DL has already proved to be able to reduce the scanning time and simultaneously improve the image reconstruction quality.
Some of the CS-DL MRI methods, use deep neural networks to learn the traditional optimization algorithms, by unrolling them ~\cite{sun2016deep,yang2017admm,zhang2020deep}.
Another category of methods uses deep networks to mitigate noise and aliasing artifacts in the MRI reconstruction. SDAE \cite{majumdar2015real}, CNNs \cite{wang2016accelerating}, residual networks \cite{7950457,han2018deep,sun2018compressed,ouchi2020reconstruction}, or GANs \cite{mardani2017deep, yu2017deep} were successfully validated as suitable architectures for CS-DL MRI and have all shown great potential, outperforming conventional CS techniques for undersampled MRI.

The main advantage of CS-DL methods in MRI lies in the capability of a DNN to capture and make use of the patterns learned within the data in both image and frequency domain, to improve the quality of the reconstruction which is of high importance for medical diagnosis. In this process of reconstruction, the deep-learning based CS solutions go beyond the standard metrics that are usually used for mathematically evaluating the quality of an image, and put more emphasis on the anatomical or pathological details of an image. Also, by shifting the computational efforts to an offline training stage, CS-DL algorithms for MRI are able to provide a fast reconstruction, up to real-time, which is  crucial in clinical practice.


Despite not being directly related to ubiquitous computing, advances in using CS-DL for MRI uncover the importance of \textit{going beyond signal sparsity} and using other, expected, structure of the signal for practical reconstruction. In addition, CS-DL approaches for MRI indicate that signal reconstruction \textit{quality should be measured from an end-to-end perspective}: the quality of the reconstruction is only good if the result is properly interpreted by a human user. With ubiquitous devices being closely integrated into human everyday environments, it is important that sensing results' usability gets precedence over the simple mathematical formalization of the reconstruction error.

Wearable computing devices enable in-situ sampling of \textit{physiological signals}, such as breathing and heartbeat activity, skin temperature, skin conductance, and others. Recent advancements in the sampling and processing of ECG and EEG signals show that \textit{CS-DL methods can be used to lower the power consumption}, with respect to the classical sampling approach, and to enable real-time signal reconstruction or high level inferences in cases where such execution was infeasible with classical sampling methods. For example, the work in \cite{majumdar2015real}, uses a stacked autoencoder (SAE), for reconstructing the ECG compressed samples, and manages to achieve real time reconstruction. 
Shrivastwa et al. ~\cite{shrivastwa2018fpga} use an MLP network for ECoG (Electrocorticography) signals compression and reconstruction.
In \cite{singhal2017semi}, the authors go a step further and directly classify ECG and EEG signals in their compressive measurements space, achieving satisfying results with minimal computation, by skipping the reconstruction phase. Mangia et al.~ \cite{9044777} use support identification through DNN-based oracles, to first guess which components are non-zero in the sparse signal to recover, and then compute their magnitudes, thus decreasing the complexity of the computation.

\textit{Wireless connectivity} is a defining characteristic of ubiquitous computing. In the field of WSNs, CS-DL based techniques are motivated by not only the sparsity of the signals, but also by the requirement of efficiency in processing in terms of energy consumption and \textit{communication bandwidth utilization}. Moreover, the compressive measurements are transmitted over wireless channels, so \textit{the impact of channel noise and fading on reconstruction are important factors to be taken into consideration}. Some of the most important contributions in the field of CS-DL based methods for WSN are the WDLReconNet, and Fast-WDLReconNet networks~\cite{lu2019wdlreconnet}, validated for the transmission of remote sensing images.
Interestingly, the number of parameters in the denoising layer accounts for about 94\% of the total number of parameters in the proposed CNN, underlying the impact of the noise on the transmitted data.  Energy efficiency is another major aspect in WSN and quantization proved to be able to ensure an efficient wireless transmission~\cite{7560597}.

\subsection{From Challenges to Opportunities}

\textbf{\textit{Distributed computing}}. A good place for using CS-DL approaches is in the field of distributed computing; instead of performing the inference either solely on edge devices or exclusively in the cloud, distributed computing proposes an alternative approach based on splitting the computation between edge devices and the cloud. Partitioning data between different devices implies using an efficient compression technique to minimize the offload transmission overhead. 
Yao et al. \cite{yao2020deep} integrated the compressive sensing theory with deep neural networks, for providing offloading functionality for various applications.
This system, called Deep Compressive Offloading or DeepCOD, includes a lightweight encoder on the mobile side to compress the to-be-transferred data and a decoder on the edge server side to reconstruct the transferred data. The data for offloading is encoded on the local device and decoded on the edge server, trading edge computing resources for data transmission time, and thus significantly reducing the offloading latency with minimal accuracy loss. 

The most important challenge related to computation offloading is to decide whether, what, and how to offload, thus finding the best splitting point with the least latency among all possible neural network partitions candidates. DeepCOD addresses this issue by using a performance predictor that estimates the execution time of the NN operations on the local device and on the edge server, and a runtime partition decision maker to find the optimal partition point (from the latency perspective) for offloading.
This system was implemented on Android mobile devices and a Linux edge server with GPUs and reduced the offloading latency by a factor of 2 to 35 with at most 1\% accuracy loss under various mobile-edge-network configurations. Adding accuracy-variable distributed NN execution, for instance in the form of early NN exiting, as proposed by SPINN~\cite{laskaridis2020spinn}, would yield an interesting compression-splitting-approximation optimization space.

Finally, new distributed computing opportunities can also arise by intertwining compressive sensing with federated learning ~\cite{yang2019federated}. In a standard federated learning scenario, multiple client devices collaboratively train a model, with local data. After decentralized local training on edge devices, the network parameters, such as weights or gradients are exchanged, which can cause communication bottlenecks and delays.
A CS-DL approach can be used to efficiently encode the network parameters on the edge device and decode them on the server side. In this manner, only the compressed version of the gradient, for example, needs to be shared, which can reduce the communication bandwidth.

Given the increasingly crucial role that NN play in IoT systems applications, further implementations that combine CS with DL for distributed computing are likely to offer solutions in domains such as task distribution in a wireless network, mobile edge deep learning, and satellite-terrestrial computation partitioning. 


\textbf{\textit{Heterogeneous architectures}}. CS-DL is ideally positioned to efficiently utilize heterogeneous architectures of today's ubiquitous computing landscape, such as the ARM big.LITTLE architecture~\cite{ARMBiglittle}, which integrate slower power-efficient processor cores with faster  power-hungry cores. By mapping an application to the optimal cores, considering the performance demands and power availability, important power savings can be achieved. Matching dynamic computational requirements with the underlying hardware capabilities is not easy, and for deep learning models, several works addressed the efficient exploitation of the hardware architecture of heterogeneous devices. 
The DeepX framework ~\cite{lane2016deepx} dynamically decomposes a neural network architecture into segments that can each be executed across different processors to maximize energy-efficiency and execution time. 
DeepSense ~\cite{huynh2016deepsense} leverages the architecture of mobile GPU devices and implies optimization techniques for optimal offloading a neural network' layers and operations on on the CPU/GPU memory and processors to achieve the best accuracy-latency trade-off.

With its adaptability, afforded by sampling rate adjustment and NN compression, CS-DL can be tuned to maximally utilize  heterogeneous hardware. An adaptable CS-DL pipeline was explored in \cite{shrivastwa2018fpga}. The authors 
aimed at porting a deep neural network for ECoG signals compression and reconstruction to a lightweight device and explores three architectural options: using a greedy algorithm (orthogonal matching pursuit), signal compression and reconstruction using a MLP with all layers implemented in the FPGA logic, and finally a heterogeneous architecture consisting of an ARM CPU and FPGA fabric, with just a single layer of the NN being deployed in the FPGA re-configurable logic.
Measurements demonstrate that the third,  heterogeneous architecture, stands out as the most efficient.
When implementing the full NN in the FPGA and connecting the layers in a cascaded fashion, the authors find that such an implementation results in significant area overhead and requires a large number of multipliers. By comparison, the heterogeneous solution is much more resource efficient -- the FPGA logic accommodates just a single layer of the network at any given time and stores the network's parameters in two BRAMs. The ARM core in this case controls the compression and reconstruction operations in the FPGA fabric. While this approach requires repeated calls to execute the entire operation (loading each layer's parameters every time in the FPGA logic), this does not impact the runtime performance given that the ECoG signals are sampled at a much lower frequency than the one that the implementation is running at. This system was realized using a Zynq processing system (ARM core) in Zedboard, and opens the door for future explorations of efficient task mapping of CS-DL implementations on heterogeneous architectures.
Perhaps the most promising avenue for research lies in energy-efficient task scheduling of a CS-DL pipeline on a mobile device equipped with a heterogeneous hardware architecture and hardware-software codesign for CS-DL. The scheduling would address the optimal task to processor assignment for achieving minimum energy consumption, which is especially important as we expect a range of advanced mobile sensing functionalities, such as speech recognition and live video processing, from our battery powered devices. The hardware-software codesign would ensure that sensors are built with compressive sensing in mind, while the processing pipeline matches the needs of the CS-DL algorithms. With respect to the latter, FPGAs stand out as likely candidate for initial implementations, due to the processing capabilities -- flexibility balance they afford.

\section{Conclusions}
\label{sec:conclusions}

\addvspace{0.2cm}
\noindent\fbox{
    \parbox{0.98\textwidth}{%
    \addvspace{0.2cm}
    \textbf{Key takeaway ideas}
    \begin{itemize}
        \item Data-driven measurement matrix proved to enhance the CS method in several cases, when compared with the conventional CS methods using random measurements.
        \item CS-DL methods showed a consistent speed up, being two orders of magnitude faster than the traditional CS algorithms.
        \item The trade-off between model performance and the number of network parameters can be addressed using residual blocks.
        \item Especially at very aggressive undersampling rates, the CS-DL methods are capable of better reconstructions than most of the classical methods.
        \item Training CS-DL pipelines requires significant computing and data resources, which might be alleviated with transfer and federated learning.
    \end{itemize}
    \addvspace{0.24cm}
    }%
}

The move from centralized storage and processing towards distributed and edge computing indicates that the  intelligence that is expected from future ubiquitous computing environments needs to be realized as close to the physical world as possible. Consequently, sensing and learning from the collected data need to be implemented directly on the ubiquitous sensing devices, and with the support for adaptive, dynamic distributed processing. Reduced rate sampling enabled by compressive sensing (CS) represents a viable solution enabling the reduction of the amount of generated sensing data, yet CS alone does not solve the issue of complex processing that may be overwhelming for ubicomp devices’ limited computational resources. Deep learning (DL) naturally complements CS in the ubicomp domain, as it reduces the computational complexity of signal reconstruction and enables full sensing-learning pipelines to be implemented. 

Despite its potential, the CS-DL area remains only sporadically explored. In this survey we identified the  current trends in the CS-DL area and reviewed some of the most significant recent efforts. We systematically examined how DL can be used to speed up CS signal reconstruction by alleviating the need for iterative algorithms. Furthermore,  classic CS methods were not designed to go beyond sparsity and exploit structure present in the data.  DL enables for the sampling matrix to be designed according to the hidden data structure that can further be exploited in the reconstruction phase. The trade-off between model performance and the number of network parameters represents a major issue in CS-DL. It has been shown that deeper network architectures can result in better network performance, yet increasing model complexity requires more intensive computational and memory requirements. Residual blocks represent a viable solution  for addressing this trade-off \cite{yao2019dr2, du2019fully}. Regarding the compression rate, studies \cite{kulkarni2016reconnet,  schlemper2017deep, yao2019dr2, shrivastwa2020brain} showed that at very aggressive undersampling rates, the DL based methods are capable of better reconstructions than most of the classical methods. For example, the ReconNet network outperforms other methods by large margins at measurement rates of up to 0.01. Finally, one of the drawbacks of accurately reconstructing signals from few measurements using DL, is the high requirements in terms of time and data for training. Transfer learning might be a solution for this issue, as shown in ~ \cite{han2018deep}.

Although compressive sensing is a relatively new field, being around for less than two decades, with deep learning being an even newer addition, CS-DL is characterized by a burgeoning community that produces a growing body of freely available online educational resources are available. A broader collection of resources ranging from conference or journal papers and tutorials to blogs, software tools and video talks can be found at \url{http://dsp.rice.edu/cs/}). In addition, novel ideas and methods in this area are often accompanied by free and open-source code of the implementations. A useful repository  containing a collection of reproducible Deep Compressive Sensing source code can be found at \url{https://github.com/ngcthuong/Reproducible-Deep-Compressive-Sensing}.  

In this survey we presented mostly academic works at the intersection of CS and DL, aiming to provide a valuable resource for future researchers and practitioners in this domain. Furthermore, the survey aims to attract new audience to CS-DL, primarily ubiquitous systems researchers. Such expansion is crucial, as challenges identified in this manuscript, including the realization of distributed CS-DL on heterogeneous architectures and with support for dynamically adaptive sampling rates need to be addressed in order to ensure further proliferation of sensing systems’ intelligence.




\bibliographystyle{ACM-Reference-Format}
\bibliography{sample-base}


\begin{thebibliography}{166}


\ifx \showCODEN    \undefined \def \showCODEN     #1{\unskip}     \fi
\ifx \showDOI      \undefined \def \showDOI       #1{#1}\fi
\ifx \showISBNx    \undefined \def \showISBNx     #1{\unskip}     \fi
\ifx \showISBNxiii \undefined \def \showISBNxiii  #1{\unskip}     \fi
\ifx \showISSN     \undefined \def \showISSN      #1{\unskip}     \fi
\ifx \showLCCN     \undefined \def \showLCCN      #1{\unskip}     \fi
\ifx \shownote     \undefined \def \shownote      #1{#1}          \fi
\ifx \showarticletitle \undefined \def \showarticletitle #1{#1}   \fi
\ifx \showURL      \undefined \def \showURL       {\relax}        \fi
\providecommand\bibfield[2]{#2}
\providecommand\bibinfo[2]{#2}
\providecommand\natexlab[1]{#1}
\providecommand\showeprint[2][]{arXiv:#2}

\bibitem[\protect\citeauthoryear{Ablin, Moreau, Massias, and Gramfort}{Ablin
  et~al\mbox{.}}{2019}]%
        {ablin2019learning}
\bibfield{author}{\bibinfo{person}{Pierre Ablin}, \bibinfo{person}{Thomas
  Moreau}, \bibinfo{person}{Mathurin Massias}, {and} \bibinfo{person}{Alexandre
  Gramfort}.} \bibinfo{year}{2019}\natexlab{}.
\newblock \showarticletitle{Learning step sizes for unfolded sparse coding}. In
  \bibinfo{booktitle}{\emph{Advances in Neural Information Processing
  Systems}}. \bibinfo{pages}{13100--13110}.
\newblock


\bibitem[\protect\citeauthoryear{Adler, Boublil, Elad, and Zibulevsky}{Adler
  et~al\mbox{.}}{2016a}]%
        {adler2016deep}
\bibfield{author}{\bibinfo{person}{Amir Adler}, \bibinfo{person}{David
  Boublil}, \bibinfo{person}{Michael Elad}, {and} \bibinfo{person}{Michael
  Zibulevsky}.} \bibinfo{year}{2016}\natexlab{a}.
\newblock \showarticletitle{A deep learning approach to block-based compressed
  sensing of images}.
\newblock \bibinfo{journal}{\emph{arXiv preprint arXiv:1606.01519}}
  (\bibinfo{year}{2016}).
\newblock


\bibitem[\protect\citeauthoryear{Adler, Elad, and Zibulevsky}{Adler
  et~al\mbox{.}}{2016b}]%
        {adler2016compressed}
\bibfield{author}{\bibinfo{person}{Amir Adler}, \bibinfo{person}{Michael Elad},
  {and} \bibinfo{person}{Michael Zibulevsky}.}
  \bibinfo{year}{2016}\natexlab{b}.
\newblock \showarticletitle{Compressed learning: A deep neural network
  approach}.
\newblock \bibinfo{journal}{\emph{arXiv preprint arXiv:1610.09615}}
  (\bibinfo{year}{2016}).
\newblock


\bibitem[\protect\citeauthoryear{Aggarwal et~al\mbox{.}}{Aggarwal
  et~al\mbox{.}}{2018}]%
        {aggarwal2018neural}
\bibfield{author}{\bibinfo{person}{Charu~C Aggarwal} {et~al\mbox{.}}}
  \bibinfo{year}{2018}\natexlab{}.
\newblock \showarticletitle{Neural networks and deep learning}.
\newblock \bibinfo{journal}{\emph{Springer}}  \bibinfo{volume}{10}
  (\bibinfo{year}{2018}), \bibinfo{pages}{978--3}.
\newblock


\bibitem[\protect\citeauthoryear{Al-Azawi and Gaze}{Al-Azawi and Gaze}{2017}]%
        {al2017combined}
\bibfield{author}{\bibinfo{person}{Maher K~Mahmood Al-Azawi} {and}
  \bibinfo{person}{Ali~M Gaze}.} \bibinfo{year}{2017}\natexlab{}.
\newblock \showarticletitle{Combined speech compression and encryption using
  chaotic compressive sensing with large key size}.
\newblock \bibinfo{journal}{\emph{IET Signal Processing}} \bibinfo{volume}{12},
  \bibinfo{number}{2} (\bibinfo{year}{2017}), \bibinfo{pages}{214--218}.
\newblock


\bibitem[\protect\citeauthoryear{Bacca, Galvis, and Arguello}{Bacca
  et~al\mbox{.}}{2020}]%
        {bacca2020coupled}
\bibfield{author}{\bibinfo{person}{Jorge Bacca}, \bibinfo{person}{Laura
  Galvis}, {and} \bibinfo{person}{Henry Arguello}.}
  \bibinfo{year}{2020}\natexlab{}.
\newblock \showarticletitle{Coupled deep learning coded aperture design for
  compressive image classification}.
\newblock \bibinfo{journal}{\emph{Optics express}} \bibinfo{volume}{28},
  \bibinfo{number}{6} (\bibinfo{year}{2020}), \bibinfo{pages}{8528--8540}.
\newblock


\bibitem[\protect\citeauthoryear{Beck and Teboulle}{Beck and Teboulle}{2009}]%
        {beck2009fast}
\bibfield{author}{\bibinfo{person}{Amir Beck} {and} \bibinfo{person}{Marc
  Teboulle}.} \bibinfo{year}{2009}\natexlab{}.
\newblock \showarticletitle{A fast iterative shrinkage-thresholding algorithm
  for linear inverse problems}.
\newblock \bibinfo{journal}{\emph{SIAM journal on imaging sciences}}
  \bibinfo{volume}{2}, \bibinfo{number}{1} (\bibinfo{year}{2009}),
  \bibinfo{pages}{183--202}.
\newblock


\bibitem[\protect\citeauthoryear{Bian and Xue}{Bian and Xue}{2012}]%
        {bian2012neural}
\bibfield{author}{\bibinfo{person}{Wei Bian} {and} \bibinfo{person}{Xiaoping
  Xue}.} \bibinfo{year}{2012}\natexlab{}.
\newblock \showarticletitle{Neural network for solving constrained convex
  optimization problems with global attractivity}.
\newblock \bibinfo{journal}{\emph{IEEE Transactions on Circuits and Systems I:
  Regular Papers}} \bibinfo{volume}{60}, \bibinfo{number}{3}
  (\bibinfo{year}{2012}), \bibinfo{pages}{710--723}.
\newblock


\bibitem[\protect\citeauthoryear{Bishop et~al\mbox{.}}{Bishop
  et~al\mbox{.}}{1995}]%
        {bishop1995neural}
\bibfield{author}{\bibinfo{person}{Christopher~M Bishop} {et~al\mbox{.}}}
  \bibinfo{year}{1995}\natexlab{}.
\newblock \bibinfo{booktitle}{\emph{Neural networks for pattern recognition}}.
\newblock \bibinfo{publisher}{Oxford university press}.
\newblock


\bibitem[\protect\citeauthoryear{Blumensath and Davies}{Blumensath and
  Davies}{2009}]%
        {blumensath2009iterative}
\bibfield{author}{\bibinfo{person}{Thomas Blumensath} {and}
  \bibinfo{person}{Mike~E Davies}.} \bibinfo{year}{2009}\natexlab{}.
\newblock \showarticletitle{Iterative hard thresholding for compressed
  sensing}.
\newblock \bibinfo{journal}{\emph{Applied and computational harmonic analysis}}
  \bibinfo{volume}{27}, \bibinfo{number}{3} (\bibinfo{year}{2009}),
  \bibinfo{pages}{265--274}.
\newblock


\bibitem[\protect\citeauthoryear{Borgerding, Schniter, and Rangan}{Borgerding
  et~al\mbox{.}}{2017}]%
        {borgerding2017amp}
\bibfield{author}{\bibinfo{person}{Mark Borgerding}, \bibinfo{person}{Philip
  Schniter}, {and} \bibinfo{person}{Sundeep Rangan}.}
  \bibinfo{year}{2017}\natexlab{}.
\newblock \showarticletitle{AMP-inspired deep networks for sparse linear
  inverse problems}.
\newblock \bibinfo{journal}{\emph{IEEE Transactions on Signal Processing}}
  \bibinfo{volume}{65}, \bibinfo{number}{16} (\bibinfo{year}{2017}),
  \bibinfo{pages}{4293--4308}.
\newblock


\bibitem[\protect\citeauthoryear{Boyd, Parikh, and Chu}{Boyd
  et~al\mbox{.}}{2011}]%
        {boyd2011distributed}
\bibfield{author}{\bibinfo{person}{Stephen Boyd}, \bibinfo{person}{Neal
  Parikh}, {and} \bibinfo{person}{Eric Chu}.} \bibinfo{year}{2011}\natexlab{}.
\newblock \bibinfo{booktitle}{\emph{Distributed optimization and statistical
  learning via the alternating direction method of multipliers}}.
\newblock \bibinfo{publisher}{Now Publishers Inc}.
\newblock


\bibitem[\protect\citeauthoryear{Calderbank, Jafarpour, and
  Schapire}{Calderbank et~al\mbox{.}}{2009}]%
        {calderbank2009compressed}
\bibfield{author}{\bibinfo{person}{Robert Calderbank}, \bibinfo{person}{Sina
  Jafarpour}, {and} \bibinfo{person}{Robert Schapire}.}
  \bibinfo{year}{2009}\natexlab{}.
\newblock \showarticletitle{Compressed learning: Universal sparse
  dimensionality reduction and learning in the measurement domain}.
\newblock \bibinfo{journal}{\emph{preprint}} (\bibinfo{year}{2009}).
\newblock


\bibitem[\protect\citeauthoryear{Candes, Tao, et~al\mbox{.}}{Candes
  et~al\mbox{.}}{2007}]%
        {candes2007dantzig}
\bibfield{author}{\bibinfo{person}{Emmanuel Candes}, \bibinfo{person}{Terence
  Tao}, {et~al\mbox{.}}} \bibinfo{year}{2007}\natexlab{}.
\newblock \showarticletitle{The Dantzig selector: Statistical estimation when p
  is much larger than n}.
\newblock \bibinfo{journal}{\emph{The annals of Statistics}}
  \bibinfo{volume}{35}, \bibinfo{number}{6} (\bibinfo{year}{2007}),
  \bibinfo{pages}{2313--2351}.
\newblock


\bibitem[\protect\citeauthoryear{Candes and Romberg}{Candes and
  Romberg}{2006}]%
        {candes2006quantitative}
\bibfield{author}{\bibinfo{person}{Emmanuel~J Candes} {and}
  \bibinfo{person}{Justin Romberg}.} \bibinfo{year}{2006}\natexlab{}.
\newblock \showarticletitle{Quantitative robust uncertainty principles and
  optimally sparse decompositions}.
\newblock \bibinfo{journal}{\emph{Foundations of Computational Mathematics}}
  \bibinfo{volume}{6}, \bibinfo{number}{2} (\bibinfo{year}{2006}),
  \bibinfo{pages}{227--254}.
\newblock


\bibitem[\protect\citeauthoryear{Cand{\`e}s, Romberg, and Tao}{Cand{\`e}s
  et~al\mbox{.}}{2006}]%
        {candes2006robust}
\bibfield{author}{\bibinfo{person}{Emmanuel~J Cand{\`e}s},
  \bibinfo{person}{Justin Romberg}, {and} \bibinfo{person}{Terence Tao}.}
  \bibinfo{year}{2006}\natexlab{}.
\newblock \showarticletitle{Robust uncertainty principles: Exact signal
  reconstruction from highly incomplete frequency information}.
\newblock \bibinfo{journal}{\emph{IEEE Transactions on information theory}}
  \bibinfo{volume}{52}, \bibinfo{number}{2} (\bibinfo{year}{2006}),
  \bibinfo{pages}{489--509}.
\newblock


\bibitem[\protect\citeauthoryear{Candes, Romberg, and Tao}{Candes
  et~al\mbox{.}}{2006}]%
        {candes2006stable}
\bibfield{author}{\bibinfo{person}{Emmanuel~J Candes},
  \bibinfo{person}{Justin~K Romberg}, {and} \bibinfo{person}{Terence Tao}.}
  \bibinfo{year}{2006}\natexlab{}.
\newblock \showarticletitle{Stable signal recovery from incomplete and
  inaccurate measurements}.
\newblock \bibinfo{journal}{\emph{Communications on Pure and Applied
  Mathematics: A Journal Issued by the Courant Institute of Mathematical
  Sciences}} \bibinfo{volume}{59}, \bibinfo{number}{8} (\bibinfo{year}{2006}),
  \bibinfo{pages}{1207--1223}.
\newblock


\bibitem[\protect\citeauthoryear{Canh and Jeon}{Canh and Jeon}{2018}]%
        {canh2018deep}
\bibfield{author}{\bibinfo{person}{Thuong~Nguyen Canh} {and}
  \bibinfo{person}{Byeungwoo Jeon}.} \bibinfo{year}{2018}\natexlab{}.
\newblock \showarticletitle{Deep learning-based Kronecker compressive imaging}.
  In \bibinfo{booktitle}{\emph{IEEE International Conference on Consumer
  Electronic-Asia (ICCE-A)}}. \bibinfo{pages}{1--4}.
\newblock


\bibitem[\protect\citeauthoryear{Chartrand and Yin}{Chartrand and Yin}{2008}]%
        {chartrand2008iteratively}
\bibfield{author}{\bibinfo{person}{Rick Chartrand} {and} \bibinfo{person}{Wotao
  Yin}.} \bibinfo{year}{2008}\natexlab{}.
\newblock \showarticletitle{Iteratively reweighted algorithms for compressive
  sensing}. In \bibinfo{booktitle}{\emph{2008 IEEE International Conference on
  Acoustics, Speech and Signal Processing}}. IEEE, \bibinfo{pages}{3869--3872}.
\newblock


\bibitem[\protect\citeauthoryear{Chen, Liu, Wan, Qiao, and Pei}{Chen
  et~al\mbox{.}}{2021}]%
        {chen2021edge}
\bibfield{author}{\bibinfo{person}{Chen Chen}, \bibinfo{person}{Bin Liu},
  \bibinfo{person}{Shaohua Wan}, \bibinfo{person}{Peng Qiao}, {and}
  \bibinfo{person}{Qingqi Pei}.} \bibinfo{year}{2021}\natexlab{}.
\newblock \showarticletitle{An edge traffic flow detection scheme based on deep
  learning in an intelligent transportation system}.
\newblock \bibinfo{journal}{\emph{IEEE Transactions on Intelligent
  Transportation Systems}} \bibinfo{volume}{22}, \bibinfo{number}{3}
  (\bibinfo{year}{2021}), \bibinfo{pages}{1840--1852}.
\newblock


\bibitem[\protect\citeauthoryear{Chen, Donoho, and Saunders}{Chen
  et~al\mbox{.}}{2001}]%
        {chen2001atomic}
\bibfield{author}{\bibinfo{person}{Scott~Shaobing Chen},
  \bibinfo{person}{David~L Donoho}, {and} \bibinfo{person}{Michael~A
  Saunders}.} \bibinfo{year}{2001}\natexlab{}.
\newblock \showarticletitle{Atomic decomposition by basis pursuit}.
\newblock \bibinfo{journal}{\emph{SIAM review}} \bibinfo{volume}{43},
  \bibinfo{number}{1} (\bibinfo{year}{2001}), \bibinfo{pages}{129--159}.
\newblock


\bibitem[\protect\citeauthoryear{Chen, Yang, Li, and Zhang}{Chen
  et~al\mbox{.}}{2020}]%
        {chen2020binarized}
\bibfield{author}{\bibinfo{person}{Yanming Chen}, \bibinfo{person}{Tianbo
  Yang}, \bibinfo{person}{Chao Li}, {and} \bibinfo{person}{Yiwen Zhang}.}
  \bibinfo{year}{2020}\natexlab{}.
\newblock \showarticletitle{A Binarized Segmented ResNet Based on Edge
  Computing for Re-Identification}.
\newblock \bibinfo{journal}{\emph{Sensors}} \bibinfo{volume}{20},
  \bibinfo{number}{23} (\bibinfo{year}{2020}), \bibinfo{pages}{6902}.
\newblock


\bibitem[\protect\citeauthoryear{Cheng, Wang, Zhou, and Zhang}{Cheng
  et~al\mbox{.}}{2017}]%
        {cheng2017survey}
\bibfield{author}{\bibinfo{person}{Yu Cheng}, \bibinfo{person}{Duo Wang},
  \bibinfo{person}{Pan Zhou}, {and} \bibinfo{person}{Tao Zhang}.}
  \bibinfo{year}{2017}\natexlab{}.
\newblock \showarticletitle{A survey of model compression and acceleration for
  deep neural networks}.
\newblock \bibinfo{journal}{\emph{arXiv preprint arXiv:1710.09282}}
  (\bibinfo{year}{2017}).
\newblock


\bibitem[\protect\citeauthoryear{Chiew, Graedel, and Miller}{Chiew
  et~al\mbox{.}}{2018}]%
        {chiew2018recovering}
\bibfield{author}{\bibinfo{person}{Mark Chiew}, \bibinfo{person}{Nadine~N
  Graedel}, {and} \bibinfo{person}{Karla~L Miller}.}
  \bibinfo{year}{2018}\natexlab{}.
\newblock \showarticletitle{Recovering task fMRI signals from highly
  under-sampled data with low-rank and temporal subspace constraints}.
\newblock \bibinfo{journal}{\emph{NeuroImage}}  \bibinfo{volume}{174}
  (\bibinfo{year}{2018}), \bibinfo{pages}{97--110}.
\newblock


\bibitem[\protect\citeauthoryear{Choudhary, Mishra, Goswami, and
  Sarangapani}{Choudhary et~al\mbox{.}}{2020}]%
        {choudhary2020comprehensive}
\bibfield{author}{\bibinfo{person}{Tejalal Choudhary}, \bibinfo{person}{Vipul
  Mishra}, \bibinfo{person}{Anurag Goswami}, {and} \bibinfo{person}{Jagannathan
  Sarangapani}.} \bibinfo{year}{2020}\natexlab{}.
\newblock \showarticletitle{A comprehensive survey on model compression and
  acceleration}.
\newblock \bibinfo{journal}{\emph{Artificial Intelligence Review}}
  (\bibinfo{year}{2020}), \bibinfo{pages}{1--43}.
\newblock


\bibitem[\protect\citeauthoryear{Davenport, Duarte, Wakin, Laska, Takhar,
  Kelly, and Baraniuk}{Davenport et~al\mbox{.}}{2007}]%
        {davenport2007smashed}
\bibfield{author}{\bibinfo{person}{Mark~A Davenport}, \bibinfo{person}{Marco~F
  Duarte}, \bibinfo{person}{Michael~B Wakin}, \bibinfo{person}{Jason~N Laska},
  \bibinfo{person}{Dharmpal Takhar}, \bibinfo{person}{Kevin~F Kelly}, {and}
  \bibinfo{person}{Richard~G Baraniuk}.} \bibinfo{year}{2007}\natexlab{}.
\newblock \showarticletitle{The smashed filter for compressive classification
  and target recognition}. In \bibinfo{booktitle}{\emph{Computational Imaging
  V}}, Vol.~\bibinfo{volume}{6498}. International Society for Optics and
  Photonics, \bibinfo{pages}{64980H}.
\newblock


\bibitem[\protect\citeauthoryear{Deglint, Jin, and Wong}{Deglint
  et~al\mbox{.}}{2019}]%
        {deglint2019investigating}
\bibfield{author}{\bibinfo{person}{Jason~L Deglint}, \bibinfo{person}{Chao
  Jin}, {and} \bibinfo{person}{Alexander Wong}.}
  \bibinfo{year}{2019}\natexlab{}.
\newblock \showarticletitle{Investigating the automatic classification of algae
  using the spectral and morphological characteristics via deep residual
  learning}. In \bibinfo{booktitle}{\emph{International Conference on Image
  Analysis and Recognition}}. Springer, \bibinfo{pages}{269--280}.
\newblock


\bibitem[\protect\citeauthoryear{Djelouat, Amira, and Bensaali}{Djelouat
  et~al\mbox{.}}{2018}]%
        {djelouat2018compressive}
\bibfield{author}{\bibinfo{person}{Hamza Djelouat}, \bibinfo{person}{Abbes
  Amira}, {and} \bibinfo{person}{Faycal Bensaali}.}
  \bibinfo{year}{2018}\natexlab{}.
\newblock \showarticletitle{Compressive sensing-based IoT applications: A
  review}.
\newblock \bibinfo{journal}{\emph{Journal of Sensor and Actuator Networks}}
  \bibinfo{volume}{7}, \bibinfo{number}{4} (\bibinfo{year}{2018}),
  \bibinfo{pages}{45}.
\newblock


\bibitem[\protect\citeauthoryear{Donoho and Tanner}{Donoho and Tanner}{2009}]%
        {donoho2009counting}
\bibfield{author}{\bibinfo{person}{David Donoho} {and} \bibinfo{person}{Jared
  Tanner}.} \bibinfo{year}{2009}\natexlab{}.
\newblock \showarticletitle{Counting faces of randomly projected polytopes when
  the projection radically lowers dimension}.
\newblock \bibinfo{journal}{\emph{Journal of the American Mathematical
  Society}} \bibinfo{volume}{22}, \bibinfo{number}{1} (\bibinfo{year}{2009}),
  \bibinfo{pages}{1--53}.
\newblock


\bibitem[\protect\citeauthoryear{Donoho}{Donoho}{2006}]%
        {donoho2006compressed}
\bibfield{author}{\bibinfo{person}{David~L Donoho}.}
  \bibinfo{year}{2006}\natexlab{}.
\newblock \showarticletitle{Compressed sensing}.
\newblock \bibinfo{journal}{\emph{IEEE Transactions on information theory}}
  \bibinfo{volume}{52}, \bibinfo{number}{4} (\bibinfo{year}{2006}),
  \bibinfo{pages}{1289--1306}.
\newblock


\bibitem[\protect\citeauthoryear{Donoho, Maleki, and Montanari}{Donoho
  et~al\mbox{.}}{2009}]%
        {donoho2009message}
\bibfield{author}{\bibinfo{person}{David~L Donoho}, \bibinfo{person}{Arian
  Maleki}, {and} \bibinfo{person}{Andrea Montanari}.}
  \bibinfo{year}{2009}\natexlab{}.
\newblock \showarticletitle{Message-passing algorithms for compressed sensing}.
\newblock \bibinfo{journal}{\emph{Proceedings of the National Academy of
  Sciences}} \bibinfo{volume}{106}, \bibinfo{number}{45}
  (\bibinfo{year}{2009}), \bibinfo{pages}{18914--18919}.
\newblock


\bibitem[\protect\citeauthoryear{Donoho, Tsaig, Drori, and Starck}{Donoho
  et~al\mbox{.}}{2012}]%
        {donoho2012sparse}
\bibfield{author}{\bibinfo{person}{David~L Donoho}, \bibinfo{person}{Yaakov
  Tsaig}, \bibinfo{person}{Iddo Drori}, {and} \bibinfo{person}{Jean-Luc
  Starck}.} \bibinfo{year}{2012}\natexlab{}.
\newblock \showarticletitle{Sparse solution of underdetermined systems of
  linear equations by stagewise orthogonal matching pursuit}.
\newblock \bibinfo{journal}{\emph{IEEE transactions on Information Theory}}
  \bibinfo{volume}{58}, \bibinfo{number}{2} (\bibinfo{year}{2012}),
  \bibinfo{pages}{1094--1121}.
\newblock


\bibitem[\protect\citeauthoryear{Du, Xie, Wang, Shi, Xu, and Wang}{Du
  et~al\mbox{.}}{2019}]%
        {du2019fully}
\bibfield{author}{\bibinfo{person}{Jiang Du}, \bibinfo{person}{Xuemei Xie},
  \bibinfo{person}{Chenye Wang}, \bibinfo{person}{Guangming Shi},
  \bibinfo{person}{Xun Xu}, {and} \bibinfo{person}{Yuxiang Wang}.}
  \bibinfo{year}{2019}\natexlab{}.
\newblock \showarticletitle{Fully convolutional measurement network for
  compressive sensing image reconstruction}.
\newblock \bibinfo{journal}{\emph{Neurocomputing}}  \bibinfo{volume}{328}
  (\bibinfo{year}{2019}), \bibinfo{pages}{105--112}.
\newblock


\bibitem[\protect\citeauthoryear{Eldar and Kutyniok}{Eldar and
  Kutyniok}{2012}]%
        {eldar2012compressed}
\bibfield{author}{\bibinfo{person}{Yonina~C Eldar} {and} \bibinfo{person}{Gitta
  Kutyniok}.} \bibinfo{year}{2012}\natexlab{}.
\newblock \bibinfo{booktitle}{\emph{Compressed sensing: theory and
  applications}}.
\newblock \bibinfo{publisher}{Cambridge university press}.
\newblock


\bibitem[\protect\citeauthoryear{Feng, Sun, and Zhu}{Feng
  et~al\mbox{.}}{2019}]%
        {feng2019robust}
\bibfield{author}{\bibinfo{person}{Lei Feng}, \bibinfo{person}{Huaijiang Sun},
  {and} \bibinfo{person}{Jun Zhu}.} \bibinfo{year}{2019}\natexlab{}.
\newblock \showarticletitle{Robust image compressive sensing based on
  half-quadratic function and weighted schatten-p norm}.
\newblock \bibinfo{journal}{\emph{Information Sciences}}  \bibinfo{volume}{477}
  (\bibinfo{year}{2019}), \bibinfo{pages}{265--280}.
\newblock


\bibitem[\protect\citeauthoryear{Figueiredo, Nowak, and Wright}{Figueiredo
  et~al\mbox{.}}{2007}]%
        {figueiredo2007gradient}
\bibfield{author}{\bibinfo{person}{M{\'a}rio~AT Figueiredo},
  \bibinfo{person}{Robert~D Nowak}, {and} \bibinfo{person}{Stephen~J Wright}.}
  \bibinfo{year}{2007}\natexlab{}.
\newblock \showarticletitle{Gradient projection for sparse reconstruction:
  Application to compressed sensing and other inverse problems}.
\newblock \bibinfo{journal}{\emph{IEEE Journal of selected topics in signal
  processing}} \bibinfo{volume}{1}, \bibinfo{number}{4} (\bibinfo{year}{2007}),
  \bibinfo{pages}{586--597}.
\newblock


\bibitem[\protect\citeauthoryear{Garg and Khandekar}{Garg and
  Khandekar}{2009}]%
        {garg2009gradient}
\bibfield{author}{\bibinfo{person}{Rahul Garg} {and} \bibinfo{person}{Rohit
  Khandekar}.} \bibinfo{year}{2009}\natexlab{}.
\newblock \showarticletitle{Gradient descent with sparsification: an iterative
  algorithm for sparse recovery with restricted isometry property}. In
  \bibinfo{booktitle}{\emph{Proceedings of the 26th annual international
  conference on machine learning}}. \bibinfo{pages}{337--344}.
\newblock


\bibitem[\protect\citeauthoryear{Goodfellow, Bengio, Courville, and
  Bengio}{Goodfellow et~al\mbox{.}}{2016}]%
        {goodfellow2016deep}
\bibfield{author}{\bibinfo{person}{Ian Goodfellow}, \bibinfo{person}{Yoshua
  Bengio}, \bibinfo{person}{Aaron Courville}, {and} \bibinfo{person}{Yoshua
  Bengio}.} \bibinfo{year}{2016}\natexlab{}.
\newblock \bibinfo{booktitle}{\emph{Deep learning}}. Vol.~\bibinfo{volume}{1}.
\newblock \bibinfo{publisher}{MIT press Cambridge}.
\newblock


\bibitem[\protect\citeauthoryear{Gorodnitsky and Rao}{Gorodnitsky and
  Rao}{1997}]%
        {gorodnitsky1997sparse}
\bibfield{author}{\bibinfo{person}{Irina~F Gorodnitsky} {and}
  \bibinfo{person}{Bhaskar~D Rao}.} \bibinfo{year}{1997}\natexlab{}.
\newblock \showarticletitle{Sparse signal reconstruction from limited data
  using FOCUSS: A re-weighted minimum norm algorithm}.
\newblock \bibinfo{journal}{\emph{IEEE Transactions on signal processing}}
  \bibinfo{volume}{45}, \bibinfo{number}{3} (\bibinfo{year}{1997}),
  \bibinfo{pages}{600--616}.
\newblock


\bibitem[\protect\citeauthoryear{Gregor and LeCun}{Gregor and LeCun}{2010}]%
        {gregor2010learning}
\bibfield{author}{\bibinfo{person}{Karol Gregor} {and} \bibinfo{person}{Yann
  LeCun}.} \bibinfo{year}{2010}\natexlab{}.
\newblock \showarticletitle{Learning fast approximations of sparse coding}. In
  \bibinfo{booktitle}{\emph{Proceedings of the 27th international conference on
  international conference on machine learning}}. \bibinfo{pages}{399--406}.
\newblock


\bibitem[\protect\citeauthoryear{Grover and Ermon}{Grover and Ermon}{2019}]%
        {grover2019uncertainty}
\bibfield{author}{\bibinfo{person}{Aditya Grover} {and}
  \bibinfo{person}{Stefano Ermon}.} \bibinfo{year}{2019}\natexlab{}.
\newblock \showarticletitle{Uncertainty autoencoders: Learning compressed
  representations via variational information maximization}. In
  \bibinfo{booktitle}{\emph{The 22nd International Conference on Artificial
  Intelligence and Statistics}}. \bibinfo{pages}{2514--2524}.
\newblock


\bibitem[\protect\citeauthoryear{Gupta, Anand, Kaushik, Chaudhury, and
  Lall}{Gupta et~al\mbox{.}}{2019}]%
        {gupta2019data}
\bibfield{author}{\bibinfo{person}{Ronak Gupta}, \bibinfo{person}{Prashant
  Anand}, \bibinfo{person}{Vinay Kaushik}, \bibinfo{person}{Santanu Chaudhury},
  {and} \bibinfo{person}{Brejesh Lall}.} \bibinfo{year}{2019}\natexlab{}.
\newblock \showarticletitle{Data Driven Sensing for Action Recognition Using
  Deep Convolutional Neural Networks}. In
  \bibinfo{booktitle}{\emph{International Conference on Pattern Recognition and
  Machine Intelligence}}. Springer, \bibinfo{pages}{250--259}.
\newblock


\bibitem[\protect\citeauthoryear{Gupta, Agrawal, Gopalakrishnan, and
  Narayanan}{Gupta et~al\mbox{.}}{2015}]%
        {gupta2015deep}
\bibfield{author}{\bibinfo{person}{Suyog Gupta}, \bibinfo{person}{Ankur
  Agrawal}, \bibinfo{person}{Kailash Gopalakrishnan}, {and}
  \bibinfo{person}{Pritish Narayanan}.} \bibinfo{year}{2015}\natexlab{}.
\newblock \showarticletitle{Deep learning with limited numerical precision}. In
  \bibinfo{booktitle}{\emph{International conference on machine learning}}.
  PMLR, \bibinfo{pages}{1737--1746}.
\newblock


\bibitem[\protect\citeauthoryear{Gurve, Delisle-Rodriguez, Bastos-Filho, and
  Krishnan}{Gurve et~al\mbox{.}}{2020}]%
        {gurve2020trends}
\bibfield{author}{\bibinfo{person}{Dharmendra Gurve}, \bibinfo{person}{Denis
  Delisle-Rodriguez}, \bibinfo{person}{Teodiano Bastos-Filho}, {and}
  \bibinfo{person}{Sridhar Krishnan}.} \bibinfo{year}{2020}\natexlab{}.
\newblock \showarticletitle{Trends in Compressive Sensing for EEG Signal
  Processing Applications}.
\newblock \bibinfo{journal}{\emph{Sensors}} \bibinfo{volume}{20},
  \bibinfo{number}{13} (\bibinfo{year}{2020}), \bibinfo{pages}{3703}.
\newblock


\bibitem[\protect\citeauthoryear{Hahn, Rosenkranz, and Zoubir}{Hahn
  et~al\mbox{.}}{2014}]%
        {hahn2014adaptive}
\bibfield{author}{\bibinfo{person}{J{\"u}rgen Hahn}, \bibinfo{person}{Simon
  Rosenkranz}, {and} \bibinfo{person}{Abdelhak~M Zoubir}.}
  \bibinfo{year}{2014}\natexlab{}.
\newblock \showarticletitle{Adaptive compressed classification for
  hyperspectral imagery}. In \bibinfo{booktitle}{\emph{2014 IEEE International
  Conference on Acoustics, Speech and Signal Processing (ICASSP)}}. IEEE,
  \bibinfo{pages}{1020--1024}.
\newblock


\bibitem[\protect\citeauthoryear{Hammernik, Klatzer, Kobler, Recht, Sodickson,
  Pock, and Knoll}{Hammernik et~al\mbox{.}}{2018}]%
        {hammernik2018learning}
\bibfield{author}{\bibinfo{person}{Kerstin Hammernik}, \bibinfo{person}{Teresa
  Klatzer}, \bibinfo{person}{Erich Kobler}, \bibinfo{person}{Michael~P Recht},
  \bibinfo{person}{Daniel~K Sodickson}, \bibinfo{person}{Thomas Pock}, {and}
  \bibinfo{person}{Florian Knoll}.} \bibinfo{year}{2018}\natexlab{}.
\newblock \showarticletitle{Learning a variational network for reconstruction
  of accelerated MRI data}.
\newblock \bibinfo{journal}{\emph{Magnetic resonance in medicine}}
  \bibinfo{volume}{79}, \bibinfo{number}{6} (\bibinfo{year}{2018}),
  \bibinfo{pages}{3055--3071}.
\newblock


\bibitem[\protect\citeauthoryear{Han, Pool, Tran, and Dally}{Han
  et~al\mbox{.}}{2015}]%
        {han2015learning}
\bibfield{author}{\bibinfo{person}{Song Han}, \bibinfo{person}{Jeff Pool},
  \bibinfo{person}{John Tran}, {and} \bibinfo{person}{William~J Dally}.}
  \bibinfo{year}{2015}\natexlab{}.
\newblock \showarticletitle{Learning both weights and connections for efficient
  neural networks}.
\newblock \bibinfo{journal}{\emph{arXiv preprint arXiv:1506.02626}}
  (\bibinfo{year}{2015}).
\newblock


\bibitem[\protect\citeauthoryear{Han, Hao, Ding, and Tang}{Han
  et~al\mbox{.}}{2017}]%
        {han2017new}
\bibfield{author}{\bibinfo{person}{Tao Han}, \bibinfo{person}{Kuangrong Hao},
  \bibinfo{person}{Yongsheng Ding}, {and} \bibinfo{person}{Xuesong Tang}.}
  \bibinfo{year}{2017}\natexlab{}.
\newblock \showarticletitle{A new multilayer LSTM method of reconstruction for
  compressed sensing in acquiring human pressure data}. In
  \bibinfo{booktitle}{\emph{2017 11th Asian Control Conference (ASCC)}}. IEEE,
  \bibinfo{pages}{2001--2006}.
\newblock


\bibitem[\protect\citeauthoryear{Han, Hao, Ding, and Tang}{Han
  et~al\mbox{.}}{2018a}]%
        {han2018sparse}
\bibfield{author}{\bibinfo{person}{Tao Han}, \bibinfo{person}{Kuangrong Hao},
  \bibinfo{person}{Yongsheng Ding}, {and} \bibinfo{person}{Xuesong Tang}.}
  \bibinfo{year}{2018}\natexlab{a}.
\newblock \showarticletitle{A sparse autoencoder compressed sensing method for
  acquiring the pressure array information of clothing}.
\newblock \bibinfo{journal}{\emph{Neurocomputing}}  \bibinfo{volume}{275}
  (\bibinfo{year}{2018}), \bibinfo{pages}{1500--1510}.
\newblock


\bibitem[\protect\citeauthoryear{Han, Yoo, Kim, Shin, Sung, and Ye}{Han
  et~al\mbox{.}}{2018b}]%
        {han2018deep}
\bibfield{author}{\bibinfo{person}{Yoseob Han}, \bibinfo{person}{Jaejun Yoo},
  \bibinfo{person}{Hak~Hee Kim}, \bibinfo{person}{Hee~Jung Shin},
  \bibinfo{person}{Kyunghyun Sung}, {and} \bibinfo{person}{Jong~Chul Ye}.}
  \bibinfo{year}{2018}\natexlab{b}.
\newblock \showarticletitle{Deep learning with domain adaptation for
  accelerated projection-reconstruction MR}.
\newblock \bibinfo{journal}{\emph{Magnetic resonance in medicine}}
  \bibinfo{volume}{80}, \bibinfo{number}{3} (\bibinfo{year}{2018}),
  \bibinfo{pages}{1189--1205}.
\newblock


\bibitem[\protect\citeauthoryear{Han, Yoo, and Ye}{Han et~al\mbox{.}}{2016}]%
        {han2016deep}
\bibfield{author}{\bibinfo{person}{Yo~Seob Han}, \bibinfo{person}{Jaejun Yoo},
  {and} \bibinfo{person}{Jong~Chul Ye}.} \bibinfo{year}{2016}\natexlab{}.
\newblock \showarticletitle{Deep residual learning for compressed sensing CT
  reconstruction via persistent homology analysis}.
\newblock \bibinfo{journal}{\emph{arXiv preprint arXiv:1611.06391}}
  (\bibinfo{year}{2016}).
\newblock


\bibitem[\protect\citeauthoryear{He, Zhang, Ren, and Sun}{He
  et~al\mbox{.}}{2016}]%
        {he2016deep}
\bibfield{author}{\bibinfo{person}{Kaiming He}, \bibinfo{person}{Xiangyu
  Zhang}, \bibinfo{person}{Shaoqing Ren}, {and} \bibinfo{person}{Jian Sun}.}
  \bibinfo{year}{2016}\natexlab{}.
\newblock \showarticletitle{Deep residual learning for image recognition}. In
  \bibinfo{booktitle}{\emph{Proceedings of the IEEE conference on computer
  vision and pattern recognition}}. \bibinfo{pages}{770--778}.
\newblock


\bibitem[\protect\citeauthoryear{Hosseini, Wang, and Hosseini}{Hosseini
  et~al\mbox{.}}{2013}]%
        {hosseini2013recurrent}
\bibfield{author}{\bibinfo{person}{Alireza Hosseini}, \bibinfo{person}{Jun
  Wang}, {and} \bibinfo{person}{S~Mohammad Hosseini}.}
  \bibinfo{year}{2013}\natexlab{}.
\newblock \showarticletitle{A recurrent neural network for solving a class of
  generalized convex optimization problems}.
\newblock \bibinfo{journal}{\emph{Neural Networks}}  \bibinfo{volume}{44}
  (\bibinfo{year}{2013}), \bibinfo{pages}{78--86}.
\newblock


\bibitem[\protect\citeauthoryear{Huynh, Balan, and Lee}{Huynh
  et~al\mbox{.}}{2016}]%
        {huynh2016deepsense}
\bibfield{author}{\bibinfo{person}{Loc~Nguyen Huynh},
  \bibinfo{person}{Rajesh~Krishna Balan}, {and} \bibinfo{person}{Youngki Lee}.}
  \bibinfo{year}{2016}\natexlab{}.
\newblock \showarticletitle{Deepsense: A gpu-based deep convolutional neural
  network framework on commodity mobile devices}. In
  \bibinfo{booktitle}{\emph{Proceedings of the 2016 Workshop on Wearable
  Systems and Applications}}. \bibinfo{pages}{25--30}.
\newblock


\bibitem[\protect\citeauthoryear{Iliadis, Spinoulas, and Katsaggelos}{Iliadis
  et~al\mbox{.}}{2016}]%
        {iliadis2016deepbinarymask}
\bibfield{author}{\bibinfo{person}{Michael Iliadis}, \bibinfo{person}{Leonidas
  Spinoulas}, {and} \bibinfo{person}{Aggelos~K Katsaggelos}.}
  \bibinfo{year}{2016}\natexlab{}.
\newblock \showarticletitle{Deepbinarymask: Learning a binary mask for video
  compressive sensing}.
\newblock \bibinfo{journal}{\emph{arXiv preprint arXiv:1607.03343}}
  (\bibinfo{year}{2016}).
\newblock


\bibitem[\protect\citeauthoryear{Iliadis, Spinoulas, and Katsaggelos}{Iliadis
  et~al\mbox{.}}{2018}]%
        {iliadis2018deep}
\bibfield{author}{\bibinfo{person}{Michael Iliadis}, \bibinfo{person}{Leonidas
  Spinoulas}, {and} \bibinfo{person}{Aggelos~K Katsaggelos}.}
  \bibinfo{year}{2018}\natexlab{}.
\newblock \showarticletitle{Deep fully-connected networks for video compressive
  sensing}.
\newblock \bibinfo{journal}{\emph{Digital Signal Processing}}
  \bibinfo{volume}{72} (\bibinfo{year}{2018}), \bibinfo{pages}{9--18}.
\newblock


\bibitem[\protect\citeauthoryear{{Ito}, {Takabe}, and {Wadayama}}{{Ito}
  et~al\mbox{.}}{2019}]%
        {8695874}
\bibfield{author}{\bibinfo{person}{D. {Ito}}, \bibinfo{person}{S. {Takabe}},
  {and} \bibinfo{person}{T. {Wadayama}}.} \bibinfo{year}{2019}\natexlab{}.
\newblock \showarticletitle{Trainable ISTA for Sparse Signal Recovery}.
\newblock \bibinfo{journal}{\emph{IEEE Transactions on Signal Processing}}
  \bibinfo{volume}{67}, \bibinfo{number}{12} (\bibinfo{year}{2019}),
  \bibinfo{pages}{3113--3125}.
\newblock


\bibitem[\protect\citeauthoryear{Ji and Carin}{Ji and Carin}{2007}]%
        {ji2007bayesian}
\bibfield{author}{\bibinfo{person}{Shihao Ji} {and} \bibinfo{person}{Lawrence
  Carin}.} \bibinfo{year}{2007}\natexlab{}.
\newblock \showarticletitle{Bayesian compressive sensing and projection
  optimization}. In \bibinfo{booktitle}{\emph{Proceedings of the 24th
  international conference on Machine learning}}. \bibinfo{pages}{377--384}.
\newblock


\bibitem[\protect\citeauthoryear{Ji, Zhu, and Champagne}{Ji
  et~al\mbox{.}}{2019}]%
        {ji2019recurrent}
\bibfield{author}{\bibinfo{person}{Yunyun Ji}, \bibinfo{person}{Wei-Ping Zhu},
  {and} \bibinfo{person}{Benoit Champagne}.} \bibinfo{year}{2019}\natexlab{}.
\newblock \showarticletitle{Recurrent Neural Network-Based Dictionary Learning
  for Compressive Speech Sensing}.
\newblock \bibinfo{journal}{\emph{Circuits, Systems, and Signal Processing}}
  \bibinfo{volume}{38}, \bibinfo{number}{8} (\bibinfo{year}{2019}),
  \bibinfo{pages}{3616--3643}.
\newblock


\bibitem[\protect\citeauthoryear{Jin, McCann, Froustey, and Unser}{Jin
  et~al\mbox{.}}{2017}]%
        {jin2017deep}
\bibfield{author}{\bibinfo{person}{Kyong~Hwan Jin}, \bibinfo{person}{Michael~T
  McCann}, \bibinfo{person}{Emmanuel Froustey}, {and} \bibinfo{person}{Michael
  Unser}.} \bibinfo{year}{2017}\natexlab{}.
\newblock \showarticletitle{Deep convolutional neural network for inverse
  problems in imaging}.
\newblock \bibinfo{journal}{\emph{IEEE Transactions on Image Processing}}
  \bibinfo{volume}{26}, \bibinfo{number}{9} (\bibinfo{year}{2017}),
  \bibinfo{pages}{4509--4522}.
\newblock


\bibitem[\protect\citeauthoryear{Kabkab, Samangouei, and Chellappa}{Kabkab
  et~al\mbox{.}}{2018}]%
        {kabkab2018task}
\bibfield{author}{\bibinfo{person}{Maya Kabkab}, \bibinfo{person}{Pouya
  Samangouei}, {and} \bibinfo{person}{Rama Chellappa}.}
  \bibinfo{year}{2018}\natexlab{}.
\newblock \showarticletitle{Task-aware compressed sensing with generative
  adversarial networks}.
\newblock \bibinfo{journal}{\emph{arXiv preprint arXiv:1802.01284}}
  (\bibinfo{year}{2018}).
\newblock


\bibitem[\protect\citeauthoryear{Kato}{Kato}{1973}]%
        {kato1973development}
\bibfield{author}{\bibinfo{person}{Ichiro Kato}.}
  \bibinfo{year}{1973}\natexlab{}.
\newblock \showarticletitle{Development of WABOT 1}.
\newblock \bibinfo{journal}{\emph{Biomechanism}}  \bibinfo{volume}{2}
  (\bibinfo{year}{1973}), \bibinfo{pages}{173--214}.
\newblock


\bibitem[\protect\citeauthoryear{Khosravy, Dey, and Duque}{Khosravy
  et~al\mbox{.}}{2020}]%
        {khosravy_dey_duque_2020}
\bibfield{author}{\bibinfo{person}{Mahdi Khosravy}, \bibinfo{person}{Nilanjan
  Dey}, {and} \bibinfo{person}{Carlos~Augusto Duque}.}
  \bibinfo{year}{2020}\natexlab{}.
\newblock \bibinfo{booktitle}{\emph{Compressive sensing in healthcare}}.
\newblock \bibinfo{publisher}{Elsevier, Academic Press}.
\newblock


\bibitem[\protect\citeauthoryear{Kim, Park, and Lee}{Kim
  et~al\mbox{.}}{2020b}]%
        {kim2020compressive}
\bibfield{author}{\bibinfo{person}{Cheolsun Kim}, \bibinfo{person}{Dongju
  Park}, {and} \bibinfo{person}{Heung-No Lee}.}
  \bibinfo{year}{2020}\natexlab{b}.
\newblock \showarticletitle{Compressive Sensing Spectroscopy Using a Residual
  Convolutional Neural Network}.
\newblock \bibinfo{journal}{\emph{Sensors}} \bibinfo{volume}{20},
  \bibinfo{number}{3} (\bibinfo{year}{2020}), \bibinfo{pages}{594}.
\newblock


\bibitem[\protect\citeauthoryear{Kim and Park}{Kim and Park}{2020}]%
        {kim2020element}
\bibfield{author}{\bibinfo{person}{Dohyun Kim} {and} \bibinfo{person}{Daeyoung
  Park}.} \bibinfo{year}{2020}\natexlab{}.
\newblock \showarticletitle{Element-Wise Adaptive Thresholds for Learned
  Iterative Shrinkage Thresholding Algorithms}.
\newblock \bibinfo{journal}{\emph{IEEE Access}}  \bibinfo{volume}{8}
  (\bibinfo{year}{2020}), \bibinfo{pages}{45874--45886}.
\newblock


\bibitem[\protect\citeauthoryear{Kim, Park, and Kim}{Kim
  et~al\mbox{.}}{2020a}]%
        {kim2020signal}
\bibfield{author}{\bibinfo{person}{Yisak Kim}, \bibinfo{person}{Juyoung Park},
  {and} \bibinfo{person}{Hyungsuk Kim}.} \bibinfo{year}{2020}\natexlab{a}.
\newblock \showarticletitle{Signal-Processing Framework for Ultrasound
  Compressed Sensing Data: Envelope Detection and Spectral Analysis}.
\newblock \bibinfo{journal}{\emph{Applied Sciences}} \bibinfo{volume}{10},
  \bibinfo{number}{19} (\bibinfo{year}{2020}), \bibinfo{pages}{6956}.
\newblock


\bibitem[\protect\citeauthoryear{Krahmer, Kruschel, and Sandbichler}{Krahmer
  et~al\mbox{.}}{2017}]%
        {krahmer2017total}
\bibfield{author}{\bibinfo{person}{Felix Krahmer}, \bibinfo{person}{Christian
  Kruschel}, {and} \bibinfo{person}{Michael Sandbichler}.}
  \bibinfo{year}{2017}\natexlab{}.
\newblock \showarticletitle{Total variation minimization in compressed
  sensing}.
\newblock In \bibinfo{booktitle}{\emph{Compressed Sensing and its
  Applications}}. \bibinfo{publisher}{Springer}, \bibinfo{pages}{333--358}.
\newblock


\bibitem[\protect\citeauthoryear{Kruizinga, van~der Meulen, Fedjajevs, Mastik,
  Springeling, de~Jong, Bosch, and Leus}{Kruizinga et~al\mbox{.}}{2017}]%
        {kruizinga2017compressive}
\bibfield{author}{\bibinfo{person}{Pieter Kruizinga}, \bibinfo{person}{Pim
  van~der Meulen}, \bibinfo{person}{Andrejs Fedjajevs}, \bibinfo{person}{Frits
  Mastik}, \bibinfo{person}{Geert Springeling}, \bibinfo{person}{Nico de Jong},
  \bibinfo{person}{Johannes~G Bosch}, {and} \bibinfo{person}{Geert Leus}.}
  \bibinfo{year}{2017}\natexlab{}.
\newblock \showarticletitle{Compressive 3D ultrasound imaging using a single
  sensor}.
\newblock \bibinfo{journal}{\emph{Science advances}} \bibinfo{volume}{3},
  \bibinfo{number}{12} (\bibinfo{year}{2017}), \bibinfo{pages}{e1701423}.
\newblock


\bibitem[\protect\citeauthoryear{Kulkarni, Lohit, Turaga, Kerviche, and
  Ashok}{Kulkarni et~al\mbox{.}}{2016}]%
        {kulkarni2016reconnet}
\bibfield{author}{\bibinfo{person}{Kuldeep Kulkarni}, \bibinfo{person}{Suhas
  Lohit}, \bibinfo{person}{Pavan Turaga}, \bibinfo{person}{Ronan Kerviche},
  {and} \bibinfo{person}{Amit Ashok}.} \bibinfo{year}{2016}\natexlab{}.
\newblock \showarticletitle{Reconnet: Non-iterative reconstruction of images
  from compressively sensed measurements}. In
  \bibinfo{booktitle}{\emph{Proceedings of the IEEE Conference on Computer
  Vision and Pattern Recognition}}. \bibinfo{pages}{449--458}.
\newblock


\bibitem[\protect\citeauthoryear{Kwan, Gribben, Rangamani, Tran, Zhang, and
  Etienne-Cummings}{Kwan et~al\mbox{.}}{2020}]%
        {kwan2020detection}
\bibfield{author}{\bibinfo{person}{Chiman Kwan}, \bibinfo{person}{David
  Gribben}, \bibinfo{person}{Akshay Rangamani}, \bibinfo{person}{Trac Tran},
  \bibinfo{person}{Jack Zhang}, {and} \bibinfo{person}{Ralph
  Etienne-Cummings}.} \bibinfo{year}{2020}\natexlab{}.
\newblock \showarticletitle{Detection and Confirmation of Multiple Human
  Targets Using Pixel-Wise Code Aperture Measurements}.
\newblock \bibinfo{journal}{\emph{Journal of Imaging}} \bibinfo{volume}{6},
  \bibinfo{number}{6} (\bibinfo{year}{2020}), \bibinfo{pages}{40}.
\newblock


\bibitem[\protect\citeauthoryear{Landweber}{Landweber}{1951}]%
        {landweber1951iteration}
\bibfield{author}{\bibinfo{person}{Louis Landweber}.}
  \bibinfo{year}{1951}\natexlab{}.
\newblock \showarticletitle{An iteration formula for Fredholm integral
  equations of the first kind}.
\newblock \bibinfo{journal}{\emph{American journal of mathematics}}
  \bibinfo{volume}{73}, \bibinfo{number}{3} (\bibinfo{year}{1951}),
  \bibinfo{pages}{615--624}.
\newblock


\bibitem[\protect\citeauthoryear{Lane, Bhattacharya, Georgiev, Forlivesi, Jiao,
  Qendro, and Kawsar}{Lane et~al\mbox{.}}{2016}]%
        {lane2016deepx}
\bibfield{author}{\bibinfo{person}{Nicholas~D Lane}, \bibinfo{person}{Sourav
  Bhattacharya}, \bibinfo{person}{Petko Georgiev}, \bibinfo{person}{Claudio
  Forlivesi}, \bibinfo{person}{Lei Jiao}, \bibinfo{person}{Lorena Qendro},
  {and} \bibinfo{person}{Fahim Kawsar}.} \bibinfo{year}{2016}\natexlab{}.
\newblock \showarticletitle{Deepx: A software accelerator for low-power deep
  learning inference on mobile devices}. In \bibinfo{booktitle}{\emph{2016 15th
  ACM/IEEE International Conference on Information Processing in Sensor
  Networks (IPSN)}}. IEEE, \bibinfo{pages}{1--12}.
\newblock


\bibitem[\protect\citeauthoryear{Laskaridis, Venieris, Almeida, Leontiadis, and
  Lane}{Laskaridis et~al\mbox{.}}{2020}]%
        {laskaridis2020spinn}
\bibfield{author}{\bibinfo{person}{Stefanos Laskaridis},
  \bibinfo{person}{Stylianos~I Venieris}, \bibinfo{person}{Mario Almeida},
  \bibinfo{person}{Ilias Leontiadis}, {and} \bibinfo{person}{Nicholas~D Lane}.}
  \bibinfo{year}{2020}\natexlab{}.
\newblock \showarticletitle{SPINN: synergistic progressive inference of neural
  networks over device and cloud}. In \bibinfo{booktitle}{\emph{Proceedings of
  the 26th Annual International Conference on Mobile Computing and
  Networking}}. \bibinfo{pages}{1--15}.
\newblock


\bibitem[\protect\citeauthoryear{Latorre-Carmona, Traver, S{\'a}nchez, and
  Tajahuerce}{Latorre-Carmona et~al\mbox{.}}{2019}]%
        {latorre2019online}
\bibfield{author}{\bibinfo{person}{Pedro Latorre-Carmona},
  \bibinfo{person}{V~Javier Traver}, \bibinfo{person}{J~Salvador S{\'a}nchez},
  {and} \bibinfo{person}{Enrique Tajahuerce}.} \bibinfo{year}{2019}\natexlab{}.
\newblock \showarticletitle{Online reconstruction-free single-pixel image
  classification}.
\newblock \bibinfo{journal}{\emph{Image and Vision Computing}}
  \bibinfo{volume}{86} (\bibinfo{year}{2019}), \bibinfo{pages}{28--37}.
\newblock


\bibitem[\protect\citeauthoryear{Lee, Lin, Chen, and Chang}{Lee
  et~al\mbox{.}}{2019}]%
        {lee2019deep}
\bibfield{author}{\bibinfo{person}{Chia-Han Lee}, \bibinfo{person}{Jia-Wei
  Lin}, \bibinfo{person}{Po-Hao Chen}, {and} \bibinfo{person}{Yu-Chieh Chang}.}
  \bibinfo{year}{2019}\natexlab{}.
\newblock \showarticletitle{Deep learning-constructed joint
  transmission-recognition for internet of things}.
\newblock \bibinfo{journal}{\emph{IEEE Access}}  \bibinfo{volume}{7}
  (\bibinfo{year}{2019}), \bibinfo{pages}{76547--76561}.
\newblock


\bibitem[\protect\citeauthoryear{Lee, Yoo, and Ye}{Lee et~al\mbox{.}}{2017}]%
        {lee2017deep}
\bibfield{author}{\bibinfo{person}{Dongwook Lee}, \bibinfo{person}{Jaejun Yoo},
  {and} \bibinfo{person}{Jong~Chul Ye}.} \bibinfo{year}{2017}\natexlab{}.
\newblock \showarticletitle{Deep residual learning for compressed sensing MRI}.
  In \bibinfo{booktitle}{\emph{2017 IEEE 14th International Symposium on
  Biomedical Imaging (ISBI 2017)}}. IEEE, \bibinfo{pages}{15--18}.
\newblock


\bibitem[\protect\citeauthoryear{{Lee}, {Yoo}, and {Ye}}{{Lee}
  et~al\mbox{.}}{2017}]%
        {7950457}
\bibfield{author}{\bibinfo{person}{D. {Lee}}, \bibinfo{person}{J. {Yoo}}, {and}
  \bibinfo{person}{J.~C. {Ye}}.} \bibinfo{year}{2017}\natexlab{}.
\newblock \showarticletitle{Deep residual learning for compressed sensing MRI}.
  In \bibinfo{booktitle}{\emph{2017 IEEE 14th International Symposium on
  Biomedical Imaging (ISBI 2017)}}. \bibinfo{pages}{15--18}.
\newblock


\bibitem[\protect\citeauthoryear{Li, Cao, Tong, Ma, Niu, and Guo}{Li
  et~al\mbox{.}}{2020}]%
        {li2020deep}
\bibfield{author}{\bibinfo{person}{Xuesong Li}, \bibinfo{person}{Tianle Cao},
  \bibinfo{person}{Yan Tong}, \bibinfo{person}{Xiaodong Ma},
  \bibinfo{person}{Zhendong Niu}, {and} \bibinfo{person}{Hua Guo}.}
  \bibinfo{year}{2020}\natexlab{}.
\newblock \showarticletitle{Deep residual network for highly accelerated fMRI
  reconstruction using variable density spiral trajectory}.
\newblock \bibinfo{journal}{\emph{Neurocomputing}}  \bibinfo{volume}{398}
  (\bibinfo{year}{2020}), \bibinfo{pages}{338--346}.
\newblock


\bibitem[\protect\citeauthoryear{Li and Wei}{Li and Wei}{2016}]%
        {li2016signal}
\bibfield{author}{\bibinfo{person}{Yuan-Min Li} {and} \bibinfo{person}{Deyun
  Wei}.} \bibinfo{year}{2016}\natexlab{}.
\newblock \showarticletitle{Signal reconstruction of compressed sensing based
  on recurrent neural networks}.
\newblock \bibinfo{journal}{\emph{Optik}} \bibinfo{volume}{127},
  \bibinfo{number}{10} (\bibinfo{year}{2016}), \bibinfo{pages}{4473--4477}.
\newblock


\bibitem[\protect\citeauthoryear{Liu, Huang, and Zhang}{Liu
  et~al\mbox{.}}{2017}]%
        {liu2017efficient}
\bibfield{author}{\bibinfo{person}{Jing Liu}, \bibinfo{person}{Kaiyu Huang},
  {and} \bibinfo{person}{Guoxian Zhang}.} \bibinfo{year}{2017}\natexlab{}.
\newblock \showarticletitle{An efficient distributed compressed sensing
  algorithm for decentralized sensor network}.
\newblock \bibinfo{journal}{\emph{Sensors}} \bibinfo{volume}{17},
  \bibinfo{number}{4} (\bibinfo{year}{2017}), \bibinfo{pages}{907}.
\newblock


\bibitem[\protect\citeauthoryear{Liu, Liu, Li, and Yang}{Liu
  et~al\mbox{.}}{2019}]%
        {liu2019compressed}
\bibfield{author}{\bibinfo{person}{Yuhong Liu}, \bibinfo{person}{Shuying Liu},
  \bibinfo{person}{Cuiran Li}, {and} \bibinfo{person}{Danfeng Yang}.}
  \bibinfo{year}{2019}\natexlab{}.
\newblock \showarticletitle{Compressed Sensing Image Reconstruction Based on
  Convolutional Neural Network}.
\newblock \bibinfo{journal}{\emph{International Journal of Computational
  Intelligence Systems}} \bibinfo{volume}{12}, \bibinfo{number}{2}
  (\bibinfo{year}{2019}), \bibinfo{pages}{873--880}.
\newblock


\bibitem[\protect\citeauthoryear{{Lohit}, {Kulkarni}, {Kerviche}, {Turaga}, and
  {Ashok}}{{Lohit} et~al\mbox{.}}{2018}]%
        {8379450}
\bibfield{author}{\bibinfo{person}{S. {Lohit}}, \bibinfo{person}{K.
  {Kulkarni}}, \bibinfo{person}{R. {Kerviche}}, \bibinfo{person}{P. {Turaga}},
  {and} \bibinfo{person}{A. {Ashok}}.} \bibinfo{year}{2018}\natexlab{}.
\newblock \showarticletitle{Convolutional Neural Networks for Noniterative
  Reconstruction of Compressively Sensed Images}.
\newblock \bibinfo{journal}{\emph{IEEE Transactions on Computational Imaging}}
  \bibinfo{volume}{4}, \bibinfo{number}{3} (\bibinfo{year}{2018}),
  \bibinfo{pages}{326--340}.
\newblock


\bibitem[\protect\citeauthoryear{Lohit, Kulkarni, and Turaga}{Lohit
  et~al\mbox{.}}{2016}]%
        {lohit2016direct}
\bibfield{author}{\bibinfo{person}{Suhas Lohit}, \bibinfo{person}{Kuldeep
  Kulkarni}, {and} \bibinfo{person}{Pavan Turaga}.}
  \bibinfo{year}{2016}\natexlab{}.
\newblock \showarticletitle{Direct inference on compressive measurements using
  convolutional neural networks}. In \bibinfo{booktitle}{\emph{2016 IEEE
  International Conference on Image Processing (ICIP)}}. IEEE,
  \bibinfo{pages}{1913--1917}.
\newblock


\bibitem[\protect\citeauthoryear{Lohit, Kulkarni, Turaga, Wang, and
  Sankaranarayanan}{Lohit et~al\mbox{.}}{2015}]%
        {lohit2015reconstruction}
\bibfield{author}{\bibinfo{person}{Suhas Lohit}, \bibinfo{person}{Kuldeep
  Kulkarni}, \bibinfo{person}{Pavan Turaga}, \bibinfo{person}{Jian Wang}, {and}
  \bibinfo{person}{Aswin~C Sankaranarayanan}.} \bibinfo{year}{2015}\natexlab{}.
\newblock \showarticletitle{Reconstruction-free inference on compressive
  measurements}. In \bibinfo{booktitle}{\emph{Proceedings of the IEEE
  Conference on Computer Vision and Pattern Recognition Workshops}}.
  \bibinfo{pages}{16--24}.
\newblock


\bibitem[\protect\citeauthoryear{Lohit, Singh, Kulkarni, and Turaga}{Lohit
  et~al\mbox{.}}{2018}]%
        {lohit2018rate}
\bibfield{author}{\bibinfo{person}{Suhas Lohit}, \bibinfo{person}{Rajhans
  Singh}, \bibinfo{person}{Kuldeep Kulkarni}, {and} \bibinfo{person}{Pavan
  Turaga}.} \bibinfo{year}{2018}\natexlab{}.
\newblock \showarticletitle{Rate-adaptive neural networks for spatial
  multiplexers}.
\newblock \bibinfo{journal}{\emph{arXiv preprint arXiv:1809.02850}}
  (\bibinfo{year}{2018}).
\newblock


\bibitem[\protect\citeauthoryear{Lorintiu, Liebgott, Alessandrini, Bernard, and
  Friboulet}{Lorintiu et~al\mbox{.}}{2015}]%
        {lorintiu2015compressed}
\bibfield{author}{\bibinfo{person}{Oana Lorintiu}, \bibinfo{person}{Herv{\'e}
  Liebgott}, \bibinfo{person}{Martino Alessandrini}, \bibinfo{person}{Olivier
  Bernard}, {and} \bibinfo{person}{Denis Friboulet}.}
  \bibinfo{year}{2015}\natexlab{}.
\newblock \showarticletitle{Compressed sensing reconstruction of 3D ultrasound
  data using dictionary learning and line-wise subsampling}.
\newblock \bibinfo{journal}{\emph{IEEE transactions on medical imaging}}
  \bibinfo{volume}{34}, \bibinfo{number}{12} (\bibinfo{year}{2015}),
  \bibinfo{pages}{2467--2477}.
\newblock


\bibitem[\protect\citeauthoryear{Ltd.}{Ltd.}{2012}]%
        {ARMBiglittle}
\bibfield{author}{\bibinfo{person}{Arm Ltd.}} \bibinfo{year}{2012}\natexlab{}.
\newblock \bibinfo{title}{{Arm big.LITTLE} technology}.
\newblock
  \bibinfo{howpublished}{\url{https://www.arm.com/why-arm/technologies/big-little}}.
\newblock
\newblock
\shownote{Accessed: 2021-04-01.}


\bibitem[\protect\citeauthoryear{Lu and Bo}{Lu and Bo}{2019}]%
        {lu2019wdlreconnet}
\bibfield{author}{\bibinfo{person}{Hancheng Lu} {and} \bibinfo{person}{Lei
  Bo}.} \bibinfo{year}{2019}\natexlab{}.
\newblock \showarticletitle{WDLReconNet: Compressive Sensing Reconstruction
  With Deep Learning Over Wireless Fading Channels}.
\newblock \bibinfo{journal}{\emph{IEEE Access}}  \bibinfo{volume}{7}
  (\bibinfo{year}{2019}), \bibinfo{pages}{24440--24451}.
\newblock


\bibitem[\protect\citeauthoryear{Lu, Dong, Wang, Shi, and Xie}{Lu
  et~al\mbox{.}}{2018}]%
        {lu2018convcsnet}
\bibfield{author}{\bibinfo{person}{Xiaotong Lu}, \bibinfo{person}{Weisheng
  Dong}, \bibinfo{person}{Peiyao Wang}, \bibinfo{person}{Guangming Shi}, {and}
  \bibinfo{person}{Xuemei Xie}.} \bibinfo{year}{2018}\natexlab{}.
\newblock \showarticletitle{Convcsnet: A convolutional compressive sensing
  framework based on deep learning}.
\newblock \bibinfo{journal}{\emph{arXiv preprint arXiv:1801.10342}}
  (\bibinfo{year}{2018}).
\newblock


\bibitem[\protect\citeauthoryear{Ma, Li, Yang, Lv, Li, Liu, Yao, and Xu}{Ma
  et~al\mbox{.}}{2017}]%
        {ma2017extraction}
\bibfield{author}{\bibinfo{person}{Teng Ma}, \bibinfo{person}{Hui Li},
  \bibinfo{person}{Hao Yang}, \bibinfo{person}{Xulin Lv},
  \bibinfo{person}{Peiyang Li}, \bibinfo{person}{Tiejun Liu},
  \bibinfo{person}{Dezhong Yao}, {and} \bibinfo{person}{Peng Xu}.}
  \bibinfo{year}{2017}\natexlab{}.
\newblock \showarticletitle{The extraction of motion-onset VEP BCI features
  based on deep learning and compressed sensing}.
\newblock \bibinfo{journal}{\emph{Journal of neuroscience methods}}
  \bibinfo{volume}{275} (\bibinfo{year}{2017}), \bibinfo{pages}{80--92}.
\newblock


\bibitem[\protect\citeauthoryear{Maimaitijiang, Sagan, Sidike, Hartling,
  Esposito, and Fritschi}{Maimaitijiang et~al\mbox{.}}{2020}]%
        {maimaitijiang2020soybean}
\bibfield{author}{\bibinfo{person}{Maitiniyazi Maimaitijiang},
  \bibinfo{person}{Vasit Sagan}, \bibinfo{person}{Paheding Sidike},
  \bibinfo{person}{Sean Hartling}, \bibinfo{person}{Flavio Esposito}, {and}
  \bibinfo{person}{Felix~B Fritschi}.} \bibinfo{year}{2020}\natexlab{}.
\newblock \showarticletitle{Soybean yield prediction from UAV using multimodal
  data fusion and deep learning}.
\newblock \bibinfo{journal}{\emph{Remote sensing of environment}}
  \bibinfo{volume}{237} (\bibinfo{year}{2020}), \bibinfo{pages}{111599}.
\newblock


\bibitem[\protect\citeauthoryear{Majumdar}{Majumdar}{2015}]%
        {majumdar2015real}
\bibfield{author}{\bibinfo{person}{Angshul Majumdar}.}
  \bibinfo{year}{2015}\natexlab{}.
\newblock \showarticletitle{Real-time dynamic MRI reconstruction using stacked
  denoising autoencoder}.
\newblock \bibinfo{journal}{\emph{arXiv preprint arXiv:1503.06383}}
  (\bibinfo{year}{2015}).
\newblock


\bibitem[\protect\citeauthoryear{Mallat and Zhang}{Mallat and Zhang}{1993}]%
        {mallat1993matching}
\bibfield{author}{\bibinfo{person}{St{\'e}phane~G Mallat} {and}
  \bibinfo{person}{Zhifeng Zhang}.} \bibinfo{year}{1993}\natexlab{}.
\newblock \showarticletitle{Matching pursuits with time-frequency
  dictionaries}.
\newblock \bibinfo{journal}{\emph{IEEE Transactions on signal processing}}
  \bibinfo{volume}{41}, \bibinfo{number}{12} (\bibinfo{year}{1993}),
  \bibinfo{pages}{3397--3415}.
\newblock


\bibitem[\protect\citeauthoryear{{Mangia}, {Marchioni}, {Prono}, {Pareschi},
  {Rovatti}, and {Setti}}{{Mangia} et~al\mbox{.}}{2020a}]%
        {9073945}
\bibfield{author}{\bibinfo{person}{M. {Mangia}}, \bibinfo{person}{A.
  {Marchioni}}, \bibinfo{person}{L. {Prono}}, \bibinfo{person}{F. {Pareschi}},
  \bibinfo{person}{R. {Rovatti}}, {and} \bibinfo{person}{G. {Setti}}.}
  \bibinfo{year}{2020}\natexlab{a}.
\newblock \showarticletitle{Low-power ECG acquisition by Compressed Sensing
  with Deep Neural Oracles}. In \bibinfo{booktitle}{\emph{2020 2nd IEEE
  International Conference on Artificial Intelligence Circuits and Systems
  (AICAS)}}. \bibinfo{pages}{158--162}.
\newblock


\bibitem[\protect\citeauthoryear{{Mangia}, {Prono}, {Marchioni}, {Pareschi},
  {Rovatti}, and {Setti}}{{Mangia} et~al\mbox{.}}{2020b}]%
        {9044777}
\bibfield{author}{\bibinfo{person}{M. {Mangia}}, \bibinfo{person}{L. {Prono}},
  \bibinfo{person}{A. {Marchioni}}, \bibinfo{person}{F. {Pareschi}},
  \bibinfo{person}{R. {Rovatti}}, {and} \bibinfo{person}{G. {Setti}}.}
  \bibinfo{year}{2020}\natexlab{b}.
\newblock \showarticletitle{Deep Neural Oracles for Short-Window Optimized
  Compressed Sensing of Biosignals}.
\newblock \bibinfo{journal}{\emph{IEEE Transactions on Biomedical Circuits and
  Systems}} \bibinfo{volume}{14}, \bibinfo{number}{3} (\bibinfo{year}{2020}),
  \bibinfo{pages}{545--557}.
\newblock


\bibitem[\protect\citeauthoryear{Mardani, Gong, Cheng, Vasanawala, Zaharchuk,
  Alley, Thakur, Han, Dally, Pauly, et~al\mbox{.}}{Mardani
  et~al\mbox{.}}{2017}]%
        {mardani2017deep}
\bibfield{author}{\bibinfo{person}{Morteza Mardani}, \bibinfo{person}{Enhao
  Gong}, \bibinfo{person}{Joseph~Y Cheng}, \bibinfo{person}{Shreyas
  Vasanawala}, \bibinfo{person}{Greg Zaharchuk}, \bibinfo{person}{Marcus
  Alley}, \bibinfo{person}{Neil Thakur}, \bibinfo{person}{Song Han},
  \bibinfo{person}{William Dally}, \bibinfo{person}{John~M Pauly},
  {et~al\mbox{.}}} \bibinfo{year}{2017}\natexlab{}.
\newblock \showarticletitle{Deep generative adversarial networks for compressed
  sensing automates MRI}.
\newblock \bibinfo{journal}{\emph{arXiv preprint arXiv:1706.00051}}
  (\bibinfo{year}{2017}).
\newblock


\bibitem[\protect\citeauthoryear{{Merhej}, {Diab}, {Khalil}, and
  {Prost}}{{Merhej} et~al\mbox{.}}{2011}]%
        {6009227}
\bibfield{author}{\bibinfo{person}{D. {Merhej}}, \bibinfo{person}{C. {Diab}},
  \bibinfo{person}{M. {Khalil}}, {and} \bibinfo{person}{R. {Prost}}.}
  \bibinfo{year}{2011}\natexlab{}.
\newblock \showarticletitle{Embedding Prior Knowledge Within Compressed Sensing
  by Neural Networks}.
\newblock \bibinfo{journal}{\emph{IEEE Transactions on Neural Networks}}
  \bibinfo{volume}{22}, \bibinfo{number}{10} (\bibinfo{year}{2011}),
  \bibinfo{pages}{1638--1649}.
\newblock


\bibitem[\protect\citeauthoryear{Metzler, Mousavi, and Baraniuk}{Metzler
  et~al\mbox{.}}{2017}]%
        {metzler2017learned}
\bibfield{author}{\bibinfo{person}{Chris Metzler}, \bibinfo{person}{Ali
  Mousavi}, {and} \bibinfo{person}{Richard Baraniuk}.}
  \bibinfo{year}{2017}\natexlab{}.
\newblock \showarticletitle{Learned D-AMP: Principled neural network based
  compressive image recovery}. In \bibinfo{booktitle}{\emph{Advances in Neural
  Information Processing Systems}}. \bibinfo{pages}{1772--1783}.
\newblock


\bibitem[\protect\citeauthoryear{Metzler}{Metzler}{2014}]%
        {metzler2014denoising}
\bibfield{author}{\bibinfo{person}{Chris~A Metzler}.}
  \bibinfo{year}{2014}\natexlab{}.
\newblock \emph{\bibinfo{title}{Denoising-based Approximate Message Passing}}.
\newblock \bibinfo{thesistype}{Ph.D. Dissertation}.
\newblock


\bibitem[\protect\citeauthoryear{{Metzler}, {Maleki}, and {Baraniuk}}{{Metzler}
  et~al\mbox{.}}{2015}]%
        {7351377}
\bibfield{author}{\bibinfo{person}{C.~A. {Metzler}}, \bibinfo{person}{A.
  {Maleki}}, {and} \bibinfo{person}{R.~G. {Baraniuk}}.}
  \bibinfo{year}{2015}\natexlab{}.
\newblock \showarticletitle{BM3D-AMP: A new image recovery algorithm based on
  BM3D denoising}. In \bibinfo{booktitle}{\emph{2015 IEEE International
  Conference on Image Processing (ICIP)}}. \bibinfo{pages}{3116--3120}.
\newblock


\bibitem[\protect\citeauthoryear{Monga, Li, and Eldar}{Monga
  et~al\mbox{.}}{2019}]%
        {monga2019algorithm}
\bibfield{author}{\bibinfo{person}{Vishal Monga}, \bibinfo{person}{Yuelong Li},
  {and} \bibinfo{person}{Yonina~C Eldar}.} \bibinfo{year}{2019}\natexlab{}.
\newblock \showarticletitle{Algorithm unrolling: Interpretable, efficient deep
  learning for signal and image processing}.
\newblock \bibinfo{journal}{\emph{arXiv preprint arXiv:1912.10557}}
  (\bibinfo{year}{2019}).
\newblock


\bibitem[\protect\citeauthoryear{Mousavi and Baraniuk}{Mousavi and
  Baraniuk}{2017}]%
        {mousavi2017learning}
\bibfield{author}{\bibinfo{person}{Ali Mousavi} {and}
  \bibinfo{person}{Richard~G Baraniuk}.} \bibinfo{year}{2017}\natexlab{}.
\newblock \showarticletitle{Learning to invert: Signal recovery via deep
  convolutional networks}. In \bibinfo{booktitle}{\emph{2017 IEEE international
  conference on acoustics, speech and signal processing (ICASSP)}}. IEEE,
  \bibinfo{pages}{2272--2276}.
\newblock


\bibitem[\protect\citeauthoryear{Mousavi, Dasarathy, and Baraniuk}{Mousavi
  et~al\mbox{.}}{2017}]%
        {mousavi2017deepcodec}
\bibfield{author}{\bibinfo{person}{Ali Mousavi}, \bibinfo{person}{Gautam
  Dasarathy}, {and} \bibinfo{person}{Richard~G Baraniuk}.}
  \bibinfo{year}{2017}\natexlab{}.
\newblock \showarticletitle{Deepcodec: Adaptive sensing and recovery via deep
  convolutional neural networks}.
\newblock \bibinfo{journal}{\emph{arXiv preprint arXiv:1707.03386}}
  (\bibinfo{year}{2017}).
\newblock


\bibitem[\protect\citeauthoryear{Mousavi, Patel, and Baraniuk}{Mousavi
  et~al\mbox{.}}{2015}]%
        {mousavi2015deep}
\bibfield{author}{\bibinfo{person}{Ali Mousavi}, \bibinfo{person}{Ankit~B
  Patel}, {and} \bibinfo{person}{Richard~G Baraniuk}.}
  \bibinfo{year}{2015}\natexlab{}.
\newblock \showarticletitle{A deep learning approach to structured signal
  recovery}. In \bibinfo{booktitle}{\emph{2015 53rd annual allerton conference
  on communication, control, and computing (Allerton)}}. IEEE,
  \bibinfo{pages}{1336--1343}.
\newblock


\bibitem[\protect\citeauthoryear{Needell and Tropp}{Needell and Tropp}{2009}]%
        {needell2009cosamp}
\bibfield{author}{\bibinfo{person}{Deanna Needell} {and}
  \bibinfo{person}{Joel~A Tropp}.} \bibinfo{year}{2009}\natexlab{}.
\newblock \showarticletitle{CoSaMP: Iterative signal recovery from incomplete
  and inaccurate samples}.
\newblock \bibinfo{journal}{\emph{Applied and computational harmonic analysis}}
  \bibinfo{volume}{26}, \bibinfo{number}{3} (\bibinfo{year}{2009}),
  \bibinfo{pages}{301--321}.
\newblock


\bibitem[\protect\citeauthoryear{Nguyen}{Nguyen}{2017}]%
        {nguyenlearning}
\bibfield{author}{\bibinfo{person}{Phan~Minh Nguyen}.}
  \bibinfo{year}{2017}\natexlab{}.
\newblock \bibinfo{booktitle}{}.
\newblock


\bibitem[\protect\citeauthoryear{Novikov, Podoprikhin, Osokin, and
  Vetrov}{Novikov et~al\mbox{.}}{2015}]%
        {novikov2015tensorizing}
\bibfield{author}{\bibinfo{person}{Alexander Novikov}, \bibinfo{person}{Dmitry
  Podoprikhin}, \bibinfo{person}{Anton Osokin}, {and} \bibinfo{person}{Dmitry
  Vetrov}.} \bibinfo{year}{2015}\natexlab{}.
\newblock \showarticletitle{Tensorizing neural networks}.
\newblock \bibinfo{journal}{\emph{arXiv preprint arXiv:1509.06569}}
  (\bibinfo{year}{2015}).
\newblock


\bibitem[\protect\citeauthoryear{Nyquist}{Nyquist}{2002}]%
        {nyquist2002certain}
\bibfield{author}{\bibinfo{person}{Harry Nyquist}.}
  \bibinfo{year}{2002}\natexlab{}.
\newblock \showarticletitle{Certain topics in telegraph transmission theory}.
\newblock \bibinfo{journal}{\emph{Proc. IEEE}} \bibinfo{volume}{90},
  \bibinfo{number}{2} (\bibinfo{year}{2002}), \bibinfo{pages}{280--305}.
\newblock


\bibitem[\protect\citeauthoryear{Ouchi and Ito}{Ouchi and Ito}{2020}]%
        {ouchi2020reconstruction}
\bibfield{author}{\bibinfo{person}{Shohei Ouchi} {and} \bibinfo{person}{Satoshi
  Ito}.} \bibinfo{year}{2020}\natexlab{}.
\newblock \showarticletitle{Reconstruction of Compressed-sensing MR Imaging
  Using Deep Residual Learning in the Image Domain}.
\newblock \bibinfo{journal}{\emph{Magnetic Resonance in Medical Sciences}}
  (\bibinfo{year}{2020}), \bibinfo{pages}{mp--2019}.
\newblock


\bibitem[\protect\citeauthoryear{Palangi, Ward, and Deng}{Palangi
  et~al\mbox{.}}{2016a}]%
        {palangi2016distributed}
\bibfield{author}{\bibinfo{person}{Hamid Palangi}, \bibinfo{person}{Rabab
  Ward}, {and} \bibinfo{person}{Li Deng}.} \bibinfo{year}{2016}\natexlab{a}.
\newblock \showarticletitle{Distributed compressive sensing: A deep learning
  approach}.
\newblock \bibinfo{journal}{\emph{IEEE Transactions on Signal Processing}}
  \bibinfo{volume}{64}, \bibinfo{number}{17} (\bibinfo{year}{2016}),
  \bibinfo{pages}{4504--4518}.
\newblock


\bibitem[\protect\citeauthoryear{Palangi, Ward, and Deng}{Palangi
  et~al\mbox{.}}{2016b}]%
        {palangi2016reconstruction}
\bibfield{author}{\bibinfo{person}{Hamid Palangi}, \bibinfo{person}{Rabab
  Ward}, {and} \bibinfo{person}{Li Deng}.} \bibinfo{year}{2016}\natexlab{b}.
\newblock \showarticletitle{Reconstruction of sparse vectors in compressive
  sensing with multiple measurement vectors using bidirectional long short-term
  memory}. In \bibinfo{booktitle}{\emph{2016 IEEE Global Conference on Signal
  and Information Processing (GlobalSIP)}}. IEEE, \bibinfo{pages}{192--196}.
\newblock


\bibitem[\protect\citeauthoryear{Pati, Rezaiifar, and Krishnaprasad}{Pati
  et~al\mbox{.}}{1993}]%
        {pati1993orthogonal}
\bibfield{author}{\bibinfo{person}{Yagyensh~Chandra Pati},
  \bibinfo{person}{Ramin Rezaiifar}, {and}
  \bibinfo{person}{Perinkulam~Sambamurthy Krishnaprasad}.}
  \bibinfo{year}{1993}\natexlab{}.
\newblock \showarticletitle{Orthogonal matching pursuit: Recursive function
  approximation with applications to wavelet decomposition}. In
  \bibinfo{booktitle}{\emph{Proceedings of 27th Asilomar conference on signals,
  systems and computers}}. IEEE, \bibinfo{pages}{40--44}.
\newblock


\bibitem[\protect\citeauthoryear{Pham and Venkatesh}{Pham and
  Venkatesh}{2013}]%
        {pham2013efficient}
\bibfield{author}{\bibinfo{person}{Duc-Son Pham} {and} \bibinfo{person}{Svetha
  Venkatesh}.} \bibinfo{year}{2013}\natexlab{}.
\newblock \showarticletitle{Efficient algorithms for robust recovery of images
  from compressed data}.
\newblock \bibinfo{journal}{\emph{IEEE transactions on image processing}}
  \bibinfo{volume}{22}, \bibinfo{number}{12} (\bibinfo{year}{2013}),
  \bibinfo{pages}{4724--4737}.
\newblock


\bibitem[\protect\citeauthoryear{Polania and Barner}{Polania and
  Barner}{2017}]%
        {polania2017exploiting}
\bibfield{author}{\bibinfo{person}{Luisa~F Polania} {and}
  \bibinfo{person}{Kenneth~E Barner}.} \bibinfo{year}{2017}\natexlab{}.
\newblock \showarticletitle{Exploiting restricted Boltzmann machines and deep
  belief networks in compressed sensing}.
\newblock \bibinfo{journal}{\emph{IEEE Transactions on Signal Processing}}
  \bibinfo{volume}{65}, \bibinfo{number}{17} (\bibinfo{year}{2017}),
  \bibinfo{pages}{4538--4550}.
\newblock


\bibitem[\protect\citeauthoryear{Qie, Hao, and Song}{Qie et~al\mbox{.}}{2020}]%
        {qie2020wireless}
\bibfield{author}{\bibinfo{person}{Youtian Qie}, \bibinfo{person}{Chuangbo
  Hao}, {and} \bibinfo{person}{Ping Song}.} \bibinfo{year}{2020}\natexlab{}.
\newblock \showarticletitle{Wireless Transmission Method for Large Data Based
  on Hierarchical Compressed Sensing and Sparse Decomposition}.
\newblock \bibinfo{journal}{\emph{Sensors}} \bibinfo{volume}{20},
  \bibinfo{number}{24} (\bibinfo{year}{2020}), \bibinfo{pages}{7146}.
\newblock


\bibitem[\protect\citeauthoryear{Qin, Schlemper, Caballero, Price, Hajnal, and
  Rueckert}{Qin et~al\mbox{.}}{2018}]%
        {qin2018convolutional}
\bibfield{author}{\bibinfo{person}{Chen Qin}, \bibinfo{person}{Jo Schlemper},
  \bibinfo{person}{Jose Caballero}, \bibinfo{person}{Anthony~N Price},
  \bibinfo{person}{Joseph~V Hajnal}, {and} \bibinfo{person}{Daniel Rueckert}.}
  \bibinfo{year}{2018}\natexlab{}.
\newblock \showarticletitle{Convolutional recurrent neural networks for dynamic
  MR image reconstruction}.
\newblock \bibinfo{journal}{\emph{IEEE transactions on medical imaging}}
  \bibinfo{volume}{38}, \bibinfo{number}{1} (\bibinfo{year}{2018}),
  \bibinfo{pages}{280--290}.
\newblock


\bibitem[\protect\citeauthoryear{Ronneberger, Fischer, and Brox}{Ronneberger
  et~al\mbox{.}}{2015}]%
        {ronneberger2015u}
\bibfield{author}{\bibinfo{person}{Olaf Ronneberger}, \bibinfo{person}{Philipp
  Fischer}, {and} \bibinfo{person}{Thomas Brox}.}
  \bibinfo{year}{2015}\natexlab{}.
\newblock \showarticletitle{U-net: Convolutional networks for biomedical image
  segmentation}. In \bibinfo{booktitle}{\emph{International Conference on
  Medical image computing and computer-assisted intervention}}. Springer,
  \bibinfo{pages}{234--241}.
\newblock


\bibitem[\protect\citeauthoryear{Rudin, Osher, and Fatemi}{Rudin
  et~al\mbox{.}}{1992}]%
        {rudin1992nonlinear}
\bibfield{author}{\bibinfo{person}{Leonid~I Rudin}, \bibinfo{person}{Stanley
  Osher}, {and} \bibinfo{person}{Emad Fatemi}.}
  \bibinfo{year}{1992}\natexlab{}.
\newblock \showarticletitle{Nonlinear total variation based noise removal
  algorithms}.
\newblock \bibinfo{journal}{\emph{Physica D: nonlinear phenomena}}
  \bibinfo{volume}{60}, \bibinfo{number}{1-4} (\bibinfo{year}{1992}),
  \bibinfo{pages}{259--268}.
\newblock


\bibitem[\protect\citeauthoryear{Schlemper, Caballero, Hajnal, Price, and
  Rueckert}{Schlemper et~al\mbox{.}}{2017}]%
        {schlemper2017deep}
\bibfield{author}{\bibinfo{person}{Jo Schlemper}, \bibinfo{person}{Jose
  Caballero}, \bibinfo{person}{Joseph~V Hajnal}, \bibinfo{person}{Anthony~N
  Price}, {and} \bibinfo{person}{Daniel Rueckert}.}
  \bibinfo{year}{2017}\natexlab{}.
\newblock \showarticletitle{A deep cascade of convolutional neural networks for
  dynamic MR image reconstruction}.
\newblock \bibinfo{journal}{\emph{IEEE transactions on Medical Imaging}}
  \bibinfo{volume}{37}, \bibinfo{number}{2} (\bibinfo{year}{2017}),
  \bibinfo{pages}{491--503}.
\newblock


\bibitem[\protect\citeauthoryear{Schniter, Rangan, and Fletcher}{Schniter
  et~al\mbox{.}}{2016}]%
        {schniter2016vector}
\bibfield{author}{\bibinfo{person}{Philip Schniter}, \bibinfo{person}{Sundeep
  Rangan}, {and} \bibinfo{person}{Alyson~K Fletcher}.}
  \bibinfo{year}{2016}\natexlab{}.
\newblock \showarticletitle{Vector approximate message passing for the
  generalized linear model}. In \bibinfo{booktitle}{\emph{2016 50th Asilomar
  Conference on Signals, Systems and Computers}}. IEEE,
  \bibinfo{pages}{1525--1529}.
\newblock


\bibitem[\protect\citeauthoryear{Sekine and Ikada}{Sekine and Ikada}{2019}]%
        {sekine2019lacsle}
\bibfield{author}{\bibinfo{person}{Masatoshi Sekine} {and}
  \bibinfo{person}{Satoshi Ikada}.} \bibinfo{year}{2019}\natexlab{}.
\newblock \showarticletitle{LACSLE: Lightweight and Adaptive Compressed Sensing
  Based on Deep Learning for Edge Devices}. In \bibinfo{booktitle}{\emph{2019
  IEEE Global Communications Conference (GLOBECOM)}}. IEEE,
  \bibinfo{pages}{1--7}.
\newblock


\bibitem[\protect\citeauthoryear{Shawky, Abd-Elnaby, Rihan, Nassar, El-Fishawy,
  and Abd El-Samie}{Shawky et~al\mbox{.}}{2017}]%
        {shawky2017efficient}
\bibfield{author}{\bibinfo{person}{Hala Shawky}, \bibinfo{person}{Mohammed
  Abd-Elnaby}, \bibinfo{person}{Mohamed Rihan}, \bibinfo{person}{MA Nassar},
  \bibinfo{person}{Adel~S El-Fishawy}, {and} \bibinfo{person}{Fathi~E Abd
  El-Samie}.} \bibinfo{year}{2017}\natexlab{}.
\newblock \showarticletitle{Efficient compression and reconstruction of speech
  signals using compressed sensing}.
\newblock \bibinfo{journal}{\emph{International Journal of Speech Technology}}
  \bibinfo{volume}{20}, \bibinfo{number}{4} (\bibinfo{year}{2017}),
  \bibinfo{pages}{851--857}.
\newblock


\bibitem[\protect\citeauthoryear{Shen, Han, Yang, Yang, Wang, Li, and Wen}{Shen
  et~al\mbox{.}}{2018}]%
        {shen2018cs}
\bibfield{author}{\bibinfo{person}{Yiran Shen}, \bibinfo{person}{Tao Han},
  \bibinfo{person}{Qing Yang}, \bibinfo{person}{Xu Yang}, \bibinfo{person}{Yong
  Wang}, \bibinfo{person}{Feng Li}, {and} \bibinfo{person}{Hongkai Wen}.}
  \bibinfo{year}{2018}\natexlab{}.
\newblock \showarticletitle{CS-CNN: Enabling robust and efficient convolutional
  neural networks inference for Internet-of-Things applications}.
\newblock \bibinfo{journal}{\emph{IEEE Access}}  \bibinfo{volume}{6}
  (\bibinfo{year}{2018}), \bibinfo{pages}{13439--13448}.
\newblock


\bibitem[\protect\citeauthoryear{{Shi}, {Jiang}, {Liu}, and {Zhao}}{{Shi}
  et~al\mbox{.}}{2020}]%
        {8765626}
\bibfield{author}{\bibinfo{person}{W. {Shi}}, \bibinfo{person}{F. {Jiang}},
  \bibinfo{person}{S. {Liu}}, {and} \bibinfo{person}{D. {Zhao}}.}
  \bibinfo{year}{2020}\natexlab{}.
\newblock \showarticletitle{Image Compressed Sensing Using Convolutional Neural
  Network}.
\newblock \bibinfo{journal}{\emph{IEEE Transactions on Image Processing}}
  \bibinfo{volume}{29} (\bibinfo{year}{2020}), \bibinfo{pages}{375--388}.
\newblock


\bibitem[\protect\citeauthoryear{Shrivastwa, Pudi, and
  Chattopadhyay}{Shrivastwa et~al\mbox{.}}{2018}]%
        {shrivastwa2018fpga}
\bibfield{author}{\bibinfo{person}{Ritu~Ranjan Shrivastwa},
  \bibinfo{person}{Vikramkumar Pudi}, {and} \bibinfo{person}{Anupam
  Chattopadhyay}.} \bibinfo{year}{2018}\natexlab{}.
\newblock \showarticletitle{An FPGA-based brain computer interfacing using
  compressive sensing and machine learning}. In \bibinfo{booktitle}{\emph{2018
  IEEE Computer Society Annual Symposium on VLSI (ISVLSI)}}. IEEE,
  \bibinfo{pages}{726--731}.
\newblock


\bibitem[\protect\citeauthoryear{Shrivastwa, Pudi, Duo, So, Chattopadhyay, and
  Cuntai}{Shrivastwa et~al\mbox{.}}{2020}]%
        {shrivastwa2020brain}
\bibfield{author}{\bibinfo{person}{Ritu~Ranjan Shrivastwa},
  \bibinfo{person}{Vikramkumar Pudi}, \bibinfo{person}{Chen Duo},
  \bibinfo{person}{Rosa So}, \bibinfo{person}{Anupam Chattopadhyay}, {and}
  \bibinfo{person}{Guan Cuntai}.} \bibinfo{year}{2020}\natexlab{}.
\newblock \showarticletitle{A Brain--Computer Interface Framework Based on
  Compressive Sensing and Deep Learning}.
\newblock \bibinfo{journal}{\emph{IEEE Consumer Electronics Magazine}}
  \bibinfo{volume}{9}, \bibinfo{number}{3} (\bibinfo{year}{2020}),
  \bibinfo{pages}{90--96}.
\newblock


\bibitem[\protect\citeauthoryear{Singhal, Majumdar, and Ward}{Singhal
  et~al\mbox{.}}{2017}]%
        {singhal2017semi}
\bibfield{author}{\bibinfo{person}{Vanika Singhal}, \bibinfo{person}{Angshul
  Majumdar}, {and} \bibinfo{person}{Rabab~K Ward}.}
  \bibinfo{year}{2017}\natexlab{}.
\newblock \showarticletitle{Semi-supervised deep blind compressed sensing for
  analysis and reconstruction of biomedical signals from compressive
  measurements}.
\newblock \bibinfo{journal}{\emph{IEEE Access}}  \bibinfo{volume}{6}
  (\bibinfo{year}{2017}), \bibinfo{pages}{545--553}.
\newblock


\bibitem[\protect\citeauthoryear{Song, Cao, Wu, Yan, and Zhang}{Song
  et~al\mbox{.}}{2020}]%
        {song2020learning}
\bibfield{author}{\bibinfo{person}{Yuhai Song}, \bibinfo{person}{Zhong Cao},
  \bibinfo{person}{Kailun Wu}, \bibinfo{person}{Ziang Yan}, {and}
  \bibinfo{person}{Changshui Zhang}.} \bibinfo{year}{2020}\natexlab{}.
\newblock \showarticletitle{Learning Fast Approximations of Sparse Nonlinear
  Regression}.
\newblock \bibinfo{journal}{\emph{arXiv preprint arXiv:2010.13490}}
  (\bibinfo{year}{2020}).
\newblock


\bibitem[\protect\citeauthoryear{{Sun}, {Feng}, {Chen}, and {Zhu}}{{Sun}
  et~al\mbox{.}}{2016}]%
        {7560597}
\bibfield{author}{\bibinfo{person}{B. {Sun}}, \bibinfo{person}{H. {Feng}},
  \bibinfo{person}{K. {Chen}}, {and} \bibinfo{person}{X. {Zhu}}.}
  \bibinfo{year}{2016}\natexlab{}.
\newblock \showarticletitle{A Deep Learning Framework of Quantized Compressed
  Sensing for Wireless Neural Recording}.
\newblock \bibinfo{journal}{\emph{IEEE Access}}  \bibinfo{volume}{4}
  (\bibinfo{year}{2016}), \bibinfo{pages}{5169--5178}.
\newblock


\bibitem[\protect\citeauthoryear{Sun, Li, Xu, et~al\mbox{.}}{Sun
  et~al\mbox{.}}{2016}]%
        {sun2016deep}
\bibfield{author}{\bibinfo{person}{Jian Sun}, \bibinfo{person}{Huibin Li},
  \bibinfo{person}{Zongben Xu}, {et~al\mbox{.}}}
  \bibinfo{year}{2016}\natexlab{}.
\newblock \showarticletitle{Deep ADMM-Net for compressive sensing MRI}.
\newblock \bibinfo{journal}{\emph{Advances in neural information processing
  systems}}  \bibinfo{volume}{29} (\bibinfo{year}{2016}),
  \bibinfo{pages}{10--18}.
\newblock


\bibitem[\protect\citeauthoryear{Sun, Fan, Huang, Ding, and Paisley}{Sun
  et~al\mbox{.}}{2018}]%
        {sun2018compressed}
\bibfield{author}{\bibinfo{person}{Liyan Sun}, \bibinfo{person}{Zhiwen Fan},
  \bibinfo{person}{Yue Huang}, \bibinfo{person}{Xinghao Ding}, {and}
  \bibinfo{person}{John~W Paisley}.} \bibinfo{year}{2018}\natexlab{}.
\newblock \showarticletitle{Compressed Sensing MRI Using a Recursive Dilated
  Network.}. In \bibinfo{booktitle}{\emph{AAAI}}. \bibinfo{pages}{2444--2451}.
\newblock


\bibitem[\protect\citeauthoryear{Tank and Hopfield}{Tank and Hopfield}{1986}]%
        {tank1986simple}
\bibfield{author}{\bibinfo{person}{Df Tank} {and} \bibinfo{person}{J
  Hopfield}.} \bibinfo{year}{1986}\natexlab{}.
\newblock \showarticletitle{Simple'neural'optimization networks: An A/D
  converter, signal decision circuit, and a linear programming circuit}.
\newblock \bibinfo{journal}{\emph{IEEE transactions on circuits and systems}}
  \bibinfo{volume}{33}, \bibinfo{number}{5} (\bibinfo{year}{1986}),
  \bibinfo{pages}{533--541}.
\newblock


\bibitem[\protect\citeauthoryear{Teerapittayanon, McDanel, and
  Kung}{Teerapittayanon et~al\mbox{.}}{2016}]%
        {teerapittayanon2016branchynet}
\bibfield{author}{\bibinfo{person}{Surat Teerapittayanon},
  \bibinfo{person}{Bradley McDanel}, {and} \bibinfo{person}{Hsiang-Tsung
  Kung}.} \bibinfo{year}{2016}\natexlab{}.
\newblock \showarticletitle{Branchynet: Fast inference via early exiting from
  deep neural networks}. In \bibinfo{booktitle}{\emph{2016 23rd International
  Conference on Pattern Recognition (ICPR)}}. IEEE,
  \bibinfo{pages}{2464--2469}.
\newblock


\bibitem[\protect\citeauthoryear{Thapaliya, Goluguri, and Suthaharan}{Thapaliya
  et~al\mbox{.}}{2020}]%
        {thapaliya2020asymptotically}
\bibfield{author}{\bibinfo{person}{Naseeb Thapaliya}, \bibinfo{person}{Lavanya
  Goluguri}, {and} \bibinfo{person}{Shan Suthaharan}.}
  \bibinfo{year}{2020}\natexlab{}.
\newblock \showarticletitle{Asymptotically Stable Privacy Protection Technique
  for fMRI Shared Data over Distributed Computer Networks}. In
  \bibinfo{booktitle}{\emph{Proceedings of the 11th ACM International
  Conference on Bioinformatics, Computational Biology and Health Informatics}}.
  \bibinfo{pages}{1--8}.
\newblock


\bibitem[\protect\citeauthoryear{Tibshirani}{Tibshirani}{1996}]%
        {tibshirani1996regression}
\bibfield{author}{\bibinfo{person}{Robert Tibshirani}.}
  \bibinfo{year}{1996}\natexlab{}.
\newblock \showarticletitle{Regression shrinkage and selection via the lasso}.
\newblock \bibinfo{journal}{\emph{Journal of the Royal Statistical Society:
  Series B (Methodological)}} \bibinfo{volume}{58}, \bibinfo{number}{1}
  (\bibinfo{year}{1996}), \bibinfo{pages}{267--288}.
\newblock


\bibitem[\protect\citeauthoryear{Vargas, Fonseca, and Arguello}{Vargas
  et~al\mbox{.}}{2018}]%
        {vargas2018object}
\bibfield{author}{\bibinfo{person}{Hector Vargas}, \bibinfo{person}{Yesid
  Fonseca}, {and} \bibinfo{person}{Henry Arguello}.}
  \bibinfo{year}{2018}\natexlab{}.
\newblock \showarticletitle{Object detection on compressive measurements using
  correlation filters and sparse representation}. In
  \bibinfo{booktitle}{\emph{2018 26th European Signal Processing Conference
  (EUSIPCO)}}. IEEE, \bibinfo{pages}{1960--1964}.
\newblock


\bibitem[\protect\citeauthoryear{Wang, Niu, and Fu}{Wang et~al\mbox{.}}{2019}]%
        {wang2019deterministic}
\bibfield{author}{\bibinfo{person}{Gang Wang}, \bibinfo{person}{Min-Yao Niu},
  {and} \bibinfo{person}{Fang-Wei Fu}.} \bibinfo{year}{2019}\natexlab{}.
\newblock \showarticletitle{Deterministic constructions of compressed sensing
  matrices based on codes}.
\newblock \bibinfo{journal}{\emph{Cryptography and Communications}}
  \bibinfo{volume}{11}, \bibinfo{number}{4} (\bibinfo{year}{2019}),
  \bibinfo{pages}{759--775}.
\newblock


\bibitem[\protect\citeauthoryear{Wang, Kwon, and Shim}{Wang
  et~al\mbox{.}}{2012}]%
        {wang2012generalized}
\bibfield{author}{\bibinfo{person}{Jian Wang}, \bibinfo{person}{Seokbeop Kwon},
  {and} \bibinfo{person}{Byonghyo Shim}.} \bibinfo{year}{2012}\natexlab{}.
\newblock \showarticletitle{Generalized orthogonal matching pursuit}.
\newblock \bibinfo{journal}{\emph{IEEE Transactions on signal processing}}
  \bibinfo{volume}{60}, \bibinfo{number}{12} (\bibinfo{year}{2012}),
  \bibinfo{pages}{6202--6216}.
\newblock


\bibitem[\protect\citeauthoryear{Wang, Su, Ying, Peng, Zhu, Liang, Feng, and
  Liang}{Wang et~al\mbox{.}}{2016}]%
        {wang2016accelerating}
\bibfield{author}{\bibinfo{person}{Shanshan Wang}, \bibinfo{person}{Zhenghang
  Su}, \bibinfo{person}{Leslie Ying}, \bibinfo{person}{Xi Peng},
  \bibinfo{person}{Shun Zhu}, \bibinfo{person}{Feng Liang},
  \bibinfo{person}{Dagan Feng}, {and} \bibinfo{person}{Dong Liang}.}
  \bibinfo{year}{2016}\natexlab{}.
\newblock \showarticletitle{Accelerating magnetic resonance imaging via deep
  learning}. In \bibinfo{booktitle}{\emph{2016 IEEE 13th International
  Symposium on Biomedical Imaging (ISBI)}}. IEEE, \bibinfo{pages}{514--517}.
\newblock


\bibitem[\protect\citeauthoryear{Wang, Lin, Zhao, Yue, Meng, and Leung}{Wang
  et~al\mbox{.}}{2017}]%
        {wang2017compressive}
\bibfield{author}{\bibinfo{person}{Yao Wang}, \bibinfo{person}{Lin Lin},
  \bibinfo{person}{Qian Zhao}, \bibinfo{person}{Tianwei Yue},
  \bibinfo{person}{Deyu Meng}, {and} \bibinfo{person}{Yee Leung}.}
  \bibinfo{year}{2017}\natexlab{}.
\newblock \showarticletitle{Compressive sensing of hyperspectral images via
  joint tensor tucker decomposition and weighted total variation
  regularization}.
\newblock \bibinfo{journal}{\emph{IEEE Geoscience and Remote Sensing Letters}}
  \bibinfo{volume}{14}, \bibinfo{number}{12} (\bibinfo{year}{2017}),
  \bibinfo{pages}{2457--2461}.
\newblock


\bibitem[\protect\citeauthoryear{Wimalajeewa and Varshney}{Wimalajeewa and
  Varshney}{2017}]%
        {wimalajeewa2017application}
\bibfield{author}{\bibinfo{person}{Thakshila Wimalajeewa} {and}
  \bibinfo{person}{Pramod~K Varshney}.} \bibinfo{year}{2017}\natexlab{}.
\newblock \showarticletitle{Application of compressive sensing techniques in
  distributed sensor networks: A survey}.
\newblock \bibinfo{journal}{\emph{arXiv preprint arXiv:1709.10401}}
  (\bibinfo{year}{2017}).
\newblock


\bibitem[\protect\citeauthoryear{Wu, Brooks, Chen, Chen, Choudhury, Dukhan,
  Hazelwood, Isaac, Jia, Jia, et~al\mbox{.}}{Wu et~al\mbox{.}}{2019a}]%
        {wu2019machine}
\bibfield{author}{\bibinfo{person}{Carole-Jean Wu}, \bibinfo{person}{David
  Brooks}, \bibinfo{person}{Kevin Chen}, \bibinfo{person}{Douglas Chen},
  \bibinfo{person}{Sy Choudhury}, \bibinfo{person}{Marat Dukhan},
  \bibinfo{person}{Kim Hazelwood}, \bibinfo{person}{Eldad Isaac},
  \bibinfo{person}{Yangqing Jia}, \bibinfo{person}{Bill Jia}, {et~al\mbox{.}}}
  \bibinfo{year}{2019}\natexlab{a}.
\newblock \showarticletitle{Machine learning at facebook: Understanding
  inference at the edge}. In \bibinfo{booktitle}{\emph{2019 IEEE International
  Symposium on High Performance Computer Architecture (HPCA)}}. IEEE,
  \bibinfo{pages}{331--344}.
\newblock


\bibitem[\protect\citeauthoryear{Wu, Guo, Li, and Zhang}{Wu
  et~al\mbox{.}}{2019b}]%
        {wu2019sparse}
\bibfield{author}{\bibinfo{person}{Kailun Wu}, \bibinfo{person}{Yiwen Guo},
  \bibinfo{person}{Ziang Li}, {and} \bibinfo{person}{Changshui Zhang}.}
  \bibinfo{year}{2019}\natexlab{b}.
\newblock \showarticletitle{Sparse Coding with Gated Learned ISTA}. In
  \bibinfo{booktitle}{\emph{International Conference on Learning
  Representations}}.
\newblock


\bibitem[\protect\citeauthoryear{Xiao, Li, Yan, and Zhuang}{Xiao
  et~al\mbox{.}}{2019}]%
        {xiao2019compressed}
\bibfield{author}{\bibinfo{person}{Shuo Xiao}, \bibinfo{person}{Tianxu Li},
  \bibinfo{person}{Yan Yan}, {and} \bibinfo{person}{Jiayu Zhuang}.}
  \bibinfo{year}{2019}\natexlab{}.
\newblock \showarticletitle{Compressed sensing in wireless sensor networks
  under complex conditions of Internet of things}.
\newblock \bibinfo{journal}{\emph{Cluster Computing}} \bibinfo{volume}{22},
  \bibinfo{number}{6} (\bibinfo{year}{2019}), \bibinfo{pages}{14145--14155}.
\newblock


\bibitem[\protect\citeauthoryear{Xu and Ren}{Xu and Ren}{2016}]%
        {xu2016csvideonet}
\bibfield{author}{\bibinfo{person}{Kai Xu} {and} \bibinfo{person}{Fengbo Ren}.}
  \bibinfo{year}{2016}\natexlab{}.
\newblock \showarticletitle{CSvideonet: A recurrent convolutional neural
  network for compressive sensing video reconstruction}.
\newblock \bibinfo{journal}{\emph{arXiv preprint arXiv:1612.05203}}
  (\bibinfo{year}{2016}).
\newblock


\bibitem[\protect\citeauthoryear{Xu, Liu, and Kelly}{Xu et~al\mbox{.}}{2020}]%
        {xu2020compressed}
\bibfield{author}{\bibinfo{person}{Yibo Xu}, \bibinfo{person}{Weidi Liu}, {and}
  \bibinfo{person}{Kevin~F Kelly}.} \bibinfo{year}{2020}\natexlab{}.
\newblock \showarticletitle{Compressed Domain Image Classification Using a
  Dynamic-Rate Neural Network}.
\newblock \bibinfo{journal}{\emph{IEEE Access}}  \bibinfo{volume}{8}
  (\bibinfo{year}{2020}), \bibinfo{pages}{217711--217722}.
\newblock


\bibitem[\protect\citeauthoryear{Xuan and Loffeld}{Xuan and Loffeld}{2018}]%
        {xuan2018deep}
\bibfield{author}{\bibinfo{person}{Vinh~Nguyen Xuan} {and}
  \bibinfo{person}{Otmar Loffeld}.} \bibinfo{year}{2018}\natexlab{}.
\newblock \showarticletitle{A deep learning framework for compressed learning
  and signal reconstruction}. In \bibinfo{booktitle}{\emph{5th International
  Workshop on Compressed Sensing applied to Radar, Multimodal Sensing, and
  Imaging (CoSeRa)}}. \bibinfo{pages}{1--5}.
\newblock


\bibitem[\protect\citeauthoryear{{Yang}, {Yu}, {Dong}, {Slabaugh}, {Dragotti},
  {Ye}, {Liu}, {Arridge}, {Keegan}, {Guo}, and {Firmin}}{{Yang}
  et~al\mbox{.}}{2018}]%
        {8233175}
\bibfield{author}{\bibinfo{person}{G. {Yang}}, \bibinfo{person}{S. {Yu}},
  \bibinfo{person}{H. {Dong}}, \bibinfo{person}{G. {Slabaugh}},
  \bibinfo{person}{P.~L. {Dragotti}}, \bibinfo{person}{X. {Ye}},
  \bibinfo{person}{F. {Liu}}, \bibinfo{person}{S. {Arridge}},
  \bibinfo{person}{J. {Keegan}}, \bibinfo{person}{Y. {Guo}}, {and}
  \bibinfo{person}{D. {Firmin}}.} \bibinfo{year}{2018}\natexlab{}.
\newblock \showarticletitle{DAGAN: Deep De-Aliasing Generative Adversarial
  Networks for Fast Compressed Sensing MRI Reconstruction}.
\newblock \bibinfo{journal}{\emph{IEEE Transactions on Medical Imaging}}
  \bibinfo{volume}{37}, \bibinfo{number}{6} (\bibinfo{year}{2018}),
  \bibinfo{pages}{1310--1321}.
\newblock


\bibitem[\protect\citeauthoryear{Yang, Liu, Chen, and Tong}{Yang
  et~al\mbox{.}}{2019}]%
        {yang2019federated}
\bibfield{author}{\bibinfo{person}{Qiang Yang}, \bibinfo{person}{Yang Liu},
  \bibinfo{person}{Tianjian Chen}, {and} \bibinfo{person}{Yongxin Tong}.}
  \bibinfo{year}{2019}\natexlab{}.
\newblock \showarticletitle{Federated machine learning: Concept and
  applications}.
\newblock \bibinfo{journal}{\emph{ACM Transactions on Intelligent Systems and
  Technology (TIST)}} \bibinfo{volume}{10}, \bibinfo{number}{2}
  (\bibinfo{year}{2019}), \bibinfo{pages}{1--19}.
\newblock


\bibitem[\protect\citeauthoryear{Yang, Sun, Li, and Xu}{Yang
  et~al\mbox{.}}{2017}]%
        {yang2017admm}
\bibfield{author}{\bibinfo{person}{Yan Yang}, \bibinfo{person}{Jian Sun},
  \bibinfo{person}{Huibin Li}, {and} \bibinfo{person}{Zongben Xu}.}
  \bibinfo{year}{2017}\natexlab{}.
\newblock \showarticletitle{ADMM-Net: A deep learning approach for compressive
  sensing MRI}.
\newblock \bibinfo{journal}{\emph{arXiv preprint arXiv:1705.06869}}
  (\bibinfo{year}{2017}).
\newblock


\bibitem[\protect\citeauthoryear{{Yang}, {Sun}, {Li}, and {Xu}}{{Yang}
  et~al\mbox{.}}{2020}]%
        {8550778}
\bibfield{author}{\bibinfo{person}{Y. {Yang}}, \bibinfo{person}{J. {Sun}},
  \bibinfo{person}{H. {Li}}, {and} \bibinfo{person}{Z. {Xu}}.}
  \bibinfo{year}{2020}\natexlab{}.
\newblock \showarticletitle{ADMM-CSNet: A Deep Learning Approach for Image
  Compressive Sensing}.
\newblock \bibinfo{journal}{\emph{IEEE Transactions on Pattern Analysis and
  Machine Intelligence}} \bibinfo{volume}{42}, \bibinfo{number}{3}
  (\bibinfo{year}{2020}), \bibinfo{pages}{521--538}.
\newblock


\bibitem[\protect\citeauthoryear{Yao, Dai, Zhang, Zhang, Tian, and Xu}{Yao
  et~al\mbox{.}}{2019}]%
        {yao2019dr2}
\bibfield{author}{\bibinfo{person}{Hantao Yao}, \bibinfo{person}{Feng Dai},
  \bibinfo{person}{Shiliang Zhang}, \bibinfo{person}{Yongdong Zhang},
  \bibinfo{person}{Qi Tian}, {and} \bibinfo{person}{Changsheng Xu}.}
  \bibinfo{year}{2019}\natexlab{}.
\newblock \showarticletitle{Dr2-net: Deep residual reconstruction network for
  image compressive sensing}.
\newblock \bibinfo{journal}{\emph{Neurocomputing}}  \bibinfo{volume}{359}
  (\bibinfo{year}{2019}), \bibinfo{pages}{483--493}.
\newblock


\bibitem[\protect\citeauthoryear{Yao, Li, Liu, Wang, Liu, Shao, and
  Abdelzaher}{Yao et~al\mbox{.}}{2020}]%
        {yao2020deep}
\bibfield{author}{\bibinfo{person}{Shuochao Yao}, \bibinfo{person}{Jinyang Li},
  \bibinfo{person}{Dongxin Liu}, \bibinfo{person}{Tianshi Wang},
  \bibinfo{person}{Shengzhong Liu}, \bibinfo{person}{Huajie Shao}, {and}
  \bibinfo{person}{Tarek Abdelzaher}.} \bibinfo{year}{2020}\natexlab{}.
\newblock \showarticletitle{Deep compressive offloading: speeding up neural
  network inference by trading edge computation for network latency}. In
  \bibinfo{booktitle}{\emph{Proceedings of the 18th Conference on Embedded
  Networked Sensor Systems}}. \bibinfo{pages}{476--488}.
\newblock


\bibitem[\protect\citeauthoryear{Yin, Yu, and Wang}{Yin et~al\mbox{.}}{2016}]%
        {yin2016compressive}
\bibfield{author}{\bibinfo{person}{Ming Yin}, \bibinfo{person}{Kai Yu}, {and}
  \bibinfo{person}{Zhi Wang}.} \bibinfo{year}{2016}\natexlab{}.
\newblock \showarticletitle{Compressive sensing based sampling and
  reconstruction for wireless sensor array network}.
\newblock \bibinfo{journal}{\emph{Mathematical Problems in Engineering}}
  \bibinfo{volume}{2016} (\bibinfo{year}{2016}).
\newblock


\bibitem[\protect\citeauthoryear{Yu, Yang, Xu, Yang, and Huang}{Yu
  et~al\mbox{.}}{2018}]%
        {yu2018slimmable}
\bibfield{author}{\bibinfo{person}{Jiahui Yu}, \bibinfo{person}{Linjie Yang},
  \bibinfo{person}{Ning Xu}, \bibinfo{person}{Jianchao Yang}, {and}
  \bibinfo{person}{Thomas Huang}.} \bibinfo{year}{2018}\natexlab{}.
\newblock \showarticletitle{Slimmable neural networks}.
\newblock \bibinfo{journal}{\emph{arXiv preprint arXiv:1812.08928}}
  (\bibinfo{year}{2018}).
\newblock


\bibitem[\protect\citeauthoryear{Yu, Dong, Yang, Slabaugh, Dragotti, Ye, Liu,
  Arridge, Keegan, Firmin, et~al\mbox{.}}{Yu et~al\mbox{.}}{2017}]%
        {yu2017deep}
\bibfield{author}{\bibinfo{person}{Simiao Yu}, \bibinfo{person}{Hao Dong},
  \bibinfo{person}{Guang Yang}, \bibinfo{person}{Greg Slabaugh},
  \bibinfo{person}{Pier~Luigi Dragotti}, \bibinfo{person}{Xujiong Ye},
  \bibinfo{person}{Fangde Liu}, \bibinfo{person}{Simon Arridge},
  \bibinfo{person}{Jennifer Keegan}, \bibinfo{person}{David Firmin},
  {et~al\mbox{.}}} \bibinfo{year}{2017}\natexlab{}.
\newblock \showarticletitle{Deep de-aliasing for fast compressive sensing MRI}.
\newblock \bibinfo{journal}{\emph{arXiv preprint arXiv:1705.07137}}
  (\bibinfo{year}{2017}).
\newblock


\bibitem[\protect\citeauthoryear{Zhang and Ghanem}{Zhang and Ghanem}{2018}]%
        {zhang2018ista}
\bibfield{author}{\bibinfo{person}{Jian Zhang} {and} \bibinfo{person}{Bernard
  Ghanem}.} \bibinfo{year}{2018}\natexlab{}.
\newblock \showarticletitle{ISTA-Net: Interpretable optimization-inspired deep
  network for image compressive sensing}. In
  \bibinfo{booktitle}{\emph{Proceedings of the IEEE conference on computer
  vision and pattern recognition}}. \bibinfo{pages}{1828--1837}.
\newblock


\bibitem[\protect\citeauthoryear{Zhang, Lian, Yang, and Su}{Zhang
  et~al\mbox{.}}{2020b}]%
        {zhang2020deep}
\bibfield{author}{\bibinfo{person}{Xiaohua Zhang}, \bibinfo{person}{Qiusheng
  Lian}, \bibinfo{person}{Yuchi Yang}, {and} \bibinfo{person}{Yueming Su}.}
  \bibinfo{year}{2020}\natexlab{b}.
\newblock \showarticletitle{A deep unrolling network inspired by total
  variation for compressed sensing MRI}.
\newblock \bibinfo{journal}{\emph{Digital Signal Processing}}
  \bibinfo{volume}{107} (\bibinfo{year}{2020}), \bibinfo{pages}{102856}.
\newblock


\bibitem[\protect\citeauthoryear{Zhang, Li, Zhao, Lu, and Cavalcante}{Zhang
  et~al\mbox{.}}{2020a}]%
        {zhang2020signal}
\bibfield{author}{\bibinfo{person}{Yanliang Zhang}, \bibinfo{person}{Xingwang
  Li}, \bibinfo{person}{Guoying Zhao}, \bibinfo{person}{Bing Lu}, {and}
  \bibinfo{person}{Charles~C Cavalcante}.} \bibinfo{year}{2020}\natexlab{a}.
\newblock \showarticletitle{Signal reconstruction of compressed sensing based
  on alternating direction method of multipliers}.
\newblock \bibinfo{journal}{\emph{Circuits, Systems, and Signal Processing}}
  \bibinfo{volume}{39}, \bibinfo{number}{1} (\bibinfo{year}{2020}),
  \bibinfo{pages}{307--323}.
\newblock


\bibitem[\protect\citeauthoryear{Zhang, Liu, Liu, Wen, and Zhu}{Zhang
  et~al\mbox{.}}{2020c}]%
        {zhang2020amp}
\bibfield{author}{\bibinfo{person}{Zhonghao Zhang}, \bibinfo{person}{Yipeng
  Liu}, \bibinfo{person}{Jiani Liu}, \bibinfo{person}{Fei Wen}, {and}
  \bibinfo{person}{Ce Zhu}.} \bibinfo{year}{2020}\natexlab{c}.
\newblock \showarticletitle{AMP-Net: Denoising based Deep Unfolding for
  Compressive Image Sensing}.
\newblock \bibinfo{journal}{\emph{arXiv preprint arXiv:2004.10078}}
  (\bibinfo{year}{2020}).
\newblock


\bibitem[\protect\citeauthoryear{Zhang, Wu, Gan, and Zhu}{Zhang
  et~al\mbox{.}}{2019}]%
        {zhang2019optimally}
\bibfield{author}{\bibinfo{person}{Zufan Zhang}, \bibinfo{person}{Yunfeng Wu},
  \bibinfo{person}{Chenquan Gan}, {and} \bibinfo{person}{Qingyi Zhu}.}
  \bibinfo{year}{2019}\natexlab{}.
\newblock \showarticletitle{The optimally designed autoencoder network for
  compressed sensing}.
\newblock \bibinfo{journal}{\emph{EURASIP Journal on Image and Video
  Processing}} \bibinfo{volume}{2019}, \bibinfo{number}{1}
  (\bibinfo{year}{2019}), \bibinfo{pages}{1--12}.
\newblock


\bibitem[\protect\citeauthoryear{Zhao, Liang, and Tang}{Zhao
  et~al\mbox{.}}{2019}]%
        {zhao2019deep}
\bibfield{author}{\bibinfo{person}{Chenkai Zhao}, \bibinfo{person}{Jing Liang},
  {and} \bibinfo{person}{Qin Tang}.} \bibinfo{year}{2019}\natexlab{}.
\newblock \showarticletitle{A Deep-Learning-Based Distributed Compressive
  Sensing in UWB Soil Signals}. In \bibinfo{booktitle}{\emph{International
  Conference in Communications, Signal Processing, and Systems}}. Springer,
  \bibinfo{pages}{1872--1879}.
\newblock


\bibitem[\protect\citeauthoryear{Zhao, Li, Zhang, Tao, and Ma}{Zhao
  et~al\mbox{.}}{2020a}]%
        {zhao2020adaptive}
\bibfield{author}{\bibinfo{person}{Xiaobin Zhao}, \bibinfo{person}{Wei Li},
  \bibinfo{person}{Mengmeng Zhang}, \bibinfo{person}{Ran Tao}, {and}
  \bibinfo{person}{Pengge Ma}.} \bibinfo{year}{2020}\natexlab{a}.
\newblock \showarticletitle{Adaptive Iterated Shrinkage Thresholding-Based
  Lp-Norm Sparse Representation for Hyperspectral Imagery Target Detection}.
\newblock \bibinfo{journal}{\emph{Remote Sensing}} \bibinfo{volume}{12},
  \bibinfo{number}{23} (\bibinfo{year}{2020}), \bibinfo{pages}{3991}.
\newblock


\bibitem[\protect\citeauthoryear{Zhao, Xie, Liu, and Pan}{Zhao
  et~al\mbox{.}}{2020b}]%
        {zhao2020hybrid}
\bibfield{author}{\bibinfo{person}{Zhifu Zhao}, \bibinfo{person}{Xuemei Xie},
  \bibinfo{person}{Wan Liu}, {and} \bibinfo{person}{Qingzhe Pan}.}
  \bibinfo{year}{2020}\natexlab{b}.
\newblock \showarticletitle{A Hybrid-3D Convolutional Network for Video
  Compressive Sensing}.
\newblock \bibinfo{journal}{\emph{IEEE Access}}  \bibinfo{volume}{8}
  (\bibinfo{year}{2020}), \bibinfo{pages}{20503--20513}.
\newblock


\bibitem[\protect\citeauthoryear{Zhou, Wu, Wu, and Zhou}{Zhou
  et~al\mbox{.}}{2015}]%
        {zhou2015exploiting}
\bibfield{author}{\bibinfo{person}{Shuchang Zhou}, \bibinfo{person}{Jia-Nan
  Wu}, \bibinfo{person}{Yuxin Wu}, {and} \bibinfo{person}{Xinyu Zhou}.}
  \bibinfo{year}{2015}\natexlab{}.
\newblock \showarticletitle{Exploiting local structures with the kronecker
  layer in convolutional networks}.
\newblock \bibinfo{journal}{\emph{arXiv preprint arXiv:1512.09194}}
  (\bibinfo{year}{2015}).
\newblock


\bibitem[\protect\citeauthoryear{Zur and Adler}{Zur and Adler}{2019}]%
        {zur2019deep}
\bibfield{author}{\bibinfo{person}{Yochai Zur} {and} \bibinfo{person}{Amir
  Adler}.} \bibinfo{year}{2019}\natexlab{}.
\newblock \showarticletitle{Deep Learning of Compressed Sensing Operators with
  Structural Similarity Loss}.
\newblock \bibinfo{journal}{\emph{arXiv preprint arXiv:1906.10411}}
  (\bibinfo{year}{2019}).
\newblock


\end{thebibliography}


\end{document}